\documentclass[aps,rmp,square]{revtex4}

\usepackage{amsmath,amssymb}
\usepackage{graphicx}
\usepackage{pst-plot}
\usepackage{mathrsfs}

\begin{document}

\setcitestyle{numbers}

\def\cs#1#2{#1_{\!{}_#2}}
\def\css#1#2#3{#1^{#2}_{\!{}_#3}}
\def\ocite#1{[\citenum{#1}]}
\def\ket#1{|#1\rangle}
\def\bra#1{\langle#1|}
\def\expac#1{\langle#1\rangle}
\def\dbl{\hbox{${1\hskip -2.4pt{\rm l}}$}}
\def\bfh#1{\bf{\hat#1}}

\newenvironment{rcase}
    {\left.\begin{aligned}}
    {\end{aligned}\right\rbrace}

\title{\sf{Disproofs of Bell, GHZ, and Hardy Type Theorems and the ${\rm I}$llusion of Entanglement}}

\author{Joy Christian}

\email{joy.christian@wolfson.ox.ac.uk}

\affiliation{Department of Physics, University of Oxford, Parks Road, Oxford OX1 3PU, United Kingdom}

\begin{abstract}
An elementary topological error in Bell's representation of the EPR elements of reality is identified.
Once recognized, it leads to a topologically correct local-realistic framework that provides exact,
deterministic, and local underpinning of at least the Bell, GHZ-3, GHZ-4, and Hardy states. The
correlations exhibited by these states are shown to be exactly the classical correlations
among the points of a 3 or 7-sphere, both of which are closed under multiplication, and hence preserve
the locality condition of Bell. The alleged non-localities of these states are thus shown to result
from misidentified topologies of the EPR elements of reality. When topologies are correctly identified,
local-realistic completion of any arbitrary entangled state is always guaranteed in our framework.
This vindicates EPR, and entails that quantum entanglement is best understood as an illusion.\break
\end{abstract}

\maketitle

\vspace{-1.26cm}

\parskip 5pt

\section{\sf{Introduction}}

No-go theorems in physics are often founded on unjustified, if tacit assumptions, and Bell's theorem is no exception.
It is no different, in this respect, from von Neumann's theorem rejecting all hidden variables \ocite{von},
or Coleman-Mandula theorem neglecting supersymmetry \ocite{Coleman}. Despite being in plain sight, the unjustified
assumptions underlying the latter two theorems seemed so innocuous to many that they escaped detection for decades.
In the case of Coleman-Mandula theorem---which concerned combining spacetime and internal symmetries---it took a
truly imaginative development of supersymmetry to finally bring about recognition of its limited veracity.
In the curious case of von Neumann's theorem, however, even an explicit counterexample---namely, the pilot wave
theory \ocite{Bohm-1927}\ocite{Bohm-1952}---did not
discourage a series of similarly misguided ``impossibility proofs'' for decades
\ocite{Bell-1982}. Thus ensued over half a century of false belief that no such completion of quantum mechanics is
possible, even in principle. Unfortunately, as is evident from the widespread belief in Bell's theorem, such examples
of institutionalized denial are not confined to the history of physics. Just as in the premises of von
Neumann and Coleman-Mandula theorems, the unjustified assumption underlying Bell's theorem is also in plain sight---in
the very first equation of Bell's paper \ocite{Bell-1964}---and yet it has received little attention. As innocent as
this equation may seem, it amounts to assuming incorrect topology for the EPR elements of reality \ocite{EPR}.
The aim of the present paper is to bring out this topological error explicitly, and demonstrate that---once recognized
and corrected---it gives way to an intrinsically local and manifestly realistic underpinning of the EPR-type
correlations, thereby providing explicit counterexamples to Bell's theorem and several of its variants
\ocite{Christian}\ocite{Further}\ocite{experiment}\ocite{GHSZ}\ocite{Hardy}.

To this end, recall that Bell begins his theorem by postulating a set of local functions ${A(\,\cdot\;,\,\cdot\,)}$,
which are equal to the numbers ${+1}$ or ${-1}$ once a unit vector ${\bf n}$ and a ``complete'' state ${\lambda}$ are
specified \ocite{Bell-1964}. He writes these functions as
\begin{equation}
A({\bf n},\,\lambda)\,=\,\pm\,1\,\in\,\{-1,\,+1\}\,\subset\,{\rm I\!R}\,,\label{Bell's}
\end{equation}
and takes them to represent the results of measuring spin components along the direction ${\bf n}$, or detecting
photons through a filter along the direction ${\bf n}$. As innocent as this equation may appear, it amounts to
presuming incorrect topology for the EPR elements of physical reality. This topological error is further
obscured by Bell in the probabilistic reformulation of his theorem, where the above function is expressed as a purely
probabilistic statement of obtaining measurement results \ocite{Bell-La}. To recognize the seriousness of this
error, let us rewrite Bell's local function as a map
\begin{equation}
A_{\bf n}(\lambda): {\rm I\!R}^3\!\times\Lambda\longrightarrow S^0,\label{Bell-map}
\end{equation}
where ${{\rm I\!R}^3}$ is the real
space of 3-vectors, ${\Lambda}$ is a space of ``complete'' states, and ${S^0}$ is a unit 0-sphere.
Now recall that a unit ${k}$-sphere, ${S^k}$, is a set of points forming a compact topological space without boundary
\ocite{Munkres}. It can be${\;}$understood as a one-point compactification, ${{\rm I\!R}^{k}\cup\{\infty\}}$, of a
non-compact
Euclidean space ${{\rm I\!R}^{k}}$. Alternatively it can be understood as a boundary of a ${k+1-}$ball, ${B^{k+1}}$,
within the higher dimensional space ${{\rm I\!R}^{k+1}}$. The sphere ${S^k}$ is then said to be embedded in
${{\rm I\!R}^{k+1}}$, and its points can be parameterized by a set of coordinates ${\{n_0,\,n_1,\,\dots,\,n_k\}}$ in
${{\rm I\!R}^{k+1}}$, satisfying
\begin{equation}
n^2_0+n^2_1+n^2_2+n^2_3+\dots+n^2_k\;=\;1.\label{consnum}
\end{equation}
Thus, a unit 0-sphere is a set of only two points, ${S^0\equiv\{-1,\,+1\}}$, whereas a unit 2-sphere, ${S^2}$, is
the boundary of a 3-ball whose antipodal points are the points ${\{-1,\,+1\}}$. In other words, a unit 0-sphere is a
great-great circle of a unit 2-sphere, which in turn is a collection of all such great-great circles forming its
surface. Thus, although ${S^0}$ and ${S^2}$ are both sets of binary numbers, ${+1}$'s and ${-1}$'s, their topological
properties are fundamentally different from one another; ${S^0}$ is parallelizable but disconnected, whereas ${S^2}$
is connected as well as simply connected, but not parallelizable. Hence Bell's postulate of equation (\ref{Bell's})
amounts to an implicit assumption of a specific topology for the EPR elements of reality. In what follows, we shall be
concerned mainly with the topologies of the spheres ${S^0}$, ${S^1}$, ${S^2}$, ${S^3}$, and ${S^7}$, each of which is a
set of binary numbers parameterized by Eq.${\,}$(\ref{consnum}), but with very different topologies from one another.
Thus, for example, the 1-sphere, ${S^1}$, is connected and parallelizable, but not simply connected. The spheres ${S^3}$
and ${S^7}$, on the other hand, are not only connected and parallelizable, but also simply connected. The crucial point
here is that---since the topological properties of different spheres are dramatically different from one another---mistaking
the points of one of them for the points of another is a serious error. But that is precisely what Bell has done.

To be sure, Bell's motivation for choosing ${S^0}$ for his purposes is quite understandable. It is trivially a space
closed under multiplication of its points, and hence naturally hosts the condition of locality he wished to
implement \ocite{Clauser-Shimony}:
\begin{equation}
(A_{\bf a}\,B_{\bf b})(\lambda)\,=\,A_{\bf a}(\lambda)\,B_{\bf b}(\lambda)\,\in\,S^0,\;\;
\forall\;\,A_{\bf a}(\lambda),\,B_{\bf b}(\lambda)\,\in\,S^0. \label{conlocbell}
\end{equation}
In other words, apart from the operational requirements of quantum mechanics, it is the requirement of locality---or
factorizability \ocite{Bell-La}---that motivated Bell to demand that the joint ``beable''
${(A_{\bf a}\,B_{\bf b})(\lambda)}$ satisfies the multiplication${\;}$map
\begin{equation}
(A_{\bf a}\,B_{\bf b})(\lambda):\,S^0\times\,S^0\,\longrightarrow\,S^0, \label{locconbell}
\end{equation}
because the converse of this map implies that any point of ${S^0}$ can be factorized into two or more points of the same
set. But, as we shall see, ${S^0}$ is only one of the many possible topological spaces that can satisfy this demand
of locality, with ${S^3}$---not ${S^0}$---being the correct choice for any two-level system.
That is to say, the group multiplication${\;}$map
\begin{equation}
({\mathscr A}_{\bf a}\,{\mathscr B}_{\bf b})(\lambda):\,S^3\times\,S^3\,\longrightarrow\,S^3,
\;\;\;\;{\rm implying}\;\;\;\;
({\mathscr A}_{\bf a}\,{\mathscr B}_{\bf b})(\lambda)\,=\,
{\mathscr A}_{\bf a}(\lambda)\,{\mathscr B}_{\bf b}(\lambda)\,\in\,S^3\,\;\;
\forall\;\,{\mathscr A}_{\bf a}(\lambda),\,{\mathscr B}_{\bf b}(\lambda)\,\in\,S^3,
\label{loclucbell}
\end{equation}
is at least as good a map as the one considered by Bell for his purposes.
In fact, {\it any} topological group \ocite{Frankel}, with measurement results as group elements, would be a perfectly
good target space in Eq.${\,}$(\ref{Bell-map}) for a locally causal theory.

More specifically, in what follows we shall demonstrate that an exact, deterministic, local, and realistic model
for the EPR correlations based on ${S^3}$ already exists \ocite{Christian}, and that it affords natural generalizations
to the rotationally non-invariant entangled states considered by GHZ and Hardy \ocite{GHSZ}\ocite{Hardy}. In particular,
within a representation-independent generalization of the local-realistic framework considered in
Refs.${\,}$\ocite{Christian}\ocite{Further}\ocite{experiment}, we shall derive the following results {\it exactly}${\,}$:
\begin{enumerate}
\item[(1)] the exact quantum mechanical expectation value for the singlet state:
${{\cal E}({\bf a},\,{\bf b})\,=\,-\,{\bf a}\cdot{\bf b}}$\,;
\item[(2)] the exact violations of Bell-CHSH inequalities:
${\,-\,2\sqrt{2}\,\;\leq\;{\cal E}({\bf a},\,{\bf b})+{\cal E}({\bf a},\,{\bf b'})+
{\cal E}({\bf a'},\,{\bf b})-{\cal E}({\bf a'},\,{\bf b'})\;\leq\;+\,2\sqrt{2}}$\;;
\item[(3)] all sixteen predictions of the Hardy state, such as
${\;\,\langle\Psi_{\bf z}\,|\,{\bf a'},\,+\rangle_1\,\otimes\,|{\bf b}\,,\,+\rangle_2\,=\,0}$\,,
\item[]${\,\;\;\;\;\;\;\;\;
\;\;\;\;\;\;\;\;\;\;\;\;\;\;\;\;\;\;\;\;\;\;\;\;\;\;\;\;\;\;\;\;\;\;\;\;
\;\;\;\;\;\;\;\;\;\;\;\;\;\;\;\;\;\;\;\;\;\;\;\;\;\;\;\;\;\;\;\;\;\;\;\;
\langle\Psi_{\bf z}\,|\,{\bf a}\,,\,+\rangle_1\,\otimes\,|{\bf b'},\,+\rangle_2\,=\,0}$\,,
\item[]${\,\;\;\;\;\;\;\;\;\;\;\;\;\;\;\;\;\;\;\;\;\;\;\;\;\;\;\;\;\;\;\;\;\;
\;\;\;\;\;\;\;\;\;\;\;\;\;\;\;\;\;\;\;\;\;\;\;\;\;\;\;\;\;\;\;\;\;\;\;\;
\;\;\;\;\;\;\;\;\;\;\;\langle\Psi_{\bf z}\,|\,{\bf a}\,,\,-\rangle_1\,\otimes\,|{\bf b}\,,\,-\rangle_2\;=\,0}$\,,
\item[]${\,\;\;\;\;\;\;\;\;\;\;\;\;\;\;\;\;\;\;\;\;\;\;\;\;\;\;\;\;\;\;\;\;\;
\;\;\;\;\;\;\;\;\;\;\;\;\;\;\;\;\;\;\;\;\;\;\;\;\;\;\;\;\;\;\;\;\;\;\;\;
\;\;\;{\rm but}\,\;\;\langle\Psi_{\bf z}\,|\,{\bf a'},\,+\rangle_1\,\otimes\,|{\bf b'},\,+\rangle_2\,=\,
\frac{\,\sin\theta\,\cos^2\theta}{\sqrt{1\,+\,\cos^2\theta\,}\,}\,\not=\,0}$\,;
\item[(4)] the exact quantum mechanical expectation value for the three-particle GHZ state:
\item[]
${\;\;\;\;\;\;{\cal E}({\bf n}_1,\,{\bf n}_2,\,{\bf n}_3)\,=\,
\cos\alpha\,\cos\theta_1\,\cos\theta_2\,\cos\theta_3\,+\,\sin\alpha\,\sin\theta_1\,
\sin\theta_2\,\sin\theta_3\,\cos\,\left(\,\phi_1\,+\,\phi_2\,+\,\phi_3\,+\,\delta\,\right)}$\,; \;and
\item[(5)] the exact quantum mechanical expectation value for the four-particle GHZ state:
\item[]
${\;\;\;\;\;\;{\cal E}({\bf n}_1,\,{\bf n}_2,\,{\bf n}_3,\,{\bf n}_4)\,=\,
\cos\theta_1\,\cos\theta_2\,\cos\theta_3\,\cos\theta_4\,-\,\sin\theta_1\,
\sin\theta_2\,\sin\theta_3\,\sin\theta_4\,\cos\,\left(\,\phi_1\,+\,\phi_2\,-\,\phi_3\,-\,\phi_4\,\right)}$.
\end{enumerate}

As we shall see, all of these correlations are strictly local correlations between correctly identified EPR
elements of reality. In other words, contrary to what is mistakenly believed, these correlations are not
exclusive manifestations of some irreducible quantum mechanical effects, but purely local-realistic, topological
effects. In the first three cases of the Bell and Hardy states, the correct topological space of the EPR elements
of reality is a unit 3-sphere, and hence
the correlations exhibited by these states are correlations among the points of a unit 3-sphere. In the last two
cases of the GHZ states, on the other hand, the correct
topological space of the EPR elements of reality is a unit 7-sphere, and hence the correlations exhibited by these
states are correlations among the points of a unit 7-sphere. More generally, we will show that, once the topological
space of the EPR elements of reality is correctly identified, an exact local-realistic underpinning of any arbitrary
entangled state is always guaranteed within our framework. The correlations exhibited by such an arbitrary state are
thus understood as local-realistic correlations among the points of this topological space,
with quantum mechanics functioning merely as a useful tool. To appreciate these results fully, however,
it would be instructive to first review the argument by Einstein, Podolsky, and Rosen \ocite{EPR}.

\begin{figure}
\hrule
\scalebox{1.0}{
\begin{pspicture}(1.2,-2.5)(4.2,2.5)

\psline[linewidth=0.1mm,dotsize=3pt 4]{*-}(-2.51,0)(-2.5,0)

\psline[linewidth=0.1mm,dotsize=3pt 4]{*-}(7.2,0)(7.15,0)

\psline[linewidth=0.4mm,arrowinset=0.3,arrowsize=3pt 3,arrowlength=2]{->}(-2.5,0)(-3,1)

\psline[linewidth=0.4mm,arrowinset=0.3,arrowsize=3pt 3,arrowlength=2]{->}(-2.5,0)(-3,-1)

\psline[linewidth=0.4mm,arrowinset=0.3,arrowsize=3pt 3,arrowlength=2]{->}(7.2,0)(8.3,0.5)

\psline[linewidth=0.4mm,arrowinset=0.3,arrowsize=3pt 3,arrowlength=2]{->}(7.2,0)(7.4,1.3)

\put(-2.4,+0.45){{\large ${\bf 1}$}}

\put(6.8,+0.45){{\large ${\bf 2}$}}

\put(-3.3,1.3){{\Large ${\bf a}$}}

\put(-3.5,-1.7){{\Large ${\bf a'}$}}

\put(8.43,0.57){{\Large ${\bf b}$}}

\put(7.3,1.5){{\Large ${\bf b'}$}}

\put(1.8,-0.65){\large Source}

\put(0.6,-1.2){\large ${\pi^0\longrightarrow\,\gamma+\,e^{-}+\,e^{+}\,}$}

\put(1.05,0.5){\large Total spin = 0}

\psline[linewidth=0.3mm,linestyle=dashed](-2.47,0)(2.1,0)

\psline[linewidth=0.4mm,arrowinset=0.3,arrowsize=3pt 3,arrowlength=2]{->}(-0.2,0)(-0.3,0)

\psline[linewidth=0.3mm,linestyle=dashed](2.6,0)(7.2,0)

\psline[linewidth=0.4mm,arrowinset=0.3,arrowsize=3pt 3,arrowlength=2]{->}(4.9,0)(5.0,0)

\psline[linewidth=0.1mm,dotsize=5pt 4]{*-}(2.35,0)(2.4,0)

\pscircle[linewidth=0.3mm,linestyle=dashed](7.2,0){1.3}

\psellipse[linewidth=0.2mm,linestyle=dashed](7.2,0)(1.28,0.3)

\pscircle[linewidth=0.3mm,linestyle=dashed](-2.51,0){1.3}

\psellipse[linewidth=0.2mm,linestyle=dashed](-2.51,0)(1.28,0.3)

\end{pspicture}}
\hrule
\caption{A spin-less neutral pion decays into an electron-positron pair. Measurements of spin components on each
separated fermion are performed at remote stations ${\bf 1}$ and ${\bf 2}$, providing binary outcomes (respectively)
along arbitrary directions ${\bf a}$ and ${\bf b}$.\break}
\hrule
\end{figure}
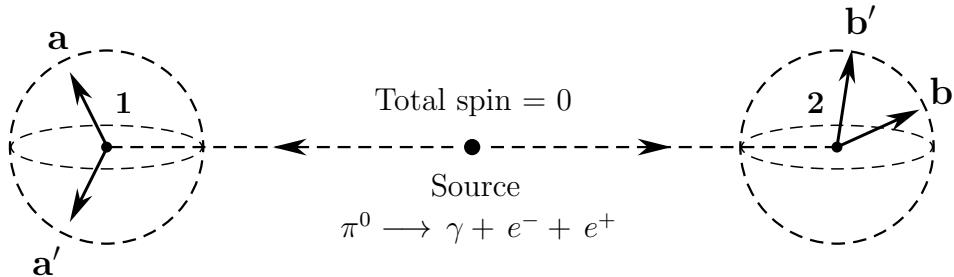

\section{EPR elements of reality are points of a 2-sphere, not 0-sphere as Bell assumed}

As Clauser and Shimony rightly emphasize \ocite{Clauser-Shimony}, the reasoning of EPR
is impeccable once the phrase ``can predict'' in their reality criterion is understood in a non-anthropocentric
sense. According to this criterion, {\it If, without in any way disturbing a system, we can
predict with certainty (i.e., with probability equal to unity) the value of a physical quantity, then there
exists an element of physical reality corresponding to this physical quantity.} We shall follow
the logic of EPR adapted to Bohm's spin version of their argument, which can be summarized as follows \ocite{GHSZ}:
\vspace{-0.7cm}
\begin{center}
\item[] ${\,\;\;\;\;\;\;}${\bf (1)} ${\;}$QM ${\!\implies\!}$ Perfect Correlations
\item[] ${\;\;\;\;\;\;\;\;\;+\,\;\;}${\bf(2)} ${\;}$Adherence to Local Causality${\,\;\;\;\;\;\;\;}$
\item[] ${\,\;\;\;\;\;\;+\;\;\;}${\bf(3)} ${\;}$Criterion of Objective Reality${\;\;\;\;\,}$
\item[] ${\;\;+\;\;\;}${\bf(4)} ${\;}$Notion of a Complete Theory${\,}$
\item[] ${\;\!\implies\!\!}$ {\bf (5)} ${\;}$QM is an Incomplete Theory.${\;}$
\end{center}

In more detail, the EPR argument proceeds as follows. Consider the physical scenario depicted in
Fig.${\,}$1. Within quantum mechanics the physical state of
such an idealized spin system is described by the entangled singlet state
\begin{equation}
|\Psi_{\bf n}\rangle=\frac{1}{\sqrt{2}\,}\,\Bigl\{|{\bf n},\,+\rangle_1\otimes
|{\bf n},\,-\rangle_2\,-\,|{\bf n},\,-\rangle_1\otimes|{\bf n},\,+\rangle_2\Bigr\}\,,\label{single}
\end{equation}
where ${\bf n}$ indicates any unit direction in the physical space ${{\rm I\!R}^{3}}$;
${{\boldsymbol\sigma}\cdot{\bf n}\,|{\bf n},\,\pm\rangle\,=\,
\pm\,|{\bf n},\,\pm\rangle}$ describes the eigenstates in which the particles have
spin ``up'' or ``down'' in units of ${\hbar=2}$; and ${\boldsymbol\sigma}$ stands for the Pauli spin ``vector.''
Now this state has two remarkable properties. First, it happens to be rotationally invariant. That is to say, it
remains the same for {\it all} directions in space, denoted by the unit vector ${\bf n}$. Second, it entails
perfect spin correlations: If the component of spin along direction ${\bf n}$ is found to be ``up'' for particle
${\bf 1}$, then with certainty it will be found to be ``down'' for particle ${\bf 2}$, and vice versa. Consequently,
one can predict with certainty the result of measuring any component of spin of particle ${\bf 2}$ by previously
measuring the same component of spin of particle ${\bf 1}$. However, local causality
demands that measurements performed on particle ${\bf 1}$ cannot bring about any real change in the remotely
situated particle ${\bf 2}$. Therefore, according to the EPR criterion of physical reality, the chosen spin
component of particle ${\bf 2}$ is an element of physical reality. But this argument goes through for any
component of spin, and hence {\it all infinitely many of the spin components of particle} ${\bf 2}$ {\it are elements
of physical reality} (in the strictly objective, non-anthropocentric sense consistently maintained by
Einstein \ocite{Einstein-1948}). However, many of these elements of physical reality have no
counterpart in the
quantum mechanical description of the system, since---as is evident from (\ref{single})---there is no quantum state
of particle ${\bf 2}$ in which all components of its spin have definite values. Consequently, by the completeness criterion
of EPR---which states that ``every element of the physical reality must have a counterpart in the physical theory'',
quantum theory cannot be a complete theory, because at least in the present example it does not provide a complete
description of the physical reality. That is to say, the notion of quantum entanglement merely conceals our
lack of knowledge.

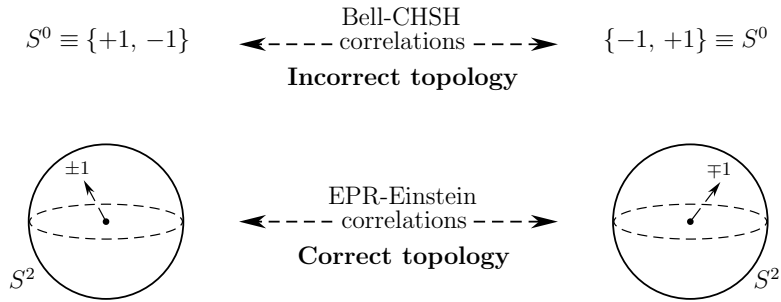
\begin{figure}
\hrule
\scalebox{0.8}{
\begin{pspicture}(1.2,-2.0)(4.2,4.4)

\psline[linewidth=0.2mm,linestyle=dashed,arrowinset=0.3,arrowsize=2pt 3,arrowlength=2]{*->}(-2.51,0)(-2.88,0.69)

\psline[linewidth=0.2mm,linestyle=dashed,arrowinset=0.3,arrowsize=2pt 3,arrowlength=2]{*->}(7.2,0)(7.7,0.66)

\put(-3.85,+2.9){{\large ${S^0\equiv}$}}

\put(-2.85,+2.9){{\large ${\{+1,\,-1\}}$}}

\put(5.75,+2.9){{\large ${\{-1,\,+1\}\equiv S^0}$}}

\put(-3.2,0.8){{${\pm1}$}}

\put(7.47,0.77){{${\mp1}$}}

\put(1.18,0.3){\large EPR-Einstein}

\put(1.4,-0.1){\large correlations}

\put(0.67,-0.7){\large {\bf Correct topology}}

\psline[linewidth=0.3mm,linestyle=dashed](-0.1,0)(1.25,0)

\psline[linewidth=0.4mm,arrowinset=0.3,arrowsize=3pt 3,arrowlength=2]{->}(-0.2,0)(-0.3,0)

\psline[linewidth=0.3mm,linestyle=dashed](3.6,0)(4.7,0)

\psline[linewidth=0.4mm,arrowinset=0.3,arrowsize=3pt 3,arrowlength=2]{->}(4.9,0)(5.0,0)

\put(1.4,3.27){\large Bell-CHSH}

\put(1.37,2.87){\large correlations}

\put(0.5,2.27){\large {\bf Incorrect topology}}

\psline[linewidth=0.3mm,linestyle=dashed](-0.1,2.95)(1.25,2.95)

\psline[linewidth=0.4mm,arrowinset=0.3,arrowsize=3pt 3,arrowlength=2]{->}(-0.2,2.95)(-0.3,2.95)

\psline[linewidth=0.3mm,linestyle=dashed](3.6,2.95)(4.7,2.95)

\psline[linewidth=0.4mm,arrowinset=0.3,arrowsize=3pt 3,arrowlength=2]{->}(4.9,2.95)(5.0,2.95)

\put(-4.125,-1.1){\large ${S^2}$}

\put(8.25,-1.1){\large ${S^2}$}

\pscircle[linewidth=0.3mm](7.2,0){1.3}

\psellipse[linewidth=0.2mm,linestyle=dashed](7.2,0)(1.28,0.3)

\pscircle[linewidth=0.3mm](-2.51,0){1.3}

\psellipse[linewidth=0.2mm,linestyle=dashed](-2.51,0)(1.28,0.3)

\end{pspicture}}
\hrule
\caption{The correlations between the EPR elements of reality are correlations between the respective points of two
2-spheres. They have nothing whatsoever to do with the correlations between the points of two 0-spheres as Bell
unjustifiably assumed.\break}
\hrule
\end{figure}

Now one of the things that clearly jumps out from this argument is that the elements of reality in question---i.e.,
the measurement results of the spin components---are points of a unit 2-sphere, not 0-sphere as Bell presumed. Indeed,
simultaneous existence of elements of reality at a remote location for {\it all} components of spin (the one that
was actually predicted as well as all the others which could have been predicted) is at the heart of the EPR argument
\ocite{EPR}\ocite{GHSZ}\ocite{Clauser-Shimony}. What is more, there are {\it infinitely many} ${\,}$spin components
that could be measured locally---one corresponding to each direction ${{\bf n}\in{\rm I\!R}^3}$ (cf. Fig.${\,}$2).
Thus there is a one-to-one correspondence between the simultaneously existing elements of reality at-a-distance and
the points of a unit 2-sphere defined by ${||{\bf n}||=1}$. Moreover, since the completeness criterion of EPR demands
that each one of these elements of reality must have a counterpart in a complete theory, any Bell-type theorem can be
applicable to the EPR argument {\it if and only if} ${\,}$it respects this 2-spherical topology of the elements. That
is to say, for the singlet state (\ref{single}), the codomain of any Bell-type map
${A_{\bf n}(\lambda): {\rm I\!R}^3\!\times\Lambda\rightarrow \Sigma}$ must be homeomorphic\break to a 2-sphere, otherwise
Bell's accounting of the elements of reality would not be complete \ocite{experiment}. Indeed, if the codomain ${\Sigma}$
of the function ${A_{\bf n}(\lambda)}$ differs from ${S^2}$ by even a single point, then it would not be homeomorphic
to a 2-sphere, and then the set of all possible values that ${A_{\bf n}(\lambda)}$ can take would fail to
be in one-to-one correspondence with the set of EPR elements of reality. In that case there would be at least one
element of reality that would not have a counterpart in the ``complete'' theory, rendering the
prescription ${A_{\bf n}(\lambda): {\rm I\!R}^3\!\times\Lambda\rightarrow \Sigma}$
quite worthless for the purposes of Bell.

Despite this danger of incompleteness, neither Bell nor his followers offer any explanation for why the topology of ${S^2}$
has been dropped from Eq.${\,}$(\ref{Bell-map}) in favor of the {\it ad hoc} choice of ${S^0}$ (or some other
${{\cal I}\subseteq{\rm I\!R}}$ for that matter). In fact, this is simply a gross error \ocite{Further}\ocite{experiment}.
It stems from adapting
the premises of EPR in too disjointed a manner than is warranted. The argument of EPR---based on their four premises---is a
package-deal, and it must be respected as such by any Bell-type theorem if it were to claim inconsistencies
within these premises. Moreover, as we
noted above, ${S^0}$ is not even connected, whereas ${S^2}$ is both connected and simply connected. If instead of ${S^0}$
some other interval of ${\rm I\!R}$ is chosen, then is this interval supposed to be open or closed? For an open interval
of ${\rm I\!R}$ is homeomorphic to ${\rm I\!R}$, but a closed one is not. It is common practice, however, to assume the
closed interval ${[-1,\,+1]}$ in the proofs of Bell's theorem, with the numbers lying within it representing the EPR
elements of reality. But why assume that these elements of reality are all ``lined up'' as points of the real line,
respecting its very specific\footnote{Recall that real line is a topological space of dimension one, with very specific
order topology. It is paracompact and second-countable as well as contractible and locally compact. It also has
a standard differentiable structure defined on it, making it a differentiable manifold.}
order topology? What justifies the presumption of such a one dimensional topological
space in the first place? Is it not evident from the above argument of EPR that the topology of the real line---or that
of any of its subsets---has nothing whatsoever to do with the topology of the EPR elements of reality? In fact, let
alone ${\rm I\!R}$, the evident topological space of these elements---namely ${S^2}$---is not even
homeomorphic to ${{\rm I\!R}^2}$, but rather a one-point compactification of ${{\rm I\!R}^2}$. Hence to reduce ${S^2}$
to ${{\rm I\!R}^2}$ requires a highly nontrivial topological operation (that of surgically removing a point), let alone
then going from ${{\rm I\!R}^2}$ to ${\rm I\!R}$, and finally from ${\rm I\!R}$ to ${S^0}$. Of course, one can also try
to reduce ${S^2}$ to ${S^0}$ directly, but that requires surgically removing an entire equatorial
circle of points from ${S^2}$. In other words, reducing ${S^2}$ to ${S^0}$ requires neglecting an entire infinity
of elements of reality from consideration \ocite{experiment}. Moreover, since each of these
spaces---${S^0}$, ${[-1,\,+1]}$, ${\rm I\!R}$, ${{\rm I\!R}^2}$,
and ${S^2}$---correspond to imposing a different topological order on the set of numbers representing the EPR
elements of reality, for each choice the correlations among these numbers would be different in general. Hence no
choice of a space can be accepted without a detailed topological justification. And yet, neither ${S^0}$ nor any other
subset of ${\rm I\!R}$ has any topological credentials to represent the EPR elements of reality, as it is abundantly
clear from the argument of EPR that it is the topology of ${S^2}$ that necessarily captures the correct order of
these elements.

Actually, since the argument of EPR goes through for the spins of both particles, the elements of reality in question
are the points of ${two\,}$ copies of ${S^2}$, one for each particle, as shown in Fig.${\,}$2. If we view these
copies as embedded in ${{\rm I\!R}^3}$, then the set of their points can be parameterized by the coordinates
${\{n_x,\,n_y,\,n_z\}}$ of ${{\rm I\!R}^3}$, satisfying the constraint
\begin{equation}
n^2_x+n^2_y+n^2_z\,=\,1.\label{2-seven}
\end{equation}
It is important to stress here that when a space such as ${S^k}$ is embedded into ${{\rm I\!R}^{k+1}}$ by a map
${f:S^k\hookrightarrow {{\rm I\!R}^{k+1}}}$, then ${f(S^k)}$ remains diffeomorphic to ${S^k}$. This is because an
embedding of a given space into a higher-dimensional space does not affect the intrinsic properties of that space,
but merely provides a useful set of coordinates on it. Moreover, it is a matter of indifference which set of
coordinates are chosen, as long as they are a well-defined set of coordinates in ${{\rm I\!R}^{k+1}}$. Thus, just as
classical physics does not care whether we choose one frame of reference or another to perform our experiments, the
topological structure of the sphere ${S^k}$---which is a totality in itself---does not care whether we choose one set
of coordinates or another in ${{\rm I\!R}^{k+1}}$ to perform our calculations. In other words, as useful as it is,
the vector ${{\bf n}\equiv(n_x,\,n_y,\,n_z)}$ corresponding to the coordinates ${\{n_x,\,n_y,\,n_z\}}$ chosen above
is not only not an intrinsic part of the 2-sphere, but is also completely dispensable, even as a tool. It merely
provides a useful pointer to the points of ${S^2}$. That is to say, each point of ${S^2}$ representing a specific
EPR element of reality is merely indexed by the vector ${\bf n}$. Consequently, in close analogy with the local
functions (\ref{Bell's}) postulated by Bell, each point of ${S^2}$ can be represented as
\begin{equation}
A({\bf n},\,\lambda)\,=\,\pm\,1\,\in\,S^2,\;\;{\rm about\;\,the\;\,direction}\;\,{\bf n}\;\,{\rm in}\,\;{\rm I\!R}^3.
\label{improveBell's}
\end{equation}
This amounts to replacing the incorrect local maps (\ref{Bell-map}) of Bell with the topologically correct local maps
\begin{equation}
A_{\bf n}(\lambda): {\rm I\!R}^3\!\times\Lambda\longrightarrow S^2.\label{my-map}
\end{equation}
Evidently, the range of these maps is still the set of points describing the binary results, ${\pm1}$, but this
set now has the topology of a 2-sphere rather than a 0-sphere. Consequently, the correlations computed using these
maps would be correlations between the points of ${S^2}$, and not between the points of some irrelevant space like
${S^0}$ or ${\rm I\!R}$ presumed by Bell. Since it is the space ${S^2}$, and not ${S^0}$, that
correctly represents the EPR elements of reality, the choice between the maps (\ref{my-map}) and (\ref{Bell-map}) is
clearly a choice between the correct and incorrect representations of the physical reality.

\section{Bell's theorem, its variants, and their spinoffs are all non-starters at best} 

Since Bell begins his theorem with a pair of incorrect maps like (\ref{Bell-map}), he forfeits his game from the start.
For the straw man he thereby knocks off has nothing to do with the EPR elements of reality. Recall that
Bell begins by considering a ``complete'' physical state ${\lambda}$ for the EPR-Bohm system, and assumes a
normalized probability measure ${\rho(\lambda)}$ on the space ${\Lambda}$ of all such states. He then postulates
the expectation value for a pair of spin measurements,
\begin{equation}
{\cal E}({\bf a},\,{\bf b})\,=\int_{\Lambda}
(A_{\bf a}\,B_{\bf b})(\lambda)\;d\rho(\lambda)\,
=\int_{\Lambda}
A_{\bf a}(\lambda)\,B_{\bf b}(\lambda)\;d\rho(\lambda)\,,\label{prob-1}
\end{equation}
and requires this (supposedly) local-realistic value to satisfy the perfect correlation constraint,
\begin{equation}
{\cal E}({\bf n},\,{\bf n})\,=\,-1\,,\label{prob-2}
\end{equation}
which he borrows from EPR, who in turn adapted it from quantum mechanics. Here the local outcome functions
\begin{equation}
A_{\bf a}(\lambda): {\rm I\!R}^3\!\times\Lambda\longrightarrow S^0 \;\;\;\;\;\;\;\;{\rm and}\;\;\;\;\;\;\;\;
B_{\bf b}(\lambda): {\rm I\!R}^3\!\times\Lambda\longrightarrow S^0\label{maps}
\end{equation}
are the functions we discussed in equations (\ref{Bell's}) and (\ref{Bell-map}) above. It should be clear by now,
however, that this expectation value has nothing whatsoever to do with the correlations between the EPR elements
of reality. For what it provides is correlations between the points of the real line, whereas EPR elements of
reality in the present case are points of a 2-sphere, not the real line, or any other interval
${{\cal I}\subseteq{\rm I\!R}}$. In fact,
it is quite astonishing that Bell thought correlations between the points of a real line have anything at all to
do with the correlations between the elements of reality.

To appreciate our amazement further, recall that Bell's ultimate goal was to conclude that
\begin{equation}
\int_{\Lambda}
A_{\bf a}(\lambda)\,B_{\bf b}(\lambda)\;d\rho(\lambda)
\;\;\,{\rm cannot\;be\;equal\;to}\;\;\,\langle\Psi_{\bf n}|\,{\boldsymbol\sigma}\cdot{\bf a}\,\otimes\,
{\boldsymbol\sigma}\cdot{\bf b}\,|\Psi_{\bf n}\rangle\label{twoobserve}
\end{equation}
for the entangled state (\ref{single}), and hence no local-realistic correlations can reproduce the quantum mechanical
correlations. But this is quite a meaningless comparison, because the right hand side of the above expression describes
correlations between the points of a 2-sphere, whereas the left hand side describes those between the points of the
real line. Indeed, the matrices ${i{\boldsymbol\sigma}\cdot{\bf n}}$ on the right hand side are elements of the Lie
algebra of the group SU(2), which is homeomorphic to a 3-sphere, and a 3-sphere is a principal U(1) bundle over a
2-sphere \ocite{Ryder}.
Hence, in their true essence, the matrices ${i{\boldsymbol\sigma}\cdot{\bf n}}$ represent nothing but the points of a
2-sphere (cf. Fig.${\,}$3 and Ref.${\,}$\ocite{experiment}). Therefore comparing the left and right hand sides of the
above expression is like comparing correlations between cars in one narrow lane of a highway with correlations between
cars on the surface of the planet. The absurdity of the comparison is breathtaking. Regardless of locality, realism,
quantum mechanics, or classical mechanics, the two sides of the above expression cannot possibly be the same, for they
describe correlations between the points of two entirely different topological spaces. Even the dimensions of these two
spaces do not match! If any meaningful conclusion about local realism is to be drawn from such a comparison, then it
can only be drawn by comparing apples with apples. That is to say, in any such comparison the two sides of the above
expression must correspond to correlations between the points of the {\it same} topological space ${\Omega}$, whatever
${\Omega}$ happens to be. In other words, if we denote the quantum mechanical expectation functional
by ${{\cal E}_{{\!}_{Q.M.}}\!}$ and the corresponding local-realistic one by ${{\cal E}_{{\!}_{L.R.}}\!}$, then the
correct Bell-type question in general should be
\begin{equation}
{\rm whether}\;\;{\cal E}_{{\!}_{L.R.}\!}(\Omega)\,\stackrel{?}{=}
\,{\cal E}_{{\!}_{Q.M.}\!}(\Omega)\,,\;\;{\rm and\;not\;whether}\;\;
{\cal E}_{{\!}_{L.R.}\!}({\cal I}\subseteq{\rm I\!R})\,\stackrel{?}{=}\,{\cal E}_{{\!}_{Q.M.}\!}(\Omega)\,.
\end{equation}
For any disagreement between ${{\cal E}_{{\!}_{L.R.}\!}({\cal I}\subseteq{\rm I\!R})}$ and
${{\cal E}_{{\!}_{Q.M.}\!}(\Omega)}$ cannot be attributed to
incompatibility between locality and realism, because it can be entirely due to the
topological differences between the spaces ${{\cal I}\subseteq{\rm I\!R}}$ and ${\Omega}$.

\begin{figure}
\hrule
\scalebox{0.75}{
\begin{pspicture}(0.3,-3.7)(4.2,2.8)

\pscircle[linewidth=0.3mm,linestyle=dashed](-1.8,-0.45){2.6}

\psellipse[linewidth=0.3mm](-0.8,-0.45)(0.7,1.4)

\psellipse[linewidth=0.3mm,border=3pt](-2.4,-0.45)(1.4,0.4)

\pscurve[linewidth=0.3mm,border=3pt](-1.485,-0.35)(-1.48,-0.25)(-1.45,0.0)

\pscircle[linewidth=0.3mm](7.0,-0.45){1.7}

\psellipse[linewidth=0.2mm,linestyle=dashed](7.0,-0.45)(1.68,0.4)

\put(-4.4,1.27){{\Large ${S^3}$}}

\put(-2.0,1.2){{\Large ${h^{-1}(q)}$}}

\put(-3.7,-1.4){{\Large ${h^{-1}(p)}$}}

\put(7.43,0.67){{\Large ${q}$}}

\psdot*(7.2,0.79)

\put(6.3,0.43){{\Large ${p}$}}

\psdot*(6.1,0.43)

\put(8.5,-1.8){{\Large ${S^2}$}}

\put(5.9,-2.7){\Large base space}

\put(1.7,0.7){\Large ${h:S^3\rightarrow S^2}$}

\pscurve[linewidth=0.3mm,arrowinset=0.3,arrowsize=3pt 3,arrowlength=2]{->}(1.2,0.25)(2.47,0.45)(3.74,0.45)(4.9,0.25)

\put(1.8,-0.45){\large Hopf fibration}

\pscurve[linewidth=0.3mm,arrowinset=0.3,arrowsize=3pt 3,arrowlength=2]{->}(4.9,-0.95)(3.74,-1.15)(2.47,-1.15)(1.2,-0.95)

\put(1.55,-1.8){\Large ${h^{-1}:S^2\rightarrow S^3}$}

\end{pspicture}}
\hrule
\caption{The tangled web of linked Hopf circles depicting the topological non-triviality of the EPR elements of
physical reality.\break}
\label{fig}
\smallskip
\hrule
\end{figure}
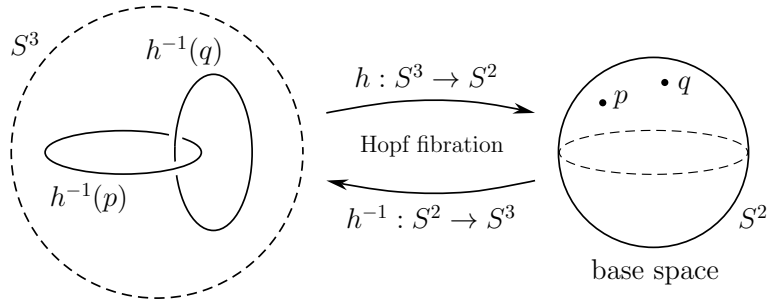

At this stage one may wonder why the topology of the EPR elements of reality should play a significant role here
when one can always reformulate the expectation functionals purely in terms of probabilities of obtaining measurement
results \ocite{Bell-La}. But no amount of probabilistic reformulation can reduce the dimensional and topological
absurdities of the comparison (\ref{twoobserve}). For, surely, if the probability of falling off the edge of the
earth turns out to be the same for both the flat-earthers (${B^2}$) and the round-earthers (${S^2}$), then there is
something seriously wrong with the way one has calculated these probabilities. That is to say, whatever scheme is used
to calculate the correlations between the EPR elements of reality, it must account for the fact that these elements
are the points of a 2-sphere, not the real line. Simply by reformulating the derivation of the Bell-CHSH inequality in
terms of conditional probabilities of obtaining measurement results (as done by Bell in Ref.${\,}$\ocite{Bell-La}) does
not alter the fact that the correlations thus being calculated are still those between the points of the real line, which
have nothing to do with the correlations between the EPR elements of reality. The same must be said about the numerous
variants and spinoffs of Bell's theorem, such as the GHZ \ocite{GHSZ}, Hardy \ocite{Hardy}, or Leggett \ocite{Leggett}
type theorems against local or nonlocal realism. As long as such theorems rely on the maps like (\ref{maps}) (or like
${A_{\bf n}(\lambda): {\rm I\!R}^3\!\times\Lambda\rightarrow {\cal I}\subseteq{\rm I\!R}}$), they are presupposing
incorrect topology for the EPR elements of reality, and hence are as irrelevant to the concerns of EPR as
the original theorem by Bell. To be sure, some of these variant theorems do not make explicit
commitment to ${\rm I\!R}$, but employ only hand-waving arguments involving incommensurate, or counterfactual
measurement results. But numbers do not exist in a mathematical void. {\it As soon as a Bell-type theorem associates
a number to an EPR element of reality, it is making a commitment to a topological space of one kind or another}.
And if there is a mismatch between this space and the space of the corresponding EPR elements
of reality, then that would inevitably lead to an illusion of inconsistency within the premises of EPR.

Suppose now we consider the familiar string of expectation functionals studied by CHSH \ocite{Clauser}, namely 
\begin{equation}
{\cal E}({\bf a},\,{\bf b})\,+\,{\cal E}({\bf a},\,{\bf b'})\,+\,
{\cal E}({\bf a'},\,{\bf b})\,-\,{\cal E}({\bf a'},\,{\bf b'})\,.
\end{equation}
With a questionable practical assumption (which we will not question) that the distribution ${\rho(\lambda)}$ remains the
same for all four of these functionals, this string can be rewritten in terms of the products of the local functions as
\begin{equation}
\int_{\Lambda}\,[\;
A_{\bf a}(\lambda)\,B_{\bf b}(\lambda)\,+\,
A_{\bf a}(\lambda)\,B_{\bf b'}(\lambda)\,+\,
A_{\bf a'}(\lambda)\,B_{\bf b}(\lambda)\,-\,
A_{\bf a'}(\lambda)\,B_{\bf b'}(\lambda)\;]
\;\,d{\rho}(\lambda)\,. \label{probnonint}
\end{equation}
And since Bell assumes that all local functions ${A_{\bf n}(\lambda)}$ and ${B_{\bf n}(\lambda)}$ involved in this
integral are elements of the real line ${\rm I\!R}$ (which of course corresponds to a commutative algebra), they
necessarily satisfy
\begin{equation}
\left[\,A_{\bf n}(\lambda),\,B_{\bf n'}(\lambda)\,\right]\,=\,0\,,
\;\;\;\forall\;\,{\bf n}\;\,{\rm and}\;\,{\bf n'}\,\in\,{\rm I\!R}^3.\label{com}
\end{equation}
If we now square the integrand of Eq.${\,}$(\ref{probnonint}), use the above commutation relations, and use the fact
that, by definition, all local functions square to unity (the algebra goes through even when the squares of the local
functions are allowed to be ${-1}$), then the absolute value of the CHSH string leads to the following form of
variance inequality \ocite{Further}:
\begin{equation}
|{\cal E}({\bf a},\,{\bf b})\,+\,{\cal E}({\bf a},\,{\bf b'})\,+\,
{\cal E}({\bf a'},\,{\bf b})\,-\,{\cal E}({\bf a'},\,{\bf b'})|\,
\leq\,\sqrt{\int_{\Lambda}\left\{\,4\,+\,\left[\,A_{\bf a}(\lambda),\,
A_{\bf a'}(\lambda)\,\right]\left[\,B_{\bf b'}(\lambda),\,
B_{\bf b}(\lambda)\,\right]\,\right\}\;d\rho(\lambda)\,}\,.\label{yever}
\end{equation}
And since all ${A_{\bf n}(\lambda)\in{\rm I\!R}}$ commute with each other, this inequality finally leads
to the Bell-CHSH inequalities:
\begin{equation}
-\,2\,\leq\,{\cal E}({\bf a},\,{\bf b})\,+\,{\cal E}({\bf a},\,{\bf b'})\,+\,
{\cal E}({\bf a'},\,{\bf b})\,-\,{\cal E}({\bf a'},\,{\bf b'})\,\leq\,+\,2\,.\label{never}
\end{equation}

But, once again, these inequalities have nothing whatsoever to do with the correlations between the EPR elements of reality.
For the correlations between the EPR elements of reality are correlations between the points of a 2-sphere, whereas the
correlations from which these inequalities are derived are those between the points of the real line.

\section{Topologically correct local-realistic framework for the EPR-type correlations}

The criticism we have presented so far constitutes ``the faulty-premise side'' of our rejection of Bell's theorem${\;}$and
its variants ({\it nego majorem, nihil ad rem}). On the counterexample side, the question then is: how should one correctly
calculate the correlations between the EPR elements of reality in general? As it turns out, the answer
to this question would have been obvious to Pauli (and even to Hamilton to some extent), but remains neglected within
the literature on Bell's theorem. It amounts to understanding the profound topological structure of the 3-sphere
(cf. Fig.${\,}$3), which we have already discussed elsewhere \ocite{experiment}. As we shall see, for any two-level
system the correct local-realistic
correlations can be computed by analyzing the interplay between the points of the corresponding 2-sphere and 3-sphere.

This can be demonstrated most transparently in the case of the original theorem by Bell, as we have done previously
in terms of a Clifford-algebraic model \ocite{Christian}\ocite{Further}\ocite{experiment}. Unfortunately, as powerful
as this Clifford-algebraic framework is, unfamiliarity with it has led some skeptics to misinterpret the model and
obfuscate its message \ocite{reply}.
The main stumbling block of the skeptics seems to be their inability to distinguish
between the concept of a point of ${S^2}$ and its bivectorial representation in our model. To spare us from the possibility
of such a spurious difficulty, in this paper we shall keep the use of Clifford algebra to a minimum. Instead, we shall
bring out the main message of our model in purely topological terms. To this end, let us stress once again that the
correct local-realistic correlations in the present case can be obtained if the incorrect local
maps (\ref{maps}) assumed by Bell are replaced with the correct local maps
\begin{equation}
A_{\bf a}(\lambda): {\rm I\!R}^3\!\times\Lambda\longrightarrow S^2 \;\;\;\;\;\;\;\;{\rm and}\;\;\;\;\;\;\;\;
B_{\bf b}(\lambda): {\rm I\!R}^3\!\times\Lambda\longrightarrow S^2,\label{s2localmaps}
\end{equation}
each of which---in the light of Eq.${\,}$(\ref{improveBell's}), and for the case of
uniform distribution ${\rho(\lambda)}$---naturally satisfies the condition
\begin{equation}
\int_{\Lambda}A_{\bf n}(\lambda)\;d\rho(\lambda)\,=\,0,\,\;{\rm about\;any\;direction}\;{\bf n}
\;\,{\rm in}\,\;{\rm I\!R}^3.\label{0}
\end{equation}
Evidently, these maps are just as local as the original maps of Bell. The measurement results ${A}$ and ${B}$
they represent do not depend either on each other, or on the settings of the remote polarizers ${\bf b}$ and ${\bf a}$
(respectively), but only on the local polarizers ${\bf a}$ and ${\bf b}$, and on the past common causes ${\lambda}$.
Moreover, just as in Bell's case, the range of these local maps is still simply a collection of binary numbers, ${+1}$'s
and ${-1}$'s, but now---instead of lying on the real line---these numbers lie on the surface of a unit ball---i.e., they
are points of a unit 2-sphere. Given these local maps, we can now define a local-realistic theory just as Bell
did---by imposing the locality condition on the joint ``beable'' ${(A_{\bf a}\,B_{\bf b})(\lambda)\,}$:
\begin{equation}
(A_{\bf a}\,B_{\bf b})(\lambda)\,=\,A_{\bf a}(\lambda)\,B_{\bf b}(\lambda).
\end{equation}
Locality thus implies that, once the state ${\lambda}$ is specified and the two particles have separated, measurements of
${A}$ can depend only upon ${\lambda}$ and ${\bf a}$, but not ${\bf b}$, and likewise measurements of ${B}$ can depend
only upon ${\lambda}$ and ${\bf b}$, but not ${\bf a}$. The correlations between the EPR elements of reality can now be
calculated by using the expectation functional
\begin{equation}
{\cal E}({\bf a},\,{\bf b})\,=\int_{\Lambda}
(A_{\bf a}\,B_{\bf b})(\lambda)\;d\rho(\lambda)\,
=\int_{\Lambda}
A_{\bf a}(\lambda)\,B_{\bf b}(\lambda)\;d\rho(\lambda)\,.\label{prob-3}
\end{equation}

It is worth recalling here that---following the pioneering analysis by von Neumann \ocite{von} and Segal
\ocite{Segal}---the above expectation functional is understood to be a strongly continuous linear functional,
${{\cal E}:\,A\mapsto {\cal E}(A)}$, defined on the set of general-valued functions such as ${A(\lambda)}$
defined above---with ${\Lambda\ni\lambda}$ being the measure space---such that
\vspace{-0.2cm}
\begin{description}
 \item[\;\;\;\;\;\;\;(a)\;\;] ${{\cal E}(e)=1}$ for the unit element ${e}$ in the set of ${A}$'s,
 \item[\;\;\;\;\;\;\;(b)\;\;] ${{\cal E}(A^*A)\geq 0\;\,\forall\;A\,}$ with ${{\cal E}(A^2)=0}$ only if ${A=0}$,
 \item[{\rm and}\;\;(c)\;\;] ${{\cal E}(B^*A\,B)\leq \alpha_{{\!\!}_{A}}{\cal E}(B^*B)}$ for some real constant
${\alpha_{{\!\!}_{A}}}$\,,
\end{description}
\vspace{-0.2cm}
where the ``${\,*\,}$'' represents an appropriate conjugation operation. The important point here is that, in this
remarkably general formulation of the standard probability theory, the codomain of the functions ${A(\lambda)}$ is
{\it not} restricted to be a subset of the real line. In particular, this formulation is quite well suited for calculating
correlations between the local functions such as those taken in Eq.${\,}$(\ref{s2localmaps}). More generally, any local
functions of the form 
\begin{equation}
A_{\bf a}(\lambda): {\rm I\!R}^3\!\times\Lambda\longrightarrow \Omega \;\;\;\;\;\;\;\;{\rm and}\;\;\;\;\;\;\;\;
B_{\bf b}(\lambda): {\rm I\!R}^3\!\times\Lambda\longrightarrow \Omega\label{omega}
\end{equation}
are admissible within this formulation, where ${\Omega}$ is an arbitrary topological space
composed of measurement results, binary or otherwise. In section VI below we shall have more to say about such
general topological codomains.

For now, we simply stress that---like Eq.${\,}$(\ref{prob-1})---the expectation functional (\ref{prob-3}) provides
perfectly adequate means for calculating the local-realistic correlations between measurement results. There is, however,
a major difference between Eqs.${\,}$(\ref{prob-1}) and (\ref{prob-3}). The product beable
${A_{\bf a}(\lambda)\,B_{\bf b}(\lambda)}$ appearing in Eq.${\,}$(\ref{prob-3}) is necessarily a point of a 3-sphere,
\begin{equation}
A_{\bf a}(\lambda)\,B_{\bf b}(\lambda): {\rm I\!R}^3\!\times {\rm I\!R}^3\!\times{\Lambda}\longrightarrow S^3,
\end{equation}
as dictated by the following elementary theorem in topology ({\it cf}. also Eqs.${\,}$(\ref{21}) and (\ref{bi-identity})).

\vspace{0.32cm}

{\parindent 0pt
{\bf The Product Point Theorem:} ${\;\;}${\it The product of any
two points of a 2-sphere is a point of a 3-sphere.}}\footnote{This theorem can also be understood purely in terms of the
algebraic representation of the parallelized
3-sphere discussed in Refs.${\,}$\ocite{Christian}, \ocite{Further}, and \ocite{experiment}.
The proof of the theorem is then simply the proof of the algebraic identity (\ref{bi-identity}) discussed below. In
particular, since
${||{\boldsymbol\mu}\cdot{\bf n}||^2 = (-\,{\boldsymbol\mu}\cdot{\bf n})(+\,{\boldsymbol\mu}\cdot{\bf n}) =
-\,{\boldsymbol\mu}^2\,{\bf n}\,{\bf n} = {\bf n}\,{\bf n}
= {\bf n}\cdot{\bf n} =||{\bf n}||^2 = 1}$ for any unit vector ${{\bf n}\in{\rm I\!R}^3}$,
the space of all bivectors ${{\boldsymbol\mu}\cdot{\bf n}}$ is isomorphic to a unit 2-sphere. The left hand
side of the identity (\ref{bi-identity}) is thus nothing but
a product of two points of this 2-sphere. On the other hand, the right hand side
of the identity represents a point, not of a 2-sphere, but 3-sphere. This can be readily seen by recognizing that
${||\,-\,{\bf a}\cdot{\bf b}\,-\,{\boldsymbol\mu}\cdot({\bf a}\times{\bf b})\;||^2 = {\bf p}\cdot{\bf p} = 1}$
for some unit vector ${{\bf p}\in{\rm I\!R}^4}$, and so the space of all multivectors
${\,-\,{\bf a}\cdot{\bf b}\,-\,{\boldsymbol\mu}\cdot({\bf a}\times{\bf b})}$ is isomorphic to a unit 3-sphere
\ocite{Christian}\ocite{Further}. And
since the identity (\ref{bi-identity}) is simply an identity in Clifford algebra \ocite{Christian}, we have the algebraic proof.}

\vspace{0.32cm}

{\parindent 0pt
{\bf Proof:}
\vspace{-0.83cm}
\begin{itemize}
\item[${\;}$] ${\;\;\;\,}$ A 3-sphere is a set of points equidistant from a fixed point in ${{\rm I\!R}^4}$. Thus
it is a boundary of a 4-ball in four dimensions. And as such, it is not the easiest space for us to visualize (although
it is not impossible to do so \ocite{AJP}). Therefore, let us first consider the theorem in one lower dimension. Consider
two copies of a unit 2-ball (i.e., two ordinary disks). The boundary of this 2-ball is a unit 1-sphere (i.e., a circle)
of points ${\pm1}$. Now glue (or identify) these boundaries together point-by-point. The resulting topological space
is a unit 2-sphere (i.e, the boundary of a unit 3-ball). Any point of this 2-sphere is thus a product of two points,
each belonging to one of the two copies of a 1-sphere analogous to the one we started out with. Thus, if ${A=+1}$ is a
point of one of the copies of a 1-sphere and ${B=-1}$ is a point of the other copy, then the product  ${AB=-1}$ is the
corresponding point of the 2-sphere constructed by identifying such points. The proof in the case of 3-sphere is exactly
analogous: Consider two copies of a unit 3-ball (i.e., two ordinary solid balls). The boundary of each of these 3-balls
is of course a unit 2-sphere. Now glue (or identify) these boundaries, point-by-point (cf. Ref.${\,}$\ocite{AJP}). The
resulting topological space is a unit 3-sphere. Each point of this 3-sphere is thus a product of the two points
belonging to two copies of a 2-sphere analogous to the one we started out with. Conversely, if ${A=+1}$ is a point
of one of the 2-spheres and ${B=-1}$ is a point of the other, then the product ${AB=-1}$ is the corresponding point
of the 3-sphere, constructed out of these 2-spheres. (Formal details of this proof can be found in
Ref.${\,}$\ocite{Frankel}.)
\end{itemize}}

Of course, the values of the points ${A}$ and ${B}$ used here are only one of the pairs of possibilities. If instead
of ${A=+1}$ and ${B=-1}$ we take ${A=-1}$ and ${B=+1}$ as our gluing points, then the
corresponding point of the 3-sphere would still be ${AB=-1}$. But if we instead take ${A=+1}$ and ${B=+1}$ or ${A=-1}$ and
${B=-1}$, then the corresponding point of the 3-sphere would be ${AB=+1}$. In other words, the points of the
3-sphere satisfy the multiplication map
\begin{equation}
{\cal F}\,(A,\,B)\,=\,AB\,,\label{Lus-map}
\end{equation}
with the following properties:
\begin{equation}
{\cal F}\,(+1,\,+1)\,=\,+1\,,\;\;\;{\cal F}\,(+1,\,-1)\,=\,-1\,,\;\;\;{\cal F}\,(-1,\,+1)\,=\,-1\,,\;\;\;{\rm and}
\;\;\;{\cal F}\,(-1,\,-1)\,=\,+1\,.\label{Lucien}
\end{equation}
These are important properties, not only because they are demanded by Bell's formulation of local causality \ocite{Further},
but also because they are in harmony with how the data from the two ends of the apparatus is usually analyzed in a typical
Bell-type experiment. And as such they must be respected by any local-realistic model for the EPR-Bohm correlations. As we
shall soon begin to appreciate, these properties are naturally satisfied by our local model of Ref.${\,}$\ocite{Christian}.
More importantly, they will soon be incorporated within our current framework in a much more cogent manner.

The upshot of the above theorem is that a 3-sphere, in the present context, is a set of all product points ${AB=\pm1}$,
equidistant from a fixed origin in the higher-dimensional space ${{\rm I\!R}^4}$. That is to say, when viewed as embedded
in ${{\rm I\!R}^4}$, a 3-sphere is best seen as parameterized by the coordinates of this embedding space, satisfying the
constraint
\begin{equation}
n^2_o+n^2_x+n^2_y+n^2_z\,=\,1.\label{2-eight}
\end{equation}
Once again we stress the obvious that the corresponding four dimensional vector ${(n_o,\,n_x,\,n_y,\,n_z)}$ is
not an intrinsic part of the 3-sphere itself, but merely provides an external index to one of its points. In fact,
it represents a non-pure quaternionic number---making ${S^3}$ a set of unit quaternions,
${\{\,q\in{\rm I\!H}\,:\,||\,q\,||=1\,\}}$---as we have discussed elsewhere \ocite{experiment}. Suppose now that
the two copies of the 2-sphere which make up these points are themselves similarly parameterized by the coordinates
of ${{\rm I\!R}^3}$; say by the vectors ${\bf a}$ and ${\bf b}$. Then the above constraint, together
with the perfect correlation condition (\ref{prob-2}), dictates that the following relation among the
points of the 2 and 3-spheres is necessarily satisfied:
\begin{align}
A_{\bf a}(\lambda)\,B_{\bf b}(\lambda)\,&=\,-\,
\cos\theta_{\bf ab}\,-\,C_{\bf c}(\lambda)\,\sin\theta_{\bf ab}\, \label{21} \\
&=\,\pm\,1\,\in\,S^3,\;\,{\rm about\;\,the\;\,direction}\;\,
(-\cos\theta_{\bf ab}\,,\;-\,c_x\sin\theta_{\bf ab}\,,\;-\,c_y\sin\theta_{\bf ab}\,,\;-\,c_z\sin\theta_{\bf ab})
\;\,{\rm in}\,\;{\rm I\!R}^4,\label{q-mappq}
\end{align}
where ${A_{\bf a}(\lambda)}$, ${B_{\bf b}(\lambda)}$, and ${C_{\bf c}(\lambda)}$ are points of the 2-sphere,
${\theta_{\bf ab}}$ is the angle between ${\bf a}$ and ${\bf b}$, and the vector ${\bf c}$ is defined as
${{\bf c}:=\frac{{\bf a}\times{\bf b}}{|{\bf a}\times{\bf b}|}}$. This expression is a well known
parameterization of the quaternionic numbers. Although it implies ${A_{\bf a}(\lambda)\,A_{\bf a}(\lambda)=-\,1}$,
there is nothing ``complex'' or ``non-real'' about it. Such nontrivial but {\it real} products are bread and butter to
the aviation engineers, and they simply reflect the fact that the projective
plane of a 2-sphere is not orientable \ocite{Penrose}. But the 2-sphere itself {\it is} orientable, with its points
furnishing the road map of a 3-sphere as above, whose projective space then turns out to be orientable \ocite{Penrose}.
Within the embedding space ${{\rm I\!R}^3}$, this map can be used to represent a counterclockwise rotation by angle
${2\,\theta_{\bf ab}}$ about the ${\bf c}$ axis. Once again, it should be noted that the vectors ${\bf a}$, ${\bf b}$,
and ${\bf c}$ in this parameterization are not intrinsic to the 2-sphere, but merely external parameters within the
embedding space ${{\rm I\!R}^3}$. That is to say, they merely index the respective points ${A_{\bf a}(\lambda)}$,
${B_{\bf b}(\lambda)}$, and ${C_{\bf c}(\lambda)}$ of the 2-sphere (each of which representing an EPR element of
reality). This allows us then to visualize these points as
\begin{align}
A_{\bf a}(\lambda)\,&=\,\pm\,1\,\in\,S^2,\;\;{\rm about\;\,the\;\,direction}\;\,{\bf a}\;\,{\rm in}\,\;{\rm I\!R}^3,
\notag \\
B_{\bf b}(\lambda)\,&=\,\pm\,1\,\in\,S^2,\;\;{\rm about\;\,the\;\,direction}\;\,{\bf b}\;\,{\rm in}\,\;{\rm I\!R}^3,
\notag \\
{\rm and}\;\;\;
C_{\bf c}(\lambda)\,&=\,\pm\,1\,\in\,S^2,\;\;{\rm about\;\,the\;\,direction}\;\,{\bf c}\;\,{\rm in}\,\;{\rm I\!R}^3,
\label{proveBell's}
\end{align}
as we have already done in the context of equations (\ref{improveBell's}) and (\ref{my-map}). Thus, if 
${C_{\bf +c}(\lambda)=-1}$ about ${\bf +c}$, then ${C_{\bf -c}(\lambda)}$ will be ${+1}$ about ${\bf -c}$,
since ${C_{\bf +c}(\lambda)=-1}$ and ${C_{\bf -c}(\lambda)=+1}$ are then antipodal points of the 2-sphere, along the
direction ${\bf c}$. On the other hand, despite appearances, the right hand side of equation (\ref{21}), namely
${\,-\,\cos\theta_{\bf ab}\,-\,C_{\bf c}(\lambda)\,\sin\theta_{\bf ab}}$, is simply a binary number in
disguise, ${+1}$ or ${-1}$, and a {\it bona fide} intrinsic point of the corresponding 3-sphere \ocite{Zulli}.

This last point is rather difficult to see at first sight, and has been a major stumbling block for several skeptics
of Ref.${\,}$\ocite{Christian}. The spurious difficulty entertained by the skeptics quickly evaporates, however,
once some intuition is developed about the interior structure of the 3-sphere. As we noted above, a 3-sphere can be
constructed by gluing the boundaries of two copies of a 3-ball (i.e., ordinary, solid, ball). Imagine then that we have
two copies of a 3-ball, and that these copies are superimposed so that their 2-spherical boundaries match---i.e., they
occupy the same space (not in the quantum mechanical sense, but in the topological sense). Imagine further that these
boundaries are then glued together, point-by-point. Thus each point of the boundary of one of the two balls is really
the same point as the corresponding point of the boundary of the other ball. The resulting space is a 3-sphere. The
points of this 3-sphere are then all of the interior points of the 3-ball counted twice, plus all of the boundary points
counted once.${\;}$Every one
of these sum total of points is then an interior point of the 3-sphere, with the boundary of the
original 3-ball${\;}$being its 2-spherical equator. Thus, a 3-sphere itself has no boundary at all. In fact, we might
actually be living in such a 3-spherical universe without being aware of it${\,}$\ocite{3-sphere}. A 3-sphere
is thus much like the 3-dimensional space we are${\;}$familiar with, but with some counterintuitive properties. 
Imagine now an insect crawling radially upwards from the center of the 3-ball (which is now one of the hemispheres of
${S^3}$), and being watched by an inhabitant of ${{\rm I\!R}^4}$. When the insect reaches the boundary of the 3-ball
(which is now the equator of ${S^3}$), it will be completely oblivious to its existence, and will carry on its journey
without noticing anything unusual. The inhabitant of ${{\rm I\!R}^4}$, however, will${\;}$witness the insect reach the
boundary, flip its ${{\rm I\!R}^3}$ direction, and continue its journey back towards the center of the${\;}$3-ball. 

It should now be easy to see how the right hand side of equation (\ref{21}), namely
${\,-\,\cos\theta_{\bf ab}\,-\,C_{\bf c}(\lambda)\,\sin\theta_{\bf ab}}$, represents all of the points of ${S^3}$
along the direction ${\bf c}$ traversed by the insect, when the vectors ${\bf a}$ and ${\bf b}$
are moved within the plane perpendicular to its radial line of motion \ocite{Zulli}. 
The point ${C_{\bf c}(\lambda)}$ then represents
the turning point of the insect at the boundary of the 3-ball, and ${-\,\cos\theta_{\bf ab}}$ plays the role of the fourth
dimension within the embedding space ${{\rm I\!R}^4}$. Each point traversed by the insect is thus either a ${+1}$ or a
${-1}$ point of ${S^3}$, depending on the angle ${\theta_{\bf ab}}$ and the complete state ${\lambda}$. The points
${A_{\bf a}(\lambda)}$, ${B_{\bf b}(\lambda)}$, and ${C_{\bf c}(\lambda)}$ are still confined to the 2-spherical
boundary of the 3-ball, but this boundary is now the equator of the 3-sphere. If the directions ${\bf a}$ and ${\bf b}$
are also allowed to move outside the confines of the perpendicular plane, then the radial direction ${\bf c}$ of the motion
of the insect too begins to vary (since ${\bf c}$ is defined as ${\frac{{\bf a}\times{\bf b}}{|{\bf a}\times{\bf b}|}}$),
providing a complete sweep of every single point of the 3-sphere. Thus, far from being mysterious, the right hand side
of equation (\ref{21}) is actually a very natural (and quite well known \ocite{Frankel}\ocite{Ryder}) representation
of the points of the 3-sphere, each one being either ${+1}$ or ${-1}$, and representing a product of the two EPR elements
of reality. Finally, as alluded to in the Fig.${\,}$3 above, an alternative, more analytical dissection of the 3-sphere
by means of the celebrated Hopf fibration, as a U(1) bundle over ${S^2}$, may also be useful in this context,
as we have suggested in Ref.${}$\ocite{experiment}.

We are now well equipped to compute the topologically correct local-realistic correlations between the EPR elements
of reality---i.e., correlations between the points of a 2-sphere rather than the real line. Using equation (\ref{21})
(which corresponds to
an embedded picture of ${S^2}$ into ${{\rm I\!R}^3}$), we immediately see that the correct correlations are given by  
\begin{equation}
{\cal E}({\bf a},\,{\bf b})\,=\int_{\Lambda}
A_{\bf a}(\lambda)\,B_{\bf b}(\lambda)\;d\rho(\lambda)\,=\,-\,\cos\theta_{\bf ab}\int_{\Lambda}d\rho(\lambda)
\,-\,\sin\theta_{\bf ab}\int_{\Lambda}\,C_{\bf c}(\lambda)\;\;d\rho(\lambda).\label{q-prob-q}
\end{equation}
Now the second term on the right hand side of this result vanishes identically for more than one reason. To begin with, it
involves an average of the functions ${C_{\bf c}(\lambda)}$, and hence is necessarily zero, thanks to the relations
(\ref{improveBell's}) and (\ref{0}). Moreover, operationally the functions ${C_{\bf c}(\lambda)}$ themselves are
necessarily zero, because they represent
measurement results along the direction that is not only orthogonal to both ${\bf a}$
and ${\bf b}$, but also strictly exclusive to them both. That is to say, any detector along the direction ${\bf c}$ would
necessarily yield a null result, provided the detectors along the directions ${\bf a}$ and ${\bf b}$ have yielded
non-null results, because the direction ${\bf c}$ is defined as ${\frac{{\bf a}\times{\bf b}}{|{\bf a}\times{\bf b}|}}$.
Consequently, by substituting the above equation in the CHSH string of expectation values, we arrive at the
topologically correct result
\begin{equation}
{\cal E}({\bf a},\,{\bf b})\,+\,{\cal E}({\bf a},\,{\bf b'})\,+\,
{\cal E}({\bf a'},\,{\bf b})\,-\,{\cal E}({\bf a'},\,{\bf b'})\,=\,
\left[\,-\,\cos\theta_{{\bf a}{\bf b}}\,-\,\cos\theta_{{\bf a}{\bf b'}}\,-\,
\cos\theta_{{\bf a'}{\bf b}}\,+\,\cos\theta_{{\bf a'}{\bf b'}}\,\right]\int_{\Lambda}d\rho(\lambda).
\label{My-CHSH}
\end{equation}
Assuming now that the distribution ${\rho(\lambda)}$ is normalized on the space ${\Lambda}$,
we finally arrive at the inequalities
\begin{equation}
-\,2\sqrt{2}\,\;\leq\;{\cal E}({\bf a},\,{\bf b})\,+\,{\cal E}({\bf a},\,{\bf b'})\,+\,
{\cal E}({\bf a'},\,{\bf b})\,-\,{\cal E}({\bf a'},\,{\bf b'})\;\leq\;+\,2\sqrt{2}.\label{My-inequa}
\end{equation}
This is of course {\it exactly} what is predicted by quantum mechanics. What is more, we have arrived at this
result without having to specify what the complete state ${\lambda}$ actually is. In fact, it is easy to see that
we can arrive at these inequalities without even having to specify whether the distribution ${\rho(\lambda)}$ is uniform
or not, or having to assume whether or not it remains the same for all four of the functionals involved in the CHSH
string of expectation values. All we need to assume, in fact, is that the distribution  ${\rho(\lambda)}$ remains
normalized on the space ${\Lambda}$ of the complete states. This implies that the
violations of Bell inequalities are purely topological phenomena that have nothing to do with quantum mechanics
{\it per se}, whether interpreted locally or ``non-locally'', realistically or ``non-realistically.''

This almost completes our refutation of Bell's theorem. Almost, because we have yet to deal with the formal part of
Bell's proof, which---as we noted after Eq.${}$(\ref{yever})---requires an assumption of commutativity. The 3-sphere,
on the other hand, containing a rather subtle form of non-commutativity within itself. Although this non-commutativity
is purely classical, it is important to demystify it to eliminate any lingering doubts about our
model. One way to demystify this non-commutativity is to note that the point ${BA}$ on the 3-sphere is a different point
from the point ${AB}$:
\begin{align}
B_{\bf b}(\lambda)\,A_{\bf a}(\lambda)\,&=\,-\,\cos\theta_{\bf ba}\,
-\,C_{\frac{{\bf b}\times{\bf a}}{|{\bf b}\times{\bf a}|}}(\lambda)\,\sin\theta_{\bf ba} \label{pope-1} \\
&=\,-\,\cos\theta_{\bf ab}\,-\,C_{({\bf -\,c})}(\lambda)\,\sin\theta_{\bf ab} \label{pope-2} \\
&=\,\pm\,1\,\in\,S^3,\;\,{\rm about\;\,the\;\,direction}\;\,
(-\cos\theta_{\bf ab}\,,\;+\,c_x\sin\theta_{\bf ab}\,,\;+\,c_y\sin\theta_{\bf ab}\,,\;+\,c_z\sin\theta_{\bf ab})
\;\,{\rm in}\,\;{\rm I\!R}^4. \label{p-maqpp}
\end{align}
Indeed, comparing equation (\ref{21}) with equation (\ref{pope-2})
we immediately see that ${AB}$ and ${BA}$ are two different
points of the 3-sphere, but with the same value. That is to say, both are equal to either ${+1}$ or ${-1}$, but located at
two different locations within ${S^3}$. Thus the factorized beable ${A_{\bf a}(\lambda)\,B_{\bf b}(\lambda)}$ in our model
would always yield the same value, regardless of the procedure used to detect it. And yet, it is easy to see that
${A_{\bf a}(\lambda)}$ and ${B_{\bf b}(\lambda)}$ do not commute:
\begin{equation}
\left[\,A_{\bf a}(\lambda),\;B_{\bf b}(\lambda)\,\right]
\,=\,-\,2\,C_{\bf c}(\lambda)\,\sin\theta_{\bf ab}
\,\equiv\,-\,2\,C_{{\bf a}\times{\bf b}}(\lambda)\,.\label{non-maqpp}
\end{equation}
How is this possible? In fact, there is nothing mysterious about this classical non-commutativity. It simply amounts to
a vector addition in the embedding space ${{\rm I\!R}^4}$. This can be easily seen by rewriting the above equation as
\begin{equation}
A_{\bf a}(\lambda)\,B_{\bf b}(\lambda)
\,=\,B_{\bf b}(\lambda)\,A_{\bf a}(\lambda)\,-\,2\,C_{\bf c}(\lambda)\,\sin\theta_{\bf ab}\,.\label{non-commm}
\end{equation}
As we noted earlier, the left hand side of this equation is a point of ${S^3}$, which can be indexed by its coordinates
in ${{\rm I\!R}^4}$, as spelt out in Eq.${\,}$(\ref{q-mappq}). The first term on the right hand side, on the other hand,
is another point of ${S^3}$, which can be indexed by its coordinates in ${{\rm I\!R}^4}$, as spelt out in
Eq.${\,}$(\ref{p-maqpp}). And the remaining term is proportional to a third point of ${S^3}$, indexed by the coordinates
${(\,0\,,\;c_x\,,\;c_y\,,\;c_z)}$ in ${{\rm I\!R}^4}$. Using these three sets of coordinates it is now easy to see that
the non-commutativity in (\ref{non-maqpp}) simply amounts to a vector addition in ${{\rm I\!R}^4}$. The embedding space
${{\rm I\!R}^4}$ thus facilitates a metric on ${S^3}$, and respecting equation (\ref{non-maqpp}) then amounts to
respecting this metric topology of ${S^3}$.

Now, as we have already noted, operationally the right hand side of Eq.${\,}$(\ref{non-maqpp}) would always vanish,
because ${C_{\bf c}(\lambda)}$ represents a measurement result along the direction that is both orthogonal and exclusive
to the directions ${\bf a}$ and ${\bf b}$ on the left hand side. In other words, operationally ${AB}$ and ${BA}$ are
identical to each other, which is consistent with the fact that they are the same numerical numbers to begin with. In
general, however, we must respect the non-commutativity in Eq.${\,}$(\ref{non-maqpp}), because otherwise we would be
making the same mistake that Bell made, and end up calculating correlations between the points of the real line instead
of a 2-sphere. This brings us to the stage of Eq.${\,}$(\ref{com}) in our derivation of the standard Bell-CHSH inequalities.
Let us then try to complete this derivation in our case, in the light of the above non-commutativity. Let us begin with
Eq.${\,}$(\ref{yever}), which we rewrite here for convenience:
\begin{equation}
|{\cal E}({\bf a},\,{\bf b})\,+\,{\cal E}({\bf a},\,{\bf b'})\,+\,
{\cal E}({\bf a'},\,{\bf b})\,-\,{\cal E}({\bf a'},\,{\bf b'})|\,
\leq\,\sqrt{\int_{\Lambda}\left\{\,4\,+\,\left[\,A_{\bf a}(\lambda),\,
A_{\bf a'}(\lambda)\,\right]\left[\,B_{\bf b'}(\lambda),\,
B_{\bf b}(\lambda)\,\right]\,\right\}\;d\rho(\lambda)\,}\,.\label{never-before}
\end{equation}
Using Eq.${\,}$(\ref{non-maqpp}) this inequality can be rewritten as:
\begin{equation}
|{\cal E}({\bf a},\,{\bf b})\,+\,{\cal E}({\bf a},\,{\bf b'})\,+\,
{\cal E}({\bf a'},\,{\bf b})\,-\,{\cal E}({\bf a'},\,{\bf b'})|\,
\leq\,\sqrt{\int_{\Lambda}\left\{\,4\,+\,\left[\,-\,2\,A_{{\bf a}\times{\bf a'}}(\lambda)\right]
\left[\,-\,2\,B_{{\bf b'}\times{\bf b}}(\lambda)\right]\,\right\}\;d\rho(\lambda)\,}\,.\label{before-op}
\end{equation}
And using Eq.${\,}$(\ref{21}) ({\it i.e.}, using the product point theorem), which we now rewrite as
\begin{equation}
A_{\bf a}(\lambda)\,B_{\bf b}(\lambda)\,=\,-\,{\bf a}\cdot{\bf b}
\,-\,C_{{\bf a}\times{\bf b}}(\lambda)\,,\label{specialcase}
\end{equation}
it can be further reduced to
\begin{align}
|{\cal E}({\bf a},\,{\bf b})\,+\,{\cal E}({\bf a},\,{\bf b'})\,+\,
{\cal E}({\bf a'},\,{\bf b})\,-\,{\cal E}({\bf a'},\,{\bf b'})|\,
&\leq\,\sqrt{\int_{\Lambda}\left\{\,4\,+\,4\,\left[\,-\,({\bf a}\times{\bf a'})\cdot({\bf b'}\times{\bf b})
\,-\,C_{({\bf a}\times{\bf a'})\times({\bf b'}\times{\bf b})}(\lambda)\,\right]\,\right\}\;d\rho(\lambda)} \\
&\leq\,\sqrt{\left\{\,4\,-\,4\;({\bf a}\times{\bf a'})\cdot({\bf b'}\times{\bf b})\right\}\!\!\int_{\Lambda}\!d\rho(\lambda)
\,-\,4\!\!\int_{\Lambda}\!\!
C_{({\bf a}\times{\bf a'})\times({\bf b'}\times{\bf b})}(\lambda)\;d\rho(\lambda)}\,.\label{before-op-2}
\end{align}
Now the last integral under the radical is proportional to the integral
\begin{equation}
\int_{\Lambda}\,C_{\bf z}(\lambda)\,\;d\rho(\lambda)\,,\,\;\;{\rm where}\;\;{\bf z}:=
\frac{({\bf a}\times{\bf a'})\times({\bf b'}\times{\bf b})}{|({\bf a}\times{\bf a'})\times({\bf b'}\times{\bf b})|}\,,
\label{q-ccc}
\end{equation}
which---as we have already noted---vanishes identically for more than one reason. If, moreover, we assume
that the distribution ${\rho(\lambda)}$ remains normalized on the space ${\Lambda}$, then the above inequality reduces to
\begin{equation}
|{\cal E}({\bf a},\,{\bf b})\,+\,{\cal E}({\bf a},\,{\bf b'})\,+\,
{\cal E}({\bf a'},\,{\bf b})\,-\,{\cal E}({\bf a'},\,{\bf b'})|\,
\leq\,2\,\sqrt{\,1-({\bf a}\times{\bf a'})\cdot({\bf b'}\times{\bf b})}\,.\label{before-op-3}
\end{equation}
Finally, by noticing that ${\,-1\leq({\bf a}\times{\bf a'})\cdot({\bf b'}\times{\bf b})\leq +1\,}$, we arrive at
the inequalities
\begin{equation}
-\,2\sqrt{2}\,\;\leq\;{\cal E}({\bf a},\,{\bf b})\,+\,{\cal E}({\bf a},\,{\bf b'})\,+\,
{\cal E}({\bf a'},\,{\bf b})\,-\,{\cal E}({\bf a'},\,{\bf b'})\;\leq\;+\,2
\sqrt{2}\,,\label{My-inequa-2}
\end{equation}
which are again {\it exactly} the inequalities predicted by quantum mechanics. We have derived these inequalities, however,
entirely local-realistically, by respecting the correct topological structure of the EPR elements of reality. Moreover,
we have derived them without specifying the complete state ${\lambda}$, without employing
any averaging procedure involving the third direction ${C_{\bf z}(\lambda)}$, and without assuming that the distribution
${\rho(\lambda)}$ remains uniform throughout the experiment. This time, however, we did have to assume that it remains
the same for all four of the expectation functionals of the CHSH string, in addition to assuming
that it remains normalized on the space ${\Lambda}$ of the complete states. 

By now readers familiar with our Clifford-algebraic model of Refs.${\,}$\ocite{Christian}\ocite{Further}\ocite{experiment}
would have recognized that it is a special case of the above described local-realistic framework. In the model we explicitly
specify the complete state ${\lambda}$, and take it to be a trivector ${\boldsymbol\mu}$ with unspecified handedness:
${{\boldsymbol\mu}:=\pm\,I}$. Here ${I}$ is the fundamental trivector that determines the entire structure of the
Euclidean space \ocite{experiment}. The sign ambiguity in ${\boldsymbol\mu}$ thus leaves the handedness of${\;}$the
Euclidean space unspecified. And it is this freedom of choice between the right and left-handed
Euclidean spaces that is the local hidden variable in our model. Given three unit directions ${\bf a}$, ${\bf b}$,
${\bf c}$ in ${{\rm I\!R}^3}$, the projections ${{\boldsymbol\mu}\cdot{\bf a}}$, ${{\boldsymbol\mu}\cdot{\bf b}}$,
and ${{\boldsymbol\mu}\cdot{\bf c}}$ are then unit bivectors (2-blades), which represent the EPR elements of reality
(i.e., measurement results) as before:
\begin{align}
{\boldsymbol\mu}\cdot{\bf a}\,&=\,\pm\,1\,\in\,S^2,\;\;
{\rm about\;\,the\;\,direction}\;\,{\bf a}\;\,{\rm in}\,\;{\rm I\!R}^3, \notag \\
{\boldsymbol\mu}\cdot{\bf b}\,&=\,\pm\,1\,\in\,S^2,\;\;
{\rm about\;\,the\;\,direction}\;\,{\bf b}\;\,{\rm in}\,\;{\rm I\!R}^3,\;\;\;\;\;\;\;\; \notag \\
{\rm and}\;\;\;{\boldsymbol\mu}\cdot{\bf c}\,&=\,\pm\,1\,\in\,S^2,\;\;
{\rm about\;\,the\;\,direction}\;\,{\bf c}\;\,{\rm in}\,\;{\rm I\!R}^3,\label{provejoy's}
\end{align}
These equations are thus special cases of the equations (\ref{proveBell's}), with the special case of equation
(\ref{specialcase}) being
\begin{equation}
(\,{\boldsymbol\mu}\cdot{\bf a})(\,{\boldsymbol\mu}\cdot{\bf b})\,
=\,-\,{\bf a}\cdot{\bf b}\,-\,{\boldsymbol\mu}\cdot({\bf a}\times{\bf b}).\label{bi-identity}
\end{equation}
Unfortunately, this model has been grossly misunderstood by the critics, mainly because of a rather spurious
issue of representation \ocite{reply}. What the critics have failed to comprehend is that the bivectors
${{\boldsymbol\mu}\cdot{\bf a}}$ and ${{\boldsymbol\mu}\cdot{\bf b}}$---no matter how they may ``look like''---simply
represent points of a 2-sphere \ocite{experiment}, with {\it no} ${\,}$properties apart from the ones specified in
the above equations. In particular, despite appearances, neither the trivector ${\boldsymbol\mu}$ nor the vector
${\bf n}$ is an intrinsic part of the bivector ${{\boldsymbol\mu}\cdot{\bf n}}$. The reader should therefore
guard against any psychological tendency to see structure within the symbol ${{\boldsymbol\mu}\cdot{\bf n}}$, other
than a binary measurement result ``${\,\pm\,1}$ about ${\bf n}$.'' Within Clifford algebra the symbol
${{\boldsymbol\mu}\cdot{\bf n}}$ stands for an abstract entity---sometimes referred to as a ``2-blade''---with quite
a distinct meaning from a similar looking symbol within vector algebra. In fact, because they are simply the
points of a unit 2-sphere, these 2-blades provide {\it the} correct representation of the EPR elements of reality.
Therefore, any suggestion of ``extracting information'' form the symbol ${{\boldsymbol\mu}\cdot{\bf n}}$---as
insisted by some of the critics \ocite{reply}---completely misses the very meaning of this two-centuries old concept.
It {\it is} the desired information. It {\it is} the correct measurement result. It {\it is} the true EPR element
of reality. Once this is understood, it is easy to see that both sides of Eq.${\,}$(\ref{bi-identity})
represent nothing but a point of a 3-sphere, which satisfies the map (\ref{Lus-map}), together with the properties
(\ref{Lucien}). Again, neither the
trivector ${\boldsymbol\mu}$ nor the vectors ${\bf a}$ and ${\bf b}$ are intrinsic parts of this point, but live
rather in the embedding space ${{\rm I\!R}^3}$. Consequently, Eq.${}$(19) of Ref.${\,}$\ocite{Christian}, namely
\begin{equation}
{\cal E}_{c.v.}({\bf a},\,{\bf b})=\int_{{\cal V}_3}
(\,{\boldsymbol\mu}\cdot{\bf a}\,)
(\,{\boldsymbol\mu}\cdot{\bf b}\,)\;\,d{\boldsymbol\rho}({\boldsymbol\mu})\,=\,-\,{\bf a}\cdot{\bf b}\,,\label{newderive}
\end{equation}
provides {\it the} correct correlations between the points of a 2-sphere, and hence between the EPR elements of reality.
The remaining details of the model are exactly the same as those of the general framework discussed above \ocite{Further}.

Despite the elegance of this model, however, we have refrained from using it in this paper, for we do not wish to
alienate the readers not familiar with the Clifford-algebraic representations of the 3-sphere. But it is worth noting that
the notations of this model allow a straightforward generalization of the theorem discussed above to arbitrary number
of points of a 2-sphere. For example the product of any three points of a 2-sphere immediately gives
\begin{equation}
(\,{\boldsymbol\mu}\cdot{\bf a})(\,{\boldsymbol\mu}\cdot{\bf b})(\,{\boldsymbol\mu}\cdot{\bf c})\,=\,
{\bf a}\cdot({\bf b}\times{\bf c})\,+\,
{\boldsymbol\mu}\cdot\left\{\,{\bf a}\times({\bf b}\times{\bf c})\,-\,
{\bf a}\,({\bf b}\cdot{\bf c})\,\right\},\label{ti-identity}
\end{equation}
which---being a scalar plus a bivector---is clearly a point of a 3-sphere. Conversely, the points of any 3-sphere can
always be represented by a sum of a scalar and a bivector. For example, the product of any four points of a 2-sphere,
\begin{align}
(\,{\boldsymbol\mu}\cdot{\bf a})(\,{\boldsymbol\mu}\cdot{\bf b})
(\,{\boldsymbol\mu}\cdot{\bf c})(\,{\boldsymbol\mu}\cdot{\bf d})&\,=\,
({\bf a}\cdot{\bf b})({\bf c}\cdot{\bf d})\,-\,({\bf a}\times{\bf b})\cdot({\bf c}\times{\bf d})\;+ \notag \\
&\;\;\;\;\;\;\;\;{\boldsymbol\mu}\cdot\left\{\,({\bf a}\cdot{\bf b})({\bf c}\times{\bf d})\,+\,
({\bf c}\cdot{\bf d})({\bf a}\times{\bf b})\,-\,
({\bf a}\times{\bf b})\times({\bf c}\times{\bf d})\,\right\},\label{bi-bi-bi-identy}
\end{align}
is clearly a scalar plus a bivector, and hence a point of a 3-sphere. In fact, the last equation can also be written as
\begin{align}
\!\!\!\left\{\,-\,{\bf a}\cdot{\bf b}\,-\,{\boldsymbol\mu}\cdot({\bf a}\times{\bf b})\,\right\}\,
\left\{\,-\,{\bf c}\cdot{\bf d}\,-\,{\boldsymbol\mu}\cdot({\bf c}\times{\bf d})\right\}&\,=\,
({\bf a}\cdot{\bf b})({\bf c}\cdot{\bf d})\,-\,({\bf a}\times{\bf b})\cdot({\bf c}\times{\bf d})\;+ \notag \\
&\;\;\;\;\;\;\;\;{\boldsymbol\mu}\cdot\left\{\,({\bf a}\cdot{\bf b})({\bf c}\times{\bf d})\,+\,
({\bf c}\cdot{\bf d})({\bf a}\times{\bf b})\,-\,
({\bf a}\times{\bf b})\times({\bf c}\times{\bf d})\,\right\},\label{bi-bi-identity}
\end{align}
which shows that the product of any two points of a 3-sphere itself remains within the 3-sphere. Indeed, it is well
known that ${S^3}$---being homeomorphic to the Lie group SU(2)---remains closed under multiplication of its points.

It is from this unique property of ${S^3}$ that the conceptual power and economy of our local-realistic framework stems.
Let us then conclude this section by using this property to further consolidate our framework. Inspired by the EPR
argument, we began this section with the maps ${A_{\bf a}(\lambda): {\rm I\!R}^3\!\times\Lambda\rightarrow S^2}$ and
${B_{\bf b}(\lambda): {\rm I\!R}^3\!\times\Lambda\rightarrow S^2}$. That is, we began by recognizing that the EPR elements
of reality for the singlet state are points of a unit 2-sphere. As we saw above, however, a 2-sphere is simply an
equator, or one of the great-spheres of the 3-sphere, just as a 1-sphere is an equator or a great-circle of a 2-sphere.
Thus, for the singlet state (\ref{single}), an EPR element of reality is just as much a point of a 3-sphere as it is
of a 2-sphere: It is a point of ${S^3}$ that has a vanishing value for one of its four embedding parameters within
${{\rm I\!R}^4}$. But this is generally not easy for us to visualize, because 3-sphere is not as intuitively accessible
to us as the 2-sphere. In fact, as we have discussed elsewhere \ocite{experiment}, there are deeper reasons why the EPR
elements of reality should be represented as points of a 3-sphere rather than a 2-sphere. To begin with, the 3-sphere is
the only sphere related to an associative algebra that is connected, simply connected, and parallelizable, with both it
and its projective space being orientable (the 2-sphere, on the other hand, is neither parallelizable, nor is its
projective plane orientable \ocite{Penrose}). In physical terms this means that every point of a 3-sphere can be
unambiguously associated with a measurement result of the spin of a spin-1/2 particle. Conversely, for a given quantum
state the correct topology of the EPR elements of reality is not difficult to determine, at least in the simple cases.
Take, for example, the general case of two-level systems we have been considering. The most general form of
such a system is of course
\begin{equation}
|\psi\rangle\,=\,\xi_1\,|\,+\,-\,\rangle\,+\,\xi_2\,|\,-\,+\,\rangle\,,\label{sch-singlet}
\end{equation}
where ${\xi_1}$ and ${\xi_2}$ are complex numbers satisfying the normalization condition
${|\,\xi_1\,|^2+|\,\xi_2\,|^2=1}$, which is equivalent to
\begin{equation}
\xi_{1r}^2\,+\,\xi_{1i}^2\,+\,\xi_{2r}^2\,+\,\xi_{2i}^2\,=\,1\,,\;\;\;\;\;\;{\rm with}\;\;\;
\xi_1\,:=\,\xi_{1r}\,+\,i\,\xi_{1i}\;\;\;{\rm and}\;\;\;\xi_2\,:=\,\xi_{2r}\,+\,i\,\xi_{2i}\,.\label{ignore-1}
\end{equation}
But this is evidently the defining equation of a unit 3-sphere, embedded in ${{\rm I\!R}^4}$. On the other hand,
${|\,\xi_1\,|^2}$ and ${|\,\xi_2\,|^2}$, respectively, are the probabilities of realizing (or actualizing) the
states ${|\,+\,-\,\rangle}$ and ${|\,-\,+\,\rangle}$. Consequently, recalling the EPR argument for the singlet
state from Section II, it is easy to see that there is a one-to-one correspondence in the present case between
the points of a 3-sphere and the EPR elements of reality. Thus the correct topological space of the EPR elements
of reality in the present case is a unit 3-sphere. More generally, for any ${n-}$level system the most likely
candidate for the correct topological space of the EPR elements of reality would be a unit ${(2n\!-\!1)}$-sphere.

These considerations then lead us to consolidate our local-realistic framework for any two-level system as follows.
We begin by replacing our tentative local maps (\ref{s2localmaps}) with more general and appropriate local maps
\begin{equation}
{\mathscr A}_{\bf a}(\lambda): {\rm I\!R}^3\!\times\Lambda\longrightarrow {S^3}, \;\;\;
{\mathscr B}_{\bf b}(\lambda): {\rm I\!R}^3\!\times\Lambda\longrightarrow {S^3}, \;\;\;
{\mathscr C}_{\bf c}(\lambda): {\rm I\!R}^3\!\times\Lambda\longrightarrow {S^3}, \;\;\;
{\mathscr D}_{\bf d}(\lambda): {\rm I\!R}^3\!\times\Lambda\longrightarrow {S^3}, \,\;\dots\,,\label{maps-s3new}
\end{equation}
(which are simply points of ${S^3}$), and impose the following locality condition on the joint beable
${({\mathscr A}_{\bf a}\,{\mathscr B}_{\bf b}\,{\mathscr C}_{\bf c}\,{\mathscr D}_{\bf d}\,\dots\,)(\lambda)}$:
\begin{equation}
({\mathscr A}_{\bf a}\,{\mathscr B}_{\bf b}\,{\mathscr C}_{\bf c}\,{\mathscr D}_{\bf d}\,\dots\,)(\lambda)\,=\,
{\mathscr A}_{\bf a}(\lambda)\,{\mathscr B}_{\bf b}(\lambda)\,{\mathscr C}_{\bf c}(\lambda)\,
{\mathscr D}_{\bf d}(\lambda)\,\dots\,=\,\pm\,1\,\in\,S^3.\label{mylocality}
\end{equation}
The crucial observation here is that, since 3-sphere happens to be closed under multiplication of its points,
the product on the right hand side of this equation is necessarily a point of the 3-sphere. Consequently, our
generalized framework respects the notion of local causality---or factorizability---just as strictly as the framework
considered by Bell. That is to say, just as in Bell's case (cf.${\;}$Eq.${\,}$(\ref{locconbell})), the joint beable
${({\mathscr A}_{\bf a}\,{\mathscr B}_{\bf b}\,{\mathscr C}_{\bf c}\,{\mathscr D}_{\bf d}\dots\,)(\lambda)}$
necessarily satisfies the map
\begin{equation}
({\mathscr A}_{\bf a}\,{\mathscr B}_{\bf b}\,{\mathscr C}_{\bf c}\,{\mathscr D}_{\bf d}\,\dots\,)(\lambda):
\,S^3\times\,S^3\times\,S^3\times\,S^3\,\dots\,\longrightarrow\,S^3,\label{3spherecause}
\end{equation}
and hence any point of a 3-sphere---such as ${{\mathscr P}_{\bf n}(\lambda)}$---can always be factorized into two or more
points of the same sphere. The analogous map considered by Bell for the points of a 0-sphere is then clearly a rather
simplistic choice, if at all a correct one. More importantly, with the above multiplication map exhibiting the
locality of our manifestly realistic framework, the internal consistency of the counterexample in
\ocite{Christian}\ocite{Further}\ocite{experiment} is now beyond the shadow of doubt.

\section{Exact, Local, and Realistic Completions of the GHZ-3, GHZ-4, and Hardy States}

Let us now turn to some variants of Bell's theorem---such as the GHZ or Hardy theorems \ocite{GHSZ}\ocite{Hardy}---to
show that the local-realistic framework introduced above has a much wider validity. In many presentations of such theorems
certain hand-waving narratives are employed to accentuate the mystique of quantum mechanics. But at the heart of these
theorems is nothing but an old-fashioned comparison of expectation values, just as it
is in the case of Bell's theorem. In this respect, Ref.${\,}$\ocite{GHSZ}---with its detailed calculations---is a
welcome exception, and we shall follow it closely.

\subsection{Exact Local-Realistic Completion of the Hardy State}

As our second example of a Bell-type theorem, let us consider Hardy's ``non-locality proof'' \ocite{Hardy}, which has
been hailed as the best version of Bell's theorem \ocite{Mermin}. The two-particle entangled state considered by Hardy
is of the form
\begin{equation}
|\Psi_{\bf z}\rangle\,=\,\frac{1}{\sqrt{1+\cos^2\theta\,}\,}\,\left\{\,\cos\theta\,\Bigl(|{\bf z},\,+\rangle_1\otimes
|{\bf z},\,-\rangle_2\,+\,|{\bf z},\,-\rangle_1\otimes|{\bf z},\,+\rangle_2\Bigr)\,-\;
\sin\theta\,\Bigl(|{\bf z},\,+\rangle_1\otimes|{\bf z},\,+\rangle_2\Bigr)\right\},\label{hardy-single}
\end{equation}
which represents an ensemble of spin-1/2 particles, just as the state (\ref{single}), but unlike (\ref{single}) it is
not a rotationally invariant state. In fact it is characterized by a single parameter ${\theta\in\{0,\,\pi/2\}}$ such
that, depending on the value of this parameter, it is either an entangled state or a product state. For instance, for
${\theta=\pi/2}$ it is a product state, whereas for ${\theta=0}$ it is a maximally entangled state (belonging to
the triplet family of the spin-1/2 particles). If we now consider measuring spin components of the particles along
the directions ${\bf a}$ and ${\bf a'}$ at the one end of the observation station and along the directions ${\bf b}$
and ${\bf b'}$ at the other end, then the above state leads to the following quantum mechanical predictions:
\begin{align}
&\langle\Psi_{\bf z}\,|\,{\bf a'},\,+\rangle_1\,\otimes\,|{\bf b}\,,\,+\rangle_2\,=\,0\,, \label{d}\\
&\langle\Psi_{\bf z}\,|\,{\bf a}\,,\,+\rangle_1\,\otimes\,|{\bf b'},\,+\rangle_2\,=\,0\,, \label{e} \\
&\langle\Psi_{\bf z}\,|\,{\bf a}\,,\,-\rangle_1\,\otimes\,|{\bf b}\,,\,-\rangle_2\,=\,0\,, \label{f} \\
&\langle\Psi_{\bf z}\,|\,{\bf a'},\,+\rangle_1\,\otimes\,|{\bf b'},\,+\rangle_2\,=\,
\frac{\,\sin\theta\,\cos^2\theta}{\sqrt{1+\cos^2\theta\,}\,}\,, \label{g}
\end{align}
\vspace{-0.5cm}
\begin{align}
\langle\Psi_{\bf z}\,|\,{\bf a'},\,+\rangle_1\,\otimes\,|{\bf b}\,,\,-\rangle_2&\,=\,
\frac{\,\cos^2\theta}{\sqrt{1+\cos^2\theta\,}\,}\,,
&\langle\Psi_{\bf z}\,|\,{\bf a}\,,\,-\rangle_1\,\otimes\,|{\bf b'},\,+\rangle_2\,=\,
\frac{\,\cos^2\theta}{\sqrt{1+\cos^2\theta\,}\,}\,,\;\;\;\;{}\label{h} \\
\langle\Psi_{\bf z}\,|\,{\bf a'},\,+\rangle_1\,\otimes\,|{\bf b'},\,-\rangle_2&\,=\,
\frac{\,\cos^3\theta}{\sqrt{1+\cos^2\theta\,}\,}\,,
&\langle\Psi_{\bf z}\,|\,{\bf a}\,,\,+\rangle_1\,\otimes\,|{\bf b}\,,\,-\rangle_2\,=\,
\frac{\,\cos\theta}{\sqrt{1+\cos^2\theta\,}\,}\,,\;\;\;\;{}\label{i} \\
\langle\Psi_{\bf z}\,|\,{\bf a'},\,-\rangle_1\,\otimes\,|{\bf b}\,,\,+\rangle_2&\,=\,
\frac{\,1\,}{\sqrt{1+\cos^2\theta\,}\,}\,,
&\langle\Psi_{\bf z}\,|\,{\bf a}\,,\,+\rangle_1\,\otimes\,|{\bf b'},\,-\rangle_2\,=\,
\frac{\,1\,}{\sqrt{1+\cos^2\theta\,}\,}\,,\;\;\;\;{}\label{j} \\
\langle\Psi_{\bf z}\,|\,{\bf a'},\,-\rangle_1\,\otimes\,|{\bf b'},\,+\rangle_2&\,=\,
\frac{\,\cos^3\theta}{\sqrt{1+\cos^2\theta\,}\,}\,,
&\langle\Psi_{\bf z}\,|\,{\bf a}\,,\,-\rangle_1\,\otimes\,|{\bf b}\,,\,+\rangle_2\,=\,
\frac{\,\cos\theta}{\sqrt{1+\cos^2\theta\,}\,}\,,\;\;\;\;{}\label{k} \\
\langle\Psi_{\bf z}\,|\,{\bf a'},\,-\rangle_1\,\otimes\,|{\bf b}\,,\,-\rangle_2&\,=\,
\frac{\,-\,\sin\theta\,\cos\theta}{\sqrt{1+\cos^2\theta\,}\,}\,,
&\langle\Psi_{\bf z}\,|\,{\bf a}\,,\,-\rangle_1\,\otimes\,|{\bf b'},\,-\rangle_2
\,=\,\frac{\,-\,\sin\theta\,\cos\theta}{\sqrt{1+\cos^2\theta\,}\,}\,,\;\;\;{}\label{l} \\
\langle\Psi_{\bf z}\,|\,{\bf a'},\,-\rangle_1\,\otimes\,|{\bf b'},\,-\rangle_2&\,=\,
\frac{-\,\sin\theta\,(1+\cos^2\theta)}{\sqrt{1+\cos^2\theta\,}}\,,
&\langle\Psi_{\bf z}\,|\,{\bf a}\,,\,+\rangle_1\,\otimes\,|{\bf b}\,,\,+\rangle_2\,=\,
\frac{\,-\,\sin\theta}{\sqrt{1+\cos^2\theta\,}\,}\,,\;\;\;\;{}\label{m}
\end{align}
provided we assume that ${{\bf a'}\cdot{\bf z}={\bf b'}\cdot{\bf z}=\cos2\theta}$ and
${{\bf a}\cdot{\bf z}={\bf b}\cdot{\bf z}=1}$. That is to say, provided we assume that
\begin{align}
|{\bf a'}\,,\,+\rangle_1\,&=\,+\,\cos\theta\,|{\bf a},\,+\rangle_1\,+\,\sin\theta\,|{\bf a},\,-\rangle_1 \\
|{\bf a'}\,,\,-\rangle_1\,&=\,-\,\sin\theta\,|{\bf a},\,+\rangle_1\,+\,\cos\theta\,|{\bf a},\,-\rangle_1 \\
|{\bf b'}\,,\,+\rangle_2\,&=\,+\,\cos\theta\,|{\bf b},\,+\rangle_2\,+\,\sin\theta\,|{\bf b},\,-\rangle_2 \\
{\rm and}\;\;\;|{\bf b'}\,,\,-\rangle_2\,&=\,-\,\sin\theta\,|{\bf b},\,+\rangle_2\,+\,
\cos\theta\,|{\bf b},\,-\rangle_2\,,\;\;\;\;\;\;\;\label{four-single}
\end{align}
with the directions ${\bf a'}$ and ${\bf b'}$ kept confined to the polar plane. Hardy's claim then is that no
local-realistic theory can reproduce these counterintuitive predictions of quantum mechanics. The claim would be
true of course, if the corresponding elements of reality were ``lined up'' as points of the real line, but by now
we know that they are not. By now we know that the elements of reality corresponding to any two-level system are
organized as points of a 3-sphere. Therefore, the correct question to ask in this context is: whether or not
local-realistic maps of the form
\begin{equation}
{\mathscr A}_{\bf a}(\theta,\,\lambda): {\rm I\!R}^3\!\times\Lambda\longrightarrow {S^3} \;\;\;\;\;{\rm and}\;\;\;\;\;
{\mathscr B}_{\bf b}(\theta,\,\lambda): {\rm I\!R}^3\!\times\Lambda\longrightarrow {S^3} \label{hardy-maps-my}
\end{equation}
(with ${\theta}$ being the known common cause \ocite{Bell-La}) can reproduce the above quantum mechanical
predictions. The answer to this question is of course in the affirmative. In fact, the above predictions are exhibiting
nothing more than correlations among various points of a 3-sphere. We can see this by specifying these points as follows
(cf.${\;}$Eq.${\,}$(\ref{21})):
\begin{align}
{\mathscr A}^{(+)}_{\bf a}(\theta,\,\lambda)&=\,\cos[\alpha(\theta)]\,+\,A_{\bf a}(\lambda)\,\sin[\alpha(\theta)]\,,\!\!\!\!
&{\mathscr A}^{(-)}_{\bf a}(\theta,\,\lambda)\,=\,\cos[\eta(\theta)]\,-\,A_{\bf a}(\lambda)\,\sin[\eta(\theta)]\,,
\label{se}\\
{\mathscr B}^{(+)}_{\bf b}(\theta,\,\lambda)&=\,\cos[\beta(\theta)]\,+\,B_{\bf b}(\lambda)\,\sin[\beta(\theta)]\,,\!\!\!\!
&{\mathscr B}^{(-)}_{\bf b}(\theta,\,\lambda)\,=\,\sin[\eta(\theta)]\,-\,B_{\bf b}(\lambda)\,\cos[\eta(\theta)]\,, \\
{\mathscr A}^{(+)}_{\bf a'}(\theta,\,\lambda)&=\,\cos[\gamma(\theta)]\,+\,A_{\bf a'}(\lambda)\,\sin[\gamma(\theta)]\,,
\!\!\!\!
&\,{\mathscr A}^{(-)}_{\bf a'}(\theta,\,\lambda)\,=\,\cos[\rho(\theta)]\,-\,A_{\bf a'}(\lambda)\,\sin[\rho(\theta)]\,,\! \\
{\mathscr B}^{(+)}_{\bf b'}(\theta,\,\lambda)&=\,\cos[\delta(\theta)]\,+\,B_{\bf b'}(\lambda)\,\sin[\delta(\theta)]\,,
\!\!\!\!
&\,{\mathscr B}^{(-)}_{\bf b'}(\theta,\,\lambda)\,=\,\cos[\nu(\theta)]\,-\,B_{\bf b'}(\lambda)\,\sin[\nu(\theta)]\,,\!\!
\label{hardybell}
\end{align}
where ${A_{\bf a}(\lambda)}$, ${A_{\bf a'}(\lambda)}$, ${B_{\bf b}(\lambda)}$, and ${B_{\bf b'}(\lambda)}$ are
points on the equator of ${S^3}$ (which, as we have noted, is a 2-sphere), and the angles
${\alpha(\theta)}$, ${\beta(\theta)}$,
${\gamma(\theta)}$, ${\delta(\theta)}$, ${\eta(\theta)}$, ${\rho(\theta)}$, and ${\nu(\theta)}$ are functions
only of the known common cause ${\theta}$, defined by
\begin{align}
\cot[\gamma(\theta)]\cot[\beta(\theta)]\,&=\,1-2\sin^2\theta\,=\,\cot[\alpha(\theta)]\cot[\delta(\theta)]\,,
\label{1-cons} \\
\cos[\alpha(\theta)+\beta(\theta)]\,&=\frac{\,-\,\sin\theta\,}{\sqrt{1+\cos^2\theta\,}\,}=\,
\cos[\rho(\theta)+\nu(\theta)]\,+\,\cos[\gamma(\theta)+\delta(\theta)]\,, \\
\cos[\gamma(\theta)+\delta(\theta)]\,&=\frac{\,\sin\theta\,\cos^2\theta}{\sqrt{1+\cos^2\theta\,}\,}\,, \label{con-gd} \\
\frac{\cos[\rho(\theta)]\,\sin[\eta(\theta)]\,-\,\cos[\nu(\theta)]\,\cos[\eta(\theta)]}{\sin[\rho(\theta)]\,
\cos[\eta(\theta)]\,-\,\sin[\nu(\theta)]\,\sin[\eta(\theta)]}\,&=\,1-2\sin^2\theta\,=\,
\frac{\cos[\gamma(\theta)]\,\sin[\eta(\theta)]\,-\,\cos[\delta(\theta)]\,\cos[\eta(\theta)]}{\sin[\delta(\theta)]\,
\sin[\eta(\theta)]\,-\,\sin[\gamma(\theta)]\,\cos[\eta(\theta)]}\,, \\
\frac{\cos[\alpha(\theta)]\,\cos[\nu(\theta)]\,-\,\cos[\rho(\theta)]\,\cos[\beta(\theta)]}{\sin[\rho(\theta)]\,
\sin[\beta(\theta)]\,-\,\sin[\alpha(\theta)]\,\sin[\nu(\theta)]}\,&=\,1-2\sin^2\theta\,=\,
\frac{\cos[\rho(\theta)]\,\sin[\eta(\theta)]\,-\,\cos[\nu(\theta)]\,\cos[\eta(\theta)]}{\sin[\rho(\theta)]\,
\cos[\eta(\theta)]\,-\,\sin[\nu(\theta)]\,\sin[\eta(\theta)]}\,, \\
\cos[\gamma(\theta)-\nu(\theta)]\,&=\frac{\,\cos^3\theta\,}{\sqrt{1+\cos^2\theta\,}\,}=
\,\cos[\rho(\theta)-\delta(\theta)]\,, \\
{\rm and}\;\;\;\sin[\alpha(\theta)+\eta(\theta)]\,&=\frac{\,\cos\theta\,}{\sqrt{1+\cos^2\theta\,}\,}=
\,\cos[\eta(\theta)-\beta(\theta)]\,.
\;\;\;\;\;\;\;\;\;\;\;\;\;\;\;\label{consangs}
\end{align}
These constraining relations among the functions ${\alpha(\theta)}$ to ${\nu(\theta)}$ arise for a number of reasons.
First, the local maps (\ref{se}) to (\ref{hardybell}), and hence the angles ${\alpha(\theta)}$ to ${\nu(\theta)}$,
must respect the topology of the 3-sphere if they are to be the genuine points of the 3-sphere. Thus, to begin with,
each point must remain normalized to unity. Moreover, since ${S^3}$ remains closed under multiplication, the end result of
a product of any number of these points (performed in any order, combination, or permutation) must also be normalized to
unity, while maintaining the general form (\ref{21}), in order to remain within the 3-sphere. Secondly,
they must also respect the fact that the state (\ref{hardy-single}) is rotationally non-invariant in a very
specific manner. Consequently, the corresponding elements of reality are not isotropically distributed
over the 3-sphere. More precisely, there is no perfect democracy among the points of the 3-sphere, and this lack of
democracy is manifested by the above constraints on the parameters of the embedding space. This can be readily seen,
for example, by setting ${\theta=0}$ and ${\theta=\pi/2}$, for which the state (\ref{hardy-single}) reduces to a maximally
entangled state and a product state, respectively. In both of these extremal cases the above constraints reduce to trivial
shifts among the angles, thereby restoring most of the symmetries of the system. The corresponding 3-sphere can
never be devoid of all asymmetries, however, because the state (\ref{hardy-single}) can never be reduced to the
rotationally invariant state such as (\ref{single}). Indeed, as we saw in the case of Bell's theorem, for the state
(\ref{single}) the analogous angles are always identically equal to ${\pi/2}$, with the corresponding elements of
reality always being on the equator of the 3-sphere. More importantly, explicit expressions of the functions
${\alpha(\theta)}$, ${\beta(\theta)}$, ${\gamma(\theta)}$, ${\delta(\theta)}$, ${\eta(\theta)}$, ${\rho(\theta)}$,
and ${\nu(\theta)}$ can be easily obtained by solving the above constraints, for example by the method of elimination.
The eight points specified in the equations (\ref{se}) to (\ref{hardybell}) are thus not only the {\it bona fide}
points of the 3-sphere, but are also completely characterized by a single known parameter ${\theta}$, just like the
quantum state (\ref{hardy-single}) itself. We will not need these explicit expressions for our calculations, however.
All we need to note for the calculations is that our local-realistic model for the Hardy state is also completely
characterized by the single known common cause ${\theta}$. Together with the complete state ${\lambda}$,
the correlations among the points (\ref{se}) to (\ref{hardybell}) of the 3-sphere would then exactly reproduce the
quantum mechanical predictions (\ref{d}) to (\ref{m}), as we now show.

The local-realistic derivation of the first of the predictions of the Hardy state, namely (\ref{d}), proceeds as follows:
\begin{align}
\!\!\int_{\Lambda}
{\mathscr A}^{(+)}_{\bf a'}(\theta,\,\lambda)\;{\mathscr B}^{(+)}_{\bf b}(\theta,\,\lambda)\,\;d\rho(\lambda)
\,&=\,\int_{\Lambda}\Bigl\{\,\cos[\gamma(\theta)]\,+\,A_{\bf a'}(\lambda)\,\sin[\gamma(\theta)]\,\Bigr\}\,
\Bigl\{\,\cos[\beta(\theta)]\,+\,B_{\bf b}(\lambda)\,\sin[\beta(\theta)]\,\Bigr\}\,\;d\rho(\lambda) \\
\,&=\,\int_{\Lambda}\Bigl\{\,\cos[\gamma(\theta)]\,\cos[\beta(\theta)]\,+\,A_{\bf a'}(\lambda)\,\sin[\gamma(\theta)]\,
\cos[\beta(\theta)] \notag \\
&\;\;\;\;\;\;\;\;\;\;
+\,B_{\bf b}(\lambda)\,\sin[\beta(\theta)]\,\cos[\gamma(\theta)]\,+\,
A_{\bf a'}(\lambda)\,B_{\bf b}(\lambda)\,\sin[\gamma(\theta)]\,\sin[\beta(\theta)]\,\Bigr\}\,\;d\rho(\lambda)\,.
\label{q-hard-1}
\end{align}
Using the product rule
${A_{\bf a'}(\lambda)\,B_{\bf b}(\lambda)\,=\,-\,{\bf a'}\cdot{\bf b}\,-\,C_{{\bf a'}\times{\bf b}}(\lambda)}$ from
Eq.${\,}$(\ref{specialcase}), this integral can be reduced to
\begin{align}
\int_{\Lambda}{\mathscr A}^{(+)}_{\bf a'}(\theta,\,\lambda)\;{\mathscr B}^{(+)}_{\bf b}(\theta,\,\lambda)\,\;d\rho(\lambda)
\,&=\,\Bigl\{\,\cos[\gamma(\theta)]\,\cos[\beta(\theta)]\,-\,({\bf a'}\cdot{\bf b})\,\sin[\gamma(\theta)]\,
\sin[\beta(\theta)]\,\Bigr\}\int_{\Lambda}d\rho(\lambda) \notag \\
&\;\;\;\;\;\;\;\;\;\;
+\int_{\Lambda}\Bigl\{\,A_{\bf a'}(\lambda)\,\sin[\gamma(\theta)]\,\cos[\beta(\theta)]\,+\,B_{\bf b}(\lambda)\,
\sin[\beta(\theta)]\,\cos[\gamma(\theta)] \notag \\
&\;\;\;\;\;\;\;\;\;\;\;\;\;\;\;\;\;\;\;
\;\;\;\;\;\;\;\;\;-\,C_{{\bf a'}\times{\bf b}}(\lambda)\,\sin[\gamma(\theta)]\,
\sin[\beta(\theta)]\,\Bigr\}\,\;d\rho(\lambda)
\,.\label{step-hard-1}
\end{align}
Now both terms on the right hand side of this equation vanish for different reasons. The first term vanishes because
of the relation (\ref{1-cons}), whereas the second term vanishes because of more than one reasons. To begin with, each
term within the integrand of the second term involves an average of the function like ${A_{\bf n}(\lambda)}$, and hence
is necessarily zero, thanks to the relations (\ref{improveBell's}) and (\ref{0}). But such an averaging procedure is
not necessary, since the entire integrand of the second term is proportional to a point of the 2-sphere about the
direction exclusive to both ${\bf a'}$ and ${\bf b}$, and hence operationally it would necessarily vanish. In
other words, with the first term vanished, the above equation reduces to
\begin{equation}
\int_{\Lambda}{\mathscr A}^{(+)}_{\bf a'}(\theta,\,\lambda)\;{\mathscr B}^{(+)}_{\bf b}(\theta,\,\lambda)\,\;d\rho(\lambda)
\,=\,0\,+\,|{\bf d}|\int_{\Lambda}D_{\frac{\bf d}{|{\bf d}|}}(\lambda)\;\, d\rho(\lambda)\,,\label{firsgonee}
\end{equation}
where ${D_{\frac{\bf d}{|{\bf d}|}}(\lambda)=\pm 1 \in S^2}$,
${\;}$and ${{\bf d}:=\{\,{\bf a'}\,\sin[\gamma(\theta)]\,\cos[\beta(\theta)]+{\bf b}\,\sin[\beta(\theta)]
\,\cos[\gamma(\theta)]-({{\bf a'}\times{\bf b}})\,\sin[\gamma(\theta)]\,\sin[\beta(\theta)]\,\}}$ is a direction
exclusive to both ${\bf a'}$ and ${\bf b}$. This can be recognized most transparently in the Clifford-algebraic
representation of Refs.${\,}$\ocite{Christian}\ocite{Further}\ocite{experiment}, in which
${D_{\bf d}({\boldsymbol\mu})\,\equiv\,
{\boldsymbol\mu}\cdot\{\,{\bf a'}\,\sin[\gamma(\theta)]\,\cos[\beta(\theta)]+{\bf b}\,\sin[\beta(\theta)]
\,\cos[\gamma(\theta)]-({{\bf a'}\times{\bf b}})\,\sin[\gamma(\theta)]\,\sin[\beta(\theta)]\,\}}$. It is now easy to
see that any detector along the direction ${\bf d}$ would necessarily yield a null result, provided the detectors
along the directions ${\bf a'}$ and ${\bf b}$ have yielded non-null results, and hence operationally the integrand on
the right hand side of the above equation would be necessarily zero. Thus, without needing to specify the complete
state ${\lambda}$ or its distribution ${\rho(\lambda)}$, and without needing to invoke any averaging procedure, we
arrive at the desired result:
\begin{equation}
\int_{\Lambda}{\mathscr A}^{(+)}_{\bf a'}(\theta,\,\lambda)\;{\mathscr B}^{(+)}_{\bf b}(\theta,\,\lambda)\,\;d\rho(\lambda)
\,=\,0\,.\label{step-hard-2}
\end{equation}
This is of course exactly the quantum mechanical prediction (\ref{d}), but we have derived it entirely local-realistically.

The next two predictions, (\ref{e}) and (\ref{f}), can be derived by using similar steps, so we simply make a note of them,
\begin{equation}
\int_{\Lambda}{\mathscr A}^{(+)}_{\bf a}(\theta,\,\lambda)\;{\mathscr B}^{(+)}_{\bf b'}(\theta,\,\lambda)\,\;d\rho(\lambda)
\,=\,0\,\;\;\;\;\;\;\;{\rm and}\;\;\;\;\;
\int_{\Lambda}{\mathscr A}^{(-)}_{\bf a}(\theta,\,\lambda)\;{\mathscr B}^{(-)}_{\bf b}(\theta,\,\lambda)\,\;d\rho(\lambda)
\,=\,0\,,
\end{equation}
and leave them as exercises. The prediction (\ref{g}), on the other hand, plays a crucial role in Hardy's
``non-locality proof'', and so we derive it here explicitly. As before, we proceed with our local-realistic derivation
as follows:
\begin{align}
\!\!\int_{\Lambda}
{\mathscr A}^{(+)}_{\bf a'}(\theta,\,\lambda)\;{\mathscr B}^{(+)}_{\bf b'}(\theta,\,\lambda)\,\;d\rho(\lambda)
\,&=\,\int_{\Lambda}\Bigl\{\,\cos[\gamma(\theta)]\,+\,A_{\bf a'}(\lambda)\,\sin[\gamma(\theta)]\,\Bigr\}\,
\Bigl\{\,\cos[\delta(\theta)]\,+\,B_{\bf b'}(\lambda)\,\sin[\delta(\theta)]\,\Bigr\}\,\;d\rho(\lambda) \\
\,&=\,\int_{\Lambda}\Bigl\{\,\cos[\gamma(\theta)]\,\cos[\delta(\theta)]\,+\,A_{\bf a'}(\lambda)\,\sin[\gamma(\theta)]\,
\cos[\delta(\theta)] \notag \\
&\;\;\;\;\;\;\;\;\;\;\;
+\,B_{\bf b'}(\lambda)\,\sin[\delta(\theta)]\,\cos[\gamma(\theta)]\,+\,
A_{\bf a'}(\lambda)\,B_{\bf b}(\lambda)\,\sin[\gamma(\theta)]\,\sin[\delta(\theta)]\,\Bigr\}\,\;d\rho(\lambda)\,.
\label{p-hard-12}
\end{align}
Using the product rule
${A_{\bf a'}(\lambda)\,B_{\bf b'}(\lambda)\,=\,-\,{\bf a'}\cdot{\bf b'}\,-\,C_{{\bf a'}\times{\bf b'}}(\lambda)}$
from Eq.${\,}$(\ref{specialcase}), this integral can be reduced to
\begin{align}
\int_{\Lambda}{\mathscr A}^{(+)}_{\bf a'}(\theta,\,\lambda)\;{\mathscr B}^{(+)}_{\bf b'}(\theta,\,\lambda)\,\;d\rho(\lambda)
\,&=\,\Bigl\{\,\cos[\gamma(\theta)]\,\cos[\delta(\theta)]\,-\,({\bf a'}\cdot{\bf b'})\,\sin[\gamma(\theta)]\,
\sin[\delta(\theta)]\,\Bigr\}\int_{\Lambda}d\rho(\lambda) \notag \\
&\;\;\;\;\;\;\;\;\;\;\;\;\;
+\int_{\Lambda}\Bigl\{\,A_{\bf a'}(\lambda)\,\sin[\gamma(\theta)]\,\cos[\delta(\theta)]\,+\,B_{\bf b'}(\lambda)\,
\sin[\delta(\theta)]\,\cos[\gamma(\theta)] \notag \\
&\;\;\;\;\;\;\;\;\;\;\;\;\;\;\;\;\;\;\;
\;\;\;\;\;\;\;\;\;\;\;\;-\,C_{{\bf a'}\times{\bf b'}}(\lambda)\,\sin[\gamma(\theta)]\,
\sin[\delta(\theta)]\,\Bigr\}\,\;d\rho(\lambda)
\,.\label{step-hnorard-13}
\end{align}
Now, in close analogy with the previous derivation, the second term on the right hand side of this equation vanishes
identically, because it involves averages of functions like ${A_{\bf n}(\lambda)}$, all of which vanish necessarily,
thanks to the relations (\ref{improveBell's}) and (\ref{0}). But once again such averaging procedures are not necessary,
since the entire integrand of the second term is proportional to a point of the 2-sphere about the direction exclusive
to both ${\bf a'}$ and ${\bf b'}$, and hence operationally it itself would vanish identically. On the other hand, thanks
to the relation (\ref{con-gd}) and the normalization condition for the distribution ${\rho(\lambda)}$, the first term
on the right hand side of the above equation is non-zero, and works out to be
\begin{equation}
\int_{\Lambda}{\mathscr A}^{(+)}_{\bf a'}(\theta,\,\lambda)\;{\mathscr B}^{(+)}_{\bf b'}(\theta,\,\lambda)
\,\;d\rho(\lambda)\,=\,\frac{\,\sin\theta\,\cos^2\theta}{\sqrt{1+\cos^2\theta\,}\,}\,. \label{gfinal}
\end{equation}
Thus, once again, without needing to specify the complete state ${\lambda}$ or its distribution ${\rho(\lambda)}$, and
without needing to invoke any averaging procedure, we have
local-realistically reproduced the {\it exact} quantum mechanical prediction.

In fact, the remaining twelve predictions of quantum mechanics---namely the predictions from (\ref{h}) to (\ref{m})---also
follow from analogous steps. (We leave their derivations as exercises since they do not play a direct role in Hardy's
theorem.) This shows that all sixteen predictions of the Hardy state, from (\ref{d}) to (\ref{m}), are simply
local-realistic correlations between the eight points (given by (\ref{se}) to (\ref{hardybell})) of a unit 3-sphere.
Moreover, since we have derived them without specifying either ${\lambda}$ or ${\rho(\lambda)}$, it is clear that these
correlations are purely topological effects. In other words, contrary to the conventional wisdom, they have nothing
whatsoever to do with ``non-locality'' or ``non-reality.''

\subsection{Exact Local-Realistic Completion of the Four-Particle GHZ State}

As our third explicit example, consider the four-particle Greenberger, Horne, Zeilinger state \ocite{GHSZ}:
\begin{equation}
|\Psi_{\bf z}\rangle\,=\,\frac{1}{\sqrt{2}\,}\,\Bigl\{|{\bf z},\,+\rangle_1\otimes
|{\bf z},\,+\rangle_2\otimes|{\bf z},\,-\rangle_3\otimes
|{\bf z},\,-\rangle_4\,
-\,|{\bf z},\,-\rangle_1\otimes|{\bf z},\,-\rangle_2\otimes
|{\bf z},\,+\rangle_3\otimes|{\bf z},\,+\rangle_4\Bigr\}.\label{ghz-single}
\end{equation}
Just like the Hardy state, this state too is rotationally non-invariant. There is a privileged direction, and
this direction is taken to be the ${\bf z}$-direction of the experimental setup \ocite{GHSZ}.
The ${\bf z}$-direction
is thus the axis of anisotropy of the system. The quantum mechanical expectation value in this state, of the product
of outcomes of the spin components---namely, the products of finding the spin of particle 1 along ${{\bf n}_1}$,
the spin of particle 2 along ${{\bf n}_2}$, etc.---is given by
\begin{equation}
{\cal E}^{\Psi_{\bf z}}_{{\!}_{Q.M.}\!}({\bf n}_1,\,{\bf n}_2,\,{\bf n}_3,\,{\bf n}_4)\,:=\,
\langle\Psi_{\bf z}|\,{\boldsymbol\sigma}\cdot{\bf n}_1\,\otimes\,
{\boldsymbol\sigma}\cdot{\bf n}_2\,\otimes\,{\boldsymbol\sigma}\cdot{\bf n}_3\,\otimes\,
{\boldsymbol\sigma}\cdot{\bf n}_4\,|\Psi_{\bf z}\rangle.\label{realobserve}
\end{equation}
This expectation value has been calculated in the Appendix F of Ref.${\,}$\ocite{GHSZ}. In the spherical
coordinates---with angles ${\theta_1}$ and ${\phi_2}$ representing the polar and azimuthal angles, respectively,
of the direction ${{\bf n}_1}$, etc.---it works out to be
\begin{equation}
{\cal E}^{\Psi_{\bf z}}_{{\!}_{Q.M.}\!}({\bf n}_1,\,{\bf n}_2,\,{\bf n}_3,\,{\bf n}_4)\,=\,
\cos\theta_1\,\cos\theta_2\,\cos\theta_3\,\cos\theta_4\,-\,\sin\theta_1\,
\sin\theta_2\,\sin\theta_3\,\sin\theta_4\,\cos\,\left(\,\phi_1\,+\,\phi_2\,-\,\phi_3\,-\,\phi_4\,\right).\label{q-preghz}
\end{equation}

Our goal now is to reproduce this result {\it exactly} within our local-realistic framework. The mystifying narratives
that usually accompany the above prediction would then become irrelevant. To this end, our first task is to determine the
correct topological space of the corresponding EPR elements of reality. As we will soon see, it cannot be a 3-sphere, for
the state (\ref{ghz-single}) represents, not a two-level, but a four-level system. Each of the two pairs of the spin-1/2
particles it represents has four alternatives available to it.
These alternatives can be represented by a state-vector of the form
\begin{equation}
|\psi\rangle\,=\,\xi_1\,|\,+\,+\,\rangle\,+\,\xi_2\,|\,+\,-\,\rangle\,+\,\xi_3\,|\,-\,+\,\rangle\,+\,
\xi_4\,|\,-\,-\,\rangle\,,\label{schmidt-single}
\end{equation}
where ${\xi_1}$, ${\xi_2}$, ${\xi_3}$, and ${\xi_4}$ are complex numbers satisfying the normalization condition
${|\,\xi_1\,|^2+|\,\xi_2\,|^2+|\,\xi_3\,|^2+|\,\xi_4\,|^2=1}$. This condition is equivalent to
\begin{equation}
\xi_{1r}^2\,+\,\xi_{1i}^2\,+\,\xi_{2r}^2\,+\,\xi_{2i}^2\,+\,\xi_{3r}^2\,+\,\xi_{3i}^2\,+\,\xi_{4r}^2\,+\,\xi_{4i}^2\,=\,1\,,
\label{ignore-2}
\end{equation}
which is the defining equation of a unit 7-sphere, embedded in ${{\rm I\!R}^8}$. And ${|\,\xi_1\,|^2}$, ${|\,\xi_2\,|^2}$,
${|\,\xi_3\,|^2}$, and ${|\,\xi_4\,|^2}$ are of course the probabilities of realizing or actualizing the states
${|\,+\,+\,\rangle}$, ${|\,+\,-\,\rangle}$, ${|\,-\,+\,\rangle}$, and ${|\,-\,-\,\rangle}$, respectively. Therefore, in
analogy with the cases we discussed earlier, there is a one-to-one correspondence in this case between the points of a
7-sphere and the EPR elements of reality. In other words, the correct topological space of the EPR elements of reality in
this case is a unit 7-sphere. Hence we begin our local-realistic description with four local maps of the form
\begin{equation}
{\mathbb A}_{{\bf n}_1}(\lambda): {\rm I\!R}^3\!\times\Lambda\longrightarrow {S^7}, \;\;\;
{\mathbb B}_{{\bf n}_2}(\lambda): {\rm I\!R}^3\!\times\Lambda\longrightarrow {S^7}, \;\;\;
{\mathbb C}_{{\bf n}_3}(\lambda): {\rm I\!R}^3\!\times\Lambda\longrightarrow {S^7}, \;\;\;{\rm and}\;\;\;
{\mathbb D}_{{\bf n}_4}(\lambda): {\rm I\!R}^3\!\times\Lambda\longrightarrow {S^7}\,.\label{n-ghz-maps-my}
\end{equation}
The crucial point to note here is that, just as the 0 and 3-spheres we discussed earlier, the 7-sphere
is also closed under multiplication of its points, and hence it too naturally preserves the locality condition of Bell:
\begin{equation}
({\mathbb A}_{{\bf n}_1}\,{\mathbb B}_{{\bf n}_2}\,{\mathbb C}_{{\bf n}_3}\,{\mathbb D}_{{\bf n}_4})(\lambda)\,=\,
{\mathbb A}_{{\bf n}_1}(\lambda)\,{\mathbb B}_{{\bf n}_2}(\lambda)\,{\mathbb C}_{{\bf n}_3}(\lambda)\,
{\mathbb D}_{{\bf n}_4}(\lambda)\,=\,\pm\,1\,\in\,S^7.\label{myourlocality}
\end{equation}
That is to say, just as in Bell's case (cf.${\;}$Eq.${\,}$(\ref{locconbell})), the joint beable
${({\mathbb A}_{{\bf n}_1}\,{\mathbb B}_{{\bf n}_2}\,{\mathbb C}_{{\bf n}_3}\,{\mathbb D}_{{\bf n}_4})(\lambda)}$
necessarily satisfies the map
\begin{equation}
({\mathbb A}_{{\bf n}_1}\,{\mathbb B}_{{\bf n}_2}\,{\mathbb C}_{{\bf n}_3}\,{\mathbb D}_{{\bf n}_4})(\lambda):
\,S^7\times\,S^7\times\,S^7\times\,S^7\,\longrightarrow\,S^7.\label{3sphereyour}
\end{equation}
Consequently, the product of any number of points of a 7-sphere is also a point of the 7-sphere, and---as would have
been demanded by Bell---any point of a 7-sphere can always be factorized into two or more points of the 7-sphere.

Equipped with this powerful mathematical constraint, we take our local beables 
${{\mathbb A}_{{\bf n}_1}(\lambda)}$, ${{\mathbb B}_{{\bf n}_2}(\lambda)}$,
${{\mathbb C}_{{\bf n}_3}(\lambda)}$, and ${{\mathbb D}_{{\bf n}_4}(\lambda)}$ to be the following four points
on the equator of a unit 7-sphere (which is of course a unit 6-sphere):
\begin{align}
{\mathbb A}_{{\bf n}_1}(\lambda)\,&=\,\pm\,1\,\in\,S^7,\;\;{\rm about\;\,the\;\,direction}\;\,
{\bf N}_1\,:=\,(\,-{n}_{1x},\,+{n}_{1y},\,-{n}_{1z},\,0,\,0,\,0,\,0\,)\,\in\,{\rm I\!R}^7, \label{defin-lll} \\
{\mathbb B}_{{\bf n}_2}(\lambda)\,&=\,\pm\,1\,\in\,S^7,\;\;{\rm about\;\,the\;\,direction}\;\,
{\bf N}_2\,:=\,(\,+{n}_{2x},\,+{n}_{2y},\,0,\,+{n}_{2z},\,0,\,0,\,0\,)\,\in\,{\rm I\!R}^7, \\
{\mathbb C}_{{\bf n}_3}(\lambda)\,&=\,\pm\,1\,\in\,S^7,\;\;{\rm about\;\,the\;\,direction}\;\,
{\bf N}_3\,:=\,(\,+{n}_{3x},\,+{n}_{3y},\,0,\,0,\,+{n}_{3z},\,0,\,0\,)\,\in\,{\rm I\!R}^7, \\
{\mathbb D}_{{\bf n}_4}(\lambda)\,&=\,\pm\,1\,\in\,S^7,\;\;{\rm about\;\,the\;\,direction}\;\,
{\bf N}_4\,:=\,(\,+{n}_{4x},\,-{n}_{4y},\,0,\,0,\,0,\,-{n}_{4z},\,0\,)\,\in\,{\rm I\!R}^7,\label{defin-nnn}
\end{align}
with ${n_{1x}}$, ${n_{1y}}$, and ${n_{1z}}$ being the components of ${{\bf n}_{1}\in{\rm I\!R}^3}$; ${n_{2x}}$, ${n_{2y}}$,
and ${n_{2z}}$ being the components of ${{\bf n}_{2}\in{\rm I\!R}^3}$; etc. Thus, with these identifications between the
points of the equators ${S^2}$ of ${S^3}$ and ${S^6}$ of ${S^7}$ (with ${S^3}$ and ${S^7}$ being the fiber and the total
space of the corresponding Hopf bundle \ocite{7-Hopf}), a specification of the experimental directions ${{\bf n}_{1}}$,
${{\bf n}_{2}}$, ${{\bf n}_{3}}$, and ${{\bf n}_{4}}$ in ${{\rm I\!R}^3}$ is equivalent to a specification of the
directions ${{\bf N}_{1}}$, ${{\bf N}_{2}}$, ${{\bf N}_{3}}$, and ${{\bf N}_{4}}$ in ${{\rm I\!R}^7}$. One may wonder
on what basis the choices of these specific four points on ${S^7}$ have been made. They have to do with the symmetries
and asymmetries of the system under consideration. For a
four-level system different from the GHZ-4 system the choices of
points would indeed have to be different. Using the above identifications, we can now rewrite the maps
(\ref{n-ghz-maps-my}) as
\begin{equation}
{\mathbb A}_{{\bf N}_1}(\lambda): {\rm I\!R}^7\!\times\Lambda\longrightarrow {S^7}, \;\;\;
{\mathbb B}_{{\bf N}_2}(\lambda): {\rm I\!R}^7\!\times\Lambda\longrightarrow {S^7}, \;\;\;
{\mathbb C}_{{\bf N}_3}(\lambda): {\rm I\!R}^7\!\times\Lambda\longrightarrow {S^7}, \;\;\;{\rm and}\;\;\;
{\mathbb D}_{{\bf N}_4}(\lambda): {\rm I\!R}^7\!\times\Lambda\longrightarrow {S^7}.\label{N-ghz-my-maps}
\end{equation}
The local-realistic expectation value for the four particle GHZ system is then given by
\begin{equation}
{\cal E}_{{\!}_{L.R.}\!}({\bf n}_1,\,{\bf n}_2,\,{\bf n}_3,\,{\bf n}_4)\,=\int_{\Lambda}
{\mathbb A}_{{\bf N}_1}(\lambda)\,{\mathbb B}_{{\bf N}_2}(\lambda)\,
{\mathbb C}_{{\bf N}_3}(\lambda)\,{\mathbb D}_{{\bf N}_4}(\lambda)\;d\rho(\lambda)\,.\label{ghzexpfun}
\end{equation}
Next, using the product rule (\ref{specialcase}) (which remains formally the same in the present case\footnote{More
generally, if ${a+{\bf X}}$ and ${b+{\bf Y}}$ are two points of ${S^7}$, then their
product is given by the point ${ab-{\bf X}\cdot{\bf Y}+a{\bf X}+b{\bf Y}-{\bf X}\times{\bf Y}\in S^7}$, where
${a}$ and ${b}$ are real numbers, and ${{\bf X}\times{\bf Y}}$ is a choice of a cross product in ${{\rm I\!R}^7}$.
For further details, see, for example, Ref.${\,}$\ocite{Lounesto}.}) we obtain the products
\begin{equation}
{\mathbb A}_{{\bf N}_1}(\lambda)\,{\mathbb B}_{{\bf N}_2}(\lambda)\,=\,-\,{\bf N}_1\cdot{\bf N}_2
\,-\,{\mathbb E}_{{\bf N}_1\times{\bf N}_2}(\lambda)\;\;\;\;{\rm and}\;\;\;\;
{\mathbb C}_{{\bf N}_3}(\lambda)\,{\mathbb D}_{{\bf N}_4}(\lambda)\,=\,-\,{\bf N}_3\cdot{\bf N}_4
\,-\,{\mathbb F}_{{\bf N}_3\times{\bf N}_4}(\lambda)\,.\label{special17}
\end{equation}
These products in turn allow us to decompose the integrand of (\ref{ghzexpfun}) into the following five terms
\begin{align}
\!\!{\mathbb A}_{{\bf N}_1}(\lambda)\,{\mathbb B}_{{\bf N}_2}(\lambda)\,{\mathbb C}_{{\bf N}_3}(\lambda)
\,{\mathbb D}_{{\bf N}_4}(\lambda)\,&=\,({\bf N}_1\cdot\,{\bf N}_2)\,({\bf N}_3\cdot\,{\bf N}_4)\,-\,
({\bf N}_1\times{\bf N}_2)\cdot({\bf N}_3\times{\bf N}_4) \notag \\
&\;\;\;\;\;+\,({\bf N}_3\cdot{\bf N}_4)\,{\mathbb E}_{{\bf N}_1\times{\bf N}_2}(\lambda)+\,({\bf N}_1\cdot{\bf N}_2)\,
{\mathbb F}_{{\bf N}_3\times{\bf N}_4}(\lambda)
\,-\,{\mathbb G}_{({\bf N}_1\times{\bf N}_2)\times({\bf N}_3\times{\bf N}_4)}(\lambda)\,.\label{special7sphere}
\end{align}
It is worth remembering here that both sides of the above expression simply represent a {\it bona fide} point of ${S^7}$,
which is equal to ${\pm\,1}$. Substituting it into equation (\ref{ghzexpfun}) then reduces the local-realistic
expectation value functional to
\begin{align}
\!\!\!{\cal E}_{{\!}_{L.R.}\!}({\bf n}_1,\,{\bf n}_2,\,{\bf n}_3,\,{\bf n}_4)
\,&=\,[\,({\bf N}_1\cdot\,{\bf N}_2)\,({\bf N}_3\cdot\,{\bf N}_4)\,-\,
({\bf N}_1\times{\bf N}_2)\cdot({\bf N}_3\times{\bf N}_4)\,]\!\int_{\Lambda}d\rho(\lambda) \notag \\
&\;\;+\!\int_{\Lambda}\Bigl\{\,({\bf N}_3\cdot{\bf N}_4)\;{\mathbb E}_{{\bf N}_1\times{\bf N}_2}(\lambda)
\,+\,({\bf N}_1\cdot{\bf N}_2)\;{\mathbb F}_{{\bf N}_3\times{\bf N}_4}(\lambda)\,-\,
{\mathbb G}_{({\bf N}_1\times{\bf N}_2)\times({\bf N}_3\times{\bf N}_4)}(\lambda)\,\Bigl\}\;d\rho(\lambda)\,.
\label{expect7sphere}
\end{align}
Now, as we recognized before in the analogous case of equation (\ref{step-hard-1}), the integrand
in the last term of this expression is proportional to a single binary point on the equator ${S^6}$ of the
sphere ${S^7}$, located about the direction ${{\bf N}\in{\rm I\!R}^7}$, which is given by
${{\bf N}:=({\bf N}_3\cdot{\bf N}_4)({\bf N}_1\times{\bf N}_2)\,+\,({\bf N}_1\cdot{\bf N}_2)
({\bf N}_3\times{\bf N}_4)\,-\,({\bf N}_1\times{\bf N}_2)\times({\bf N}_3\times{\bf N}_4)}$. Once again, this
can be seen most transparently in the elegant language of Clifford algebra ${Cl_{7,0}}$ (see, for example,
Ref.${\,}$\ocite{Lounesto}), but we will avoid this language here, because of the inability of some critics of
Refs.${\,}$\ocite{Christian}\ocite{Further}\ocite{experiment} to understand the Clifford-algebraic concepts.
But whichever language is used, the integrand in question can be easily simplified to a binary point on ${S^6}$, yielding
\begin{equation}
{\cal E}_{{\!}_{L.R.}\!}({\bf n}_1,\,{\bf n}_2,\,{\bf n}_3,\,{\bf n}_4)
\,=\,[\,({\bf N}_1\cdot\,{\bf N}_2)\,({\bf N}_3\cdot\,{\bf N}_4)\,-\,
({\bf N}_1\times{\bf N}_2)\cdot({\bf N}_3\times{\bf N}_4)\,]\!\int_{\Lambda}d\rho(\lambda)
\;+\,|{\bf N}|\int_{\Lambda}\,{\mathbb P}_{\frac{\bf N}{|{\bf N}|}}(\lambda)\,\;d\rho(\lambda)\,,
\label{expeconepoint}
\end{equation}
with ${{\mathbb P}_{\frac{\bf N}{|{\bf N}|}}(\lambda)=\pm\,1\in S^6}$. Moreover, it is evident from its definition
that ${\bf N}$ is a direction in ${{\rm I\!R}^7}$ that is exclusive to all four experimentally relevant directions
${{\bf N}_1}$, ${{\bf N}_2}$, ${{\bf N}_3}$, and ${{\bf N}_4}$ in ${{\rm I\!R}^7}$. But the latter four directions, by
construction, are equivalent to the directions ${{\bf n}_{1}}$, ${{\bf n}_{2}}$, ${{\bf n}_{3}}$, and ${{\bf n}_{4}}$
in ${{\rm I\!R}^3}$ (cf. Eqs.${}$(\ref{defin-lll}) to (\ref{defin-nnn})), and therefore the direction ${\bf n}$ in
${{\rm I\!R}^3}$ (defined to be equivalent to the direction ${\bf N}$ in ${{\rm I\!R}^7}$ by reverse construction)
will also be exclusive to the directions ${{\bf n}_{1}}$, ${{\bf n}_{2}}$, ${{\bf n}_{3}}$, and ${{\bf n}_{4}}$ in
${{\rm I\!R}^3}$. In other words, a detector along the direction ${\bf n}$---and hence equivalently along the direction
${\bf N}$---would necessarily yield a null result, provided the detectors along any pair of the directions ${{\bf n}_{1}}$,
${{\bf n}_{2}}$, ${{\bf n}_{3}}$, and ${{\bf n}_{4}}$ have yielded non-null results. As a result, the last term on the
right hand side of the above equation is zero for more than one reason. To begin with, it involves an average of a function
analogous to the function ${C_{\bf c}(\lambda)}$ of equation (\ref{q-prob-q}), and hence is necessarily zero, thanks to
the relations analogous to the relations (\ref{improveBell's}) and (\ref{0}). Moreover, operationally the integrand of the
term itself is necessarily zero, for the reasons we just spelled out. Consequently, without specifying the complete state
${\lambda}$ or its distribution ${\rho(\lambda)}$, and without invoking any averaging procedure, we arrive at the result
\begin{equation}
{\cal E}_{{\!}_{L.R.}\!}({\bf n}_1,\,{\bf n}_2,\,{\bf n}_3,\,{\bf n}_4)\,=\,
({\bf N}_1\cdot\,{\bf N}_2)\,({\bf N}_3\cdot\,{\bf N}_4)\,-\,({\bf N}_1\times{\bf N}_2)\cdot({\bf N}_3\times{\bf N}_4)\,.
\label{almostexpt}
\end{equation}

The explicit evaluation of the above equation requires understanding of the vector product in seven dimensions. So far
we have employed vector products in our equations without mentioning the profound fact that they exist only in three and
seven dimensions \ocite{crossproduct}\ocite{Lounesto}. In seven dimensions, however, they are not invariant under the full
rotation group SO(7). An additional vector---say ${\bf Z}$---is required to characterize the vector space ${{\rm I\!R}^7}$,
and the vector product is then defined only with respect to this characterizing vector. The essential reason for this has to
do with the fact that in dimensions greater than four there are more than one planes orthogonal to a given direction. Once
the vector ${\bf Z}$ is chosen, however, the algebraic structure of the space ${{\rm I\!R}^7}$ is
complete \ocite{crossproduct}\ocite{Lounesto}. The
corresponding vector product is then invariant under the subgroup ${G_2}$ of the group SO(7), which is an exceptional
Lie group that preserves the structure of the 7-sphere. More precisely, the group ${G_2}$ is the automorphism group of the
vectors in ${{\rm I\!R}^8}$, and hence is homeomorphic to ${S^7}$.
All this information can be neatly captured by a generalized form of Lagrange's identity:
\begin{equation}
({\bf N}_1\times{\bf N}_2)\cdot({\bf N}_3\times{\bf N}_4)\,=\,
({\bf N}_1\cdot\,{\bf N}_3)\,({\bf N}_2\cdot\,{\bf N}_4)\,-\,({\bf N}_1\cdot\,{\bf N}_4)\,({\bf N}_2\cdot\,{\bf N}_3)\,+\,
{\bf N}_1\cdot\,{\bf Z}\,.\label{lagidn}
\end{equation}
In three dimensions the last term on the right hand side of this identity vanishes identically, because
${{\bf Z}\equiv 0}$, and we recover the familiar Lagrange's identity \ocite{crossproduct}. In seven
dimensions, however, the last term is non-vanishing in general, because
${{\bf Z}:={\bf N}_2\times({\bf N}_3\times{\bf N}_4)-{\bf N}_3\,({\bf N}_2\cdot{\bf N}_4)+
{\bf N}_4\,({\bf N}_2\cdot{\bf N}_3)}$ is a non-zero function of ${{\bf N}_2}$, ${{\bf N}_3}$, and
${{\bf N}_4}$ such that
\begin{equation}
{\bf Z}\cdot{\bf N}_2\,=\,{\bf Z}\cdot{\bf N}_3\,=\,{\bf Z}\cdot{\bf N}_4\,=\,{\bf Z}\cdot({\bf N}_2\times{\bf N}_3)\,=\,
{\bf Z}\cdot({\bf N}_3\times{\bf N}_4)\,=\,{\bf Z}\cdot({\bf N}_4\times{\bf N}_2)\,=\,0.
\end{equation}
Note that the very definition of the vector product in all of the above equations depends on the choice of ${\bf Z}$ itself.
Given such a choice of ${\bf Z}$ and the corresponding vector product, the expectation value (\ref{almostexpt}),
via (\ref{lagidn}), simplifies to
\begin{equation}
{\cal E}_{{\!}_{L.R.}\!}({\bf n}_1,\,{\bf n}_2,\,{\bf n}_3,\,{\bf n}_4)\,=\,
({\bf N}_1\cdot\,{\bf N}_2)\,({\bf N}_3\cdot\,{\bf N}_4)\,-\,
({\bf N}_1\cdot\,{\bf N}_3)\,({\bf N}_2\cdot\,{\bf N}_4)\,+\,({\bf N}_1\cdot\,{\bf N}_4)\,({\bf N}_2\cdot\,{\bf N}_3)\,-\,
{\bf N}_1\cdot\,{\bf Z}\,,\label{expnewred}
\end{equation}
which involves only a sequence of benign scalar products in seven dimensions. More importantly, this expression
of the expectation value is now guaranteed to be invariant under the group ${G_2\subset{\rm SO(7)}}$, thereby respecting
the structure of ${S^7}$. Considering that the ${{\bf z}\in{\rm I\!R}^3}$ is a privileged direction for our GHZ-4 system,
the natural choice of ${\bf Z}$ for us is
\begin{equation}
{\bf Z}\,:=\,{\bf \hat e}_3\,[\,{n}_{2z}\,{n}_{3z}\,{n}_{4z}\,]\,+\,
{\bf \hat e}_7\,[\,f(\,{\bf N}_1,\,{\bf N}_2,\,{\bf N}_3,\,{\bf N}_4\,)\,]\,,
\end{equation}
where ${\{\,{\bf \hat e}_1,\,{\bf \hat e}_2,\,{\bf \hat e}_3,\,{\bf \hat e}_4,\,{\bf \hat e}_5,\,{\bf \hat e}_6,
\,{\bf \hat e}_7\}}$ are the basis vectors in ${{\rm I\!R}^7}$, and
${f(\,{\bf N}_1,\,{\bf N}_2,\,{\bf N}_3,\,{\bf N}_4\,)}$ is a scalar function. With a little bit of intuition
developed for both the GHZ-4 system (easy) and the 7-sphere (not easy), it is not difficult to appreciate that this
choice of ${\bf Z}$ correctly characterizes the anisotropy or the rotational non-invariance of the GHZ-4 system
within the space ${{\rm I\!R}^7}$. With this ${\bf Z}$, and the definitions (\ref{defin-lll}) to (\ref{defin-nnn}),
the expectation value (\ref{expnewred}) reduces to
\begin{align}
{\cal E}_{{\!}_{L.R.}\!}({\bf n}_1,\,{\bf n}_2,\,{\bf n}_3,\,{\bf n}_4)\,=
&\,+\,{n}_{1z}\,{n}_{2z}\,{n}_{3z}\,{n}_{4z} \notag \\
&\,-\,{n}_{1y}\,{n}_{2y}\,{n}_{3y}\,{n}_{4y}
\,-\,{n}_{1x}\,{n}_{2y}\,{n}_{3x}\,{n}_{4y}
\,-\,{n}_{1y}\,{n}_{2x}\,{n}_{3y}\,{n}_{4x}
\,-\,{n}_{1x}\,{n}_{2x}\,{n}_{3x}\,{n}_{4x} \notag \\
&\,+\,{n}_{1x}\,{n}_{2x}\,{n}_{3y}\,{n}_{4y}
\,-\,{n}_{1x}\,{n}_{2y}\,{n}_{3y}\,{n}_{4x}
\,-\,{n}_{1y}\,{n}_{2x}\,{n}_{3x}\,{n}_{4y}
\,+\,{n}_{1y}\,{n}_{2y}\,{n}_{3x}\,{n}_{4x}\,.
\end{align}
In the spherical coordinates---with angles ${\theta_1}$ and ${\phi_2}$ representing respectively
the polar and azimuthal angles of the direction ${{\bf n}_1}$, etc.---this expression can be further simplified to
\begin{equation}
{\cal E}_{{\!}_{L.R.}\!}({\bf n}_1,\,{\bf n}_2,\,{\bf n}_3,\,{\bf n}_4)\,=\,
\cos\theta_1\,\cos\theta_2\,\cos\theta_3\,\cos\theta_4\,-\,\sin\theta_1\,
\sin\theta_2\,\sin\theta_3\,\sin\theta_4\,\cos\,\left(\,\phi_1\,+\,\phi_2\,-\,\phi_3\,-\,\phi_4\,\right)\,.
\end{equation}
This is of course exactly the quantum mechanical prediction (\ref{q-preghz}) for the GHZ-4 state (\ref{ghz-single}).
We have, however, derived this result within our purely local-realistic framework. Moreover, we have derived it without
specifying what the complete state ${\lambda}$, or the distribution ${\rho(\lambda)}$ is, and without employing any
averaging procedure over this complete state. This shows that the correlations exhibited by this expectation value
are purely topological effects. They are simply the classical, deterministic, local, and realistic correlations
among four points of a unit 7-sphere.

\subsection{Exact Local-Realistic Completion of the Three-Particle GHZ State}

As our fourth explicit example, consider the three-particle Greenberger, Horne, Zeilinger state \ocite{GHSZ}\ocite{Wu}:
\begin{equation}
|\Psi_{\bf z}\rangle\,=\,\cos\left(\frac{\alpha}{2}\right)\,\Bigl\{\,|{\bf z},\,+\rangle_1\otimes
|{\bf z},\,+\rangle_2\otimes|{\bf z},\,+\rangle_3\,\Bigr\}\,+\;\sin\left(\frac{\alpha}{2}\right)\,
\exp\bigl(-\,i\,\delta\bigr)\,\Bigl\{\,|{\bf z},\,-\rangle_1\otimes|{\bf z},\,-\rangle_2\otimes
|{\bf z},\,-\rangle_3\,\Bigr\}.\label{ghz-three}
\end{equation}
Just like the Hardy state and the four-particle GHZ state, this state too is rotationally non-invariant.
There is a privileged direction, and this direction is taken to be the ${\bf z}$-direction of the experimental
setup \ocite{GHSZ}. In their discussion of this case GHZ and Shimony \ocite{GHSZ} emphasize that Bell's theorem
does not hinge on spin. And neither do our refutations of the theorem. By now it should be clear that all of the
so-called quantum correlations in such examples are in fact purely topological effects, and hence they should not---and
do not---hinge on spin. It is less cumbersome, however, to formulate Bell's theorem and its refutations in terms
of spin, and so we continue this trend. The quantum mechanical expectation value
in the above state, of the product of outcomes of the spin components---namely, the products of finding the spin
of particle 1 along ${{\bf n}_1}$, the spin of particle 2 along ${{\bf n}_2}$, etc.---is given by
\begin{equation}
{\cal E}^{\Psi_{\bf z}}_{{\!}_{Q.M.}\!}({\bf n}_1,\,{\bf n}_2,\,{\bf n}_3)\,:=\,
\langle\Psi_{\bf z}|\,{\boldsymbol\sigma}\cdot{\bf n}_1\,\otimes\,
{\boldsymbol\sigma}\cdot{\bf n}_2\,\otimes
\,{\boldsymbol\sigma}\cdot{\bf n}_3\,|\Psi_{\bf z}\rangle.\label{nonrealobse}
\end{equation}
For a special case, this expectation value has been calculated in Appendix G of Ref.${\,}$\ocite{GHSZ},
but we shall use the most general quantum mechanical prediction, as stated, for example, in Ref.${\,}$\ocite{Wu}.
In the spherical coordinates---with angles ${\theta_1}$ and
${\phi_2}$ representing the polar and azimuthal angles, respectively,
of the direction ${{\bf n}_1}$, etc.---it can be expressed as
\begin{equation}
{\cal E}^{\Psi_{\bf z}}_{{\!}_{Q.M.}\!}({\bf n}_1,\,{\bf n}_2,\,{\bf n}_3)\,=\,
\cos\alpha\,\cos\theta_1\,\cos\theta_2\,\cos\theta_3\,+\,\sin\alpha\,\sin\theta_1\,
\sin\theta_2\,\sin\theta_3\,\cos\,\left(\,\phi_1\,+\,\phi_2\,+\,\phi_3\,+\,\delta\,\right).\label{pppreghz}
\end{equation}

Once again, this result can be derived {\it exactly} within our local-realistic framework. To accomplish this, however,
we must first determine the correct topological space of the corresponding EPR elements of reality. It turn out
that this space is again a unit 7-sphere. The most general form of the three-particle state such as (\ref{ghz-three})
can be written as
\begin{equation}
|\psi\rangle\,=\,\zeta_1\,|\,+\,+\,+\,\rangle\,+\;\zeta_2\,
|-\,+\,+\,\rangle\,+\;\zeta_3\,|+\,-\,+\,\rangle\,+\;\zeta_4\,|+\,+\,-\,\rangle\,+\;\zeta_5
\,|+\,-\,-\,\rangle\,+\;\zeta_6\,|-\,+\,-\,\rangle\,+\;\zeta_7\,|-\,-\,+\,\rangle\,+\;
\zeta_8\,|\,-\,-\,-\,\rangle.\label{scnnndt-single}
\end{equation}
From this state it is easy to see that when one of the particles is in a definite spin state, the remaining pair of
particles has four alternative states available to it. These alternative states can be represented by a state-vector
of the form
\begin{equation}
|\chi\rangle\,=\,\xi_1\,|\,+\,+\,\rangle\,+\,\xi_2\,|\,+\,-\,\rangle\,+\,\xi_3\,|\,-\,+\,\rangle\,+\,
\xi_4\,|\,-\,-\,\rangle\,,\label{schmidt-sonion}
\end{equation}
where ${\xi_1}$, ${\xi_2}$, ${\xi_3}$, and ${\xi_4}$ are complex numbers satisfying the normalization condition
${|\,\xi_1\,|^2+|\,\xi_2\,|^2+|\,\xi_3\,|^2+|\,\xi_4\,|^2=1}$. This condition is equivalent to
\begin{equation}
\xi_{1r}^2\,+\,\xi_{1i}^2\,+\,\xi_{2r}^2\,+\,\xi_{2i}^2\,+\,\xi_{3r}^2\,+\,\xi_{3i}^2\,+\,\xi_{4r}^2\,+\,\xi_{4i}^2\,=\,1\,,
\label{ignore-3}
\end{equation}
which is the defining equation of a unit 7-sphere, embedded in ${{\rm I\!R}^8}$. And ${|\,\xi_1\,|^2}$, ${|\,\xi_2\,|^2}$,
${|\,\xi_3\,|^2}$, and ${|\,\xi_4\,|^2}$ are of course the probabilities of realizing or actualizing the states
${|\,+\,+\,\rangle}$, ${|\,+\,-\,\rangle}$, ${|\,-\,+\,\rangle}$, and ${|\,-\,-\,\rangle}$, respectively. Therefore, in
analogy with the previous case, in this case too there is a one-to-one correspondence between the points of a
7-sphere and the EPR elements of reality. In other words, the correct topological space of the EPR elements of reality in
the present case is again a unit 7-sphere. Hence we begin our local-realistic description with three local maps of the form
\begin{equation}
{\mathbb A}_{{\bf n}_1}(\lambda): {\rm I\!R}^3\!\times\Lambda\longrightarrow {S^7}, \;\;\;
{\mathbb B}_{{\bf n}_2}(\lambda): {\rm I\!R}^3\!\times\Lambda\longrightarrow {S^7}, \;\;\;{\rm and}\;\;\;
{\mathbb C}_{{\bf n}_3}(\lambda): {\rm I\!R}^3\!\times\Lambda\longrightarrow {S^7}\,,\label{n-gz-ms-my}
\end{equation}
and take them to be the following three points on the equator of a unit 7-sphere (which is of course a unit 6-sphere):
\begin{align}
{\mathbb A}_{{\bf n}_1}(\lambda)\,&=\,\pm\,1\,\in\,S^7,\;\;{\rm about\;\,the\;\,direction}\;\,
{\bf N}_1\,:=\,(\,+{n}_{1x},\,+{n}_{1y},\,0,\,+{n}_{1z},\,0,\,0,\,0\,)\,\in\,{\rm I\!R}^7,\label{defoplin-lll} \\
{\mathbb B}_{{\bf n}_2}(\lambda)\,&=\,\pm\,1\,\in\,S^7,\;\;{\rm about\;\,the\;\,direction}\;\,
{\bf N}_2\,:=\,(\,+{n}_{2x},\,-{n}_{2y},\,0,\,0,\,-{n}_{2z},\,0,\,0\,)\,\in\,{\rm I\!R}^7, \\
{\mathbb C}_{{\bf n}_3}(\lambda)\,&=\,\pm\,1\,\in\,S^7,\;\;{\rm about\;\,the\;\,direction}\;\,
{\bf N}_3\,:=\,(\,-{n}_{3x},\,-{n}_{3y},\,0,\,0,\,0,\,+{n}_{3z},\,0\,)\,\in\,{\rm I\!R}^7.\label{defoplin-nnn}
\end{align}
Here ${n_{1x}}$, ${n_{1y}}$, and ${n_{1z}}$ are the components of ${{\bf n}_{1}\in{\rm I\!R}^3}$; ${n_{2x}}$, ${n_{2y}}$,
and ${n_{2z}}$ are the components of ${{\bf n}_{2}\in{\rm I\!R}^3}$; and so on. Thus, just as in the previous case, a
specification of the directions ${{\bf n}_{1}}$,
${{\bf n}_{2}}$, and ${{\bf n}_{3}}$ in ${{\rm I\!R}^3}$ is equivalent to a specification of the
directions ${{\bf N}_{1}}$, ${{\bf N}_{2}}$, and ${{\bf N}_{3}}$ in ${{\rm I\!R}^7}$. Once again the choice of these
three points are motivated by the symmetries and asymmetries of the state under consideration. Using these identifications,
we can now rewrite the maps (\ref{n-gz-ms-my}) as
\begin{equation}
{\mathbb A}_{{\bf N}_1}(\lambda): {\rm I\!R}^7\!\times\Lambda\longrightarrow {S^7}, \;\;\;
{\mathbb B}_{{\bf N}_2}(\lambda): {\rm I\!R}^7\!\times\Lambda\longrightarrow {S^7}, \;\;\;{\rm and}\;\;\;
{\mathbb C}_{{\bf N}_3}(\lambda): {\rm I\!R}^7\!\times\Lambda\longrightarrow {S^7}.\label{N-gz-m-ma}
\end{equation}
The local-realistic expectation value for the three particle GHZ system is then given by
\begin{equation}
{\cal E}_{{\!}_{L.R.}\!}({\bf n}_1,\,{\bf n}_2,\,{\bf n}_3)\,=\int_{\Lambda}
{\mathbb P}_{{\bf N}_0}\,
{\mathbb A}_{{\bf N}_1}(\lambda)\,{\mathbb B}_{{\bf N}_2}(\lambda)\,
{\mathbb C}_{{\bf N}_3}(\lambda)\;d\rho(\lambda)\,,\label{hzepfun}
\end{equation}
where ${{\mathbb P}_{{\bf N}_0}\in S^7}$ about the direction
${{\bf N}_0:=(\,-{n}_{0x},\,+{n}_{0y},\,-{n}_{0z},\,0,\,0,\,0,\,0\,)\in{\rm I\!R}^7}$
is a fixed reference point on ${S^7}$, with ${{n}_{0x}=\sin\alpha\,\cos\delta}$,
${{n}_{0y}=\sin\alpha\,\sin\delta}$, and ${{n}_{0z}=\cos\alpha}$ being its Cartesian and spherical
coordinates in ${{\rm I\!R}^3}$. In other words, the point ${{\mathbb P}_{{\bf N}_0}\in S^7}$ in the
expression (\ref{hzepfun}) specifies the relative and overall phases of the state (\ref{ghz-three}).

Next, using the product rule (\ref{specialcase}) (which again remains the same in the present case${{}^3}$)
we obtain the products
\begin{equation}
{\mathbb P}_{{\bf N}_0}\,{\mathbb A}_{{\bf N}_1}(\lambda)\,=\,-\,{\bf N}_0\cdot{\bf N}_1
\,-\,{\mathbb E}_{{\bf N}_0\times{\bf N}_1}(\lambda)\;\;\;\;{\rm and}\;\;\;\;
{\mathbb B}_{{\bf N}_2}(\lambda)\,{\mathbb C}_{{\bf N}_3}(\lambda)\,=\,-\,{\bf N}_2\cdot{\bf N}_3
\,-\,{\mathbb F}_{{\bf N}_2\times{\bf N}_3}(\lambda)\,.\label{spec7}
\end{equation}
These products in turn allow us to decompose the integrand of (\ref{hzepfun}) into the following five terms
\begin{align}
{\mathbb P}_{{\bf N}_0}\,{\mathbb A}_{{\bf N}_1}(\lambda)\,{\mathbb B}_{{\bf N}_2}(\lambda)
\,{\mathbb C}_{{\bf N}_3}(\lambda)\,&=\,({\bf N}_0\cdot\,{\bf N}_1)\,({\bf N}_2\cdot\,{\bf N}_3)\,-\,
({\bf N}_0\times{\bf N}_1)\cdot({\bf N}_2\times{\bf N}_3) \notag \\
&\;\;\;\;\;+\,({\bf N}_2\cdot{\bf N}_3)\,{\mathbb E}_{{\bf N}_0\times{\bf N}_1}(\lambda)+\,({\bf N}_0\cdot{\bf N}_1)\,
{\mathbb F}_{{\bf N}_2\times{\bf N}_3}(\lambda)
\,-\,{\mathbb G}_{({\bf N}_0\times{\bf N}_1)\times({\bf N}_2\times{\bf N}_3)}(\lambda)\,.\label{spal7spere}
\end{align}
It is worth remembering here that both sides of the above expression simply represent a {\it bona fide} point of ${S^7}$,
which is equal to ${\pm\,1}$. Substituting it into equation (\ref{hzepfun}) then reduces the local-realistic
expectation value functional to
\begin{align}
{\cal E}_{{\!}_{L.R.}\!}({\bf n}_1,\,{\bf n}_2,\,{\bf n}_3)
\,&=\,[\,({\bf N}_0\cdot\,{\bf N}_1)\,({\bf N}_2\cdot\,{\bf N}_3)\,-\,
({\bf N}_0\times{\bf N}_1)\cdot({\bf N}_2\times{\bf N}_3)\,] \int_{\Lambda}d\rho(\lambda) \notag \\
&\;\;\;\;\;+\!\int_{\Lambda}\Bigl\{\,({\bf N}_2\cdot{\bf N}_3)\;{\mathbb E}_{{\bf N}_0\times{\bf N}_1}(\lambda)\,
\,+\,({\bf N}_0\cdot{\bf N}_1)\;{\mathbb F}_{{\bf N}_2\times{\bf N}_3}(\lambda)\,-\,
{\mathbb G}_{({\bf N}_0\times{\bf N}_1)\times({\bf N}_2\times{\bf N}_3)}(\lambda)\,\Bigl\}\;d\rho(\lambda)\,.
\label{exct7sphe}
\end{align}
Now, just as we saw in the case of equation (\ref{expect7sphere}), we see that the integrand in the last term of this
equation is proportional to a point on the equator ${S^6}$ of the sphere ${S^7}$, located about the direction
${{\bf N}\in{\rm I\!R}^7}$, which is given by
${{\bf N}:=({\bf N}_2\cdot{\bf N}_3)({\bf N}_0\times{\bf N}_1)\,+\,({\bf N}_0\cdot{\bf N}_1)
({\bf N}_2\times{\bf N}_3)\,-\,({\bf N}_0\times{\bf N}_1)\times({\bf N}_2\times{\bf N}_3)}$. And hence this
equation simplifies to
\begin{equation}
{\cal E}_{{\!}_{L.R.}\!}({\bf n}_1,\,{\bf n}_2,\,{\bf n}_3)
\,=\,[\,({\bf N}_0\cdot\,{\bf N}_1)\,({\bf N}_2\cdot\,{\bf N}_3)\,-\,
({\bf N}_0\times{\bf N}_1)\cdot({\bf N}_2\times{\bf N}_3)\,]\!\int_{\Lambda}d\rho(\lambda)
\;+\,|{\bf N}|\int_{\Lambda}\,{\mathbb P}_{\frac{\bf N}{|{\bf N}|}}(\lambda)\,\;d\rho(\lambda)\,,
\label{notrealonepoint}
\end{equation}
with ${{\mathbb P}_{\frac{\bf N}{|{\bf N}|}}(\lambda)=\pm\,1\in S^6}$. Moreover, it is evident from the above definition
that ${\bf N}$ is a direction in ${{\rm I\!R}^7}$ that is exclusive to all three experimentally relevant directions
${{\bf N}_1}$, ${{\bf N}_2}$, and ${{\bf N}_3}$ in ${{\rm I\!R}^7}$. But the latter three directions, by construction,
are equivalent to the directions ${{\bf n}_{1}}$, ${{\bf n}_{2}}$, and ${{\bf n}_{3}}$ in ${{\rm I\!R}^3}$
(cf. Eqs.${}$(\ref{defoplin-lll}) to (\ref{defoplin-nnn})), and therefore the direction ${\bf n}$ in ${{\rm I\!R}^3}$
(defined to be equivalent to the direction ${\bf N}$ in ${{\rm I\!R}^7}$ by reverse construction)
will also be exclusive to the directions ${{\bf n}_{1}}$, ${{\bf n}_{2}}$, and ${{\bf n}_{3}}$ in
${{\rm I\!R}^3}$. In other words, a detector along the direction ${\bf n}$---and hence equivalently along the direction
${\bf N}$---would necessarily yield a null result, provided the detectors along any pair of the directions ${{\bf n}_{1}}$,
${{\bf n}_{2}}$, and ${{\bf n}_{3}}$ have yielded non-null results. As a result, the last term on the right
hand side of the above equation is zero for more than one reason. To begin with, it involves an average of a function
analogous to the function ${C_{\bf c}(\lambda)}$ of equation (\ref{q-prob-q}), and hence is necessarily zero, thanks to
the relations analogous to the relations (\ref{improveBell's}) and (\ref{0}). Moreover, operationally the integrand of the
term itself is necessarily zero, for the reasons we just spelled out. Consequently, without specifying the complete state
${\lambda}$ or its distribution ${\rho(\lambda)}$, and without invoking any averaging procedure, we arrive at the result
\begin{equation}
{\cal E}_{{\!}_{L.R.}\!}({\bf n}_1,\,{\bf n}_2,\,{\bf n}_3)\,=\,
({\bf N}_0\cdot\,{\bf N}_1)\,({\bf N}_2\cdot\,{\bf N}_3)\,-\,({\bf N}_0\times{\bf N}_1)\cdot({\bf N}_2\times{\bf N}_3)\,.
\label{mosexpt}
\end{equation}

This equation can now be explicitly evaluated just as before. First, using the generalized Lagrange's identity
\begin{equation}
({\bf N}_0\times{\bf N}_1)\cdot({\bf N}_2\times{\bf N}_3)\,=\,
({\bf N}_0\cdot\,{\bf N}_2)\,({\bf N}_1\cdot\,{\bf N}_3)\,-\,({\bf N}_0\cdot\,{\bf N}_3)\,({\bf N}_1\cdot\,{\bf N}_2)\,+\,
{\bf N}_0\cdot\,{\bf Z}\label{hjglagidn}
\end{equation}
it can be simplified to
\begin{equation}
{\cal E}_{{\!}_{L.R.}\!}({\bf n}_1,\,{\bf n}_2,\,{\bf n}_3)\,=\,
({\bf N}_0\cdot\,{\bf N}_1)\,({\bf N}_2\cdot\,{\bf N}_3)\,-\,
({\bf N}_0\cdot\,{\bf N}_2)\,({\bf N}_1\cdot\,{\bf N}_3)\,+\,({\bf N}_0\cdot\,{\bf N}_3)\,({\bf N}_1\cdot\,{\bf N}_2)\,-\,
{\bf N}_0\cdot\,{\bf Z}\,.\label{exphgrenewred}
\end{equation}
Next, noting that the
${{\bf z}\in{\rm I\!R}^3}$ is a privileged direction for the GHZ-3 system, ${\bf Z}$ can be recognized to be of the form
\begin{equation}
{\bf Z}\,:=\,{\bf \hat e}_3\,[\,{n}_{1z}\,{n}_{2z}\,{n}_{3z}\,]\,+\,
{\bf \hat e}_7\,[\,f(\,{\bf N}_0,\,{\bf N}_1,\,{\bf N}_2,\,{\bf N}_3\,)\,]\,,
\end{equation}
where ${\{\,{\bf \hat e}_1,\,{\bf \hat e}_2,\,{\bf \hat e}_3,\,{\bf \hat e}_4,\,{\bf \hat e}_5,\,{\bf \hat e}_6,
\,{\bf \hat e}_7\}}$ are the basis vectors in ${{\rm I\!R}^7}$, and
${f(\,{\bf N}_0,\,{\bf N}_1,\,{\bf N}_2,\,{\bf N}_3\,)}$ is a scalar function.
Then, using this ${\bf Z}$, the vector ${{\bf N}_0}$, and the definitions (\ref{defoplin-lll}) to
(\ref{defoplin-nnn}), the expectation value (\ref{exphgrenewred}) can be simplified to
\begin{align}
{\cal E}_{{\!}_{L.R.}\!}({\bf n}_1,\,{\bf n}_2,\,{\bf n}_3)\,=
&\,+\,{n}_{0z}\,{n}_{1z}\,{n}_{2z}\,{n}_{3z} \notag \\
&\,+\,{n}_{0y}\,{n}_{1y}\,{n}_{2y}\,{n}_{3y}
\,-\,{n}_{0x}\,{n}_{1y}\,{n}_{2x}\,{n}_{3y}
\,-\,{n}_{0y}\,{n}_{1x}\,{n}_{2y}\,{n}_{3x}
\,+\,{n}_{0x}\,{n}_{1x}\,{n}_{2x}\,{n}_{3x} \notag \\
&\,-\,{n}_{0x}\,{n}_{1x}\,{n}_{2y}\,{n}_{3y}
\,-\,{n}_{0x}\,{n}_{1y}\,{n}_{2y}\,{n}_{3x}
\,-\,{n}_{0y}\,{n}_{1x}\,{n}_{2x}\,{n}_{3y}
\,-\,{n}_{0y}\,{n}_{1y}\,{n}_{2x}\,{n}_{3x}\,.
\end{align}
In the spherical coordinates---with angles ${\theta_1}$ and ${\phi_2}$ representing respectively the polar
and azimuthal angles of the direction ${{\bf n}_1}$, etc.---this expression can then be further reduced to
\begin{equation}
{\cal E}_{{\!}_{L.R.}\!}({\bf n}_1,\,{\bf n}_2,\,{\bf n}_3)\,=\,
\cos\alpha\,\cos\theta_1\,\cos\theta_2\,\cos\theta_3\,+\,\sin\alpha\,
\sin\theta_1\,\sin\theta_2\,\sin\theta_3\,\cos\,\left(\,\phi_1\,+\,\phi_2\,+\,\phi_3\,+\,\delta\,\right)\,.
\end{equation}
This is of course exactly the quantum mechanical prediction (\ref{pppreghz}) for the GHZ-3 state (\ref{ghz-three}).
We have, however, derived this result within our purely local-realistic framework. Moreover, we have derived it without
specifying what the complete state ${\lambda}$, or the distribution ${\rho(\lambda)}$ is, and without employing any
averaging procedure over the complete state. This shows that the correlations expressed in this expectation value
are purely topological effects. They are simply the classical, deterministic, local, and realistic correlations
among three points of a unit 7-sphere.

\section{Exact Local-Realistic completion of any arbitrary entangled state is always possible}

Amidst all the details, we learned in the previous sections that correlations between the points of two different
topological spaces cannot be the same in general. For example, correlations between the points of the real line are
necessarily linear, whereas correlations between the points of a 3-sphere, or a 7-sphere are sinusoidal. Moreover,
we learned that local-realistic completions of the quantum states of any two, three, or four-level systems---such as
the Bell, Hardy, or GHZ states---are easy to achieve, as long as the topological spaces of the corresponding EPR
elements of reality are correctly identified. Now for the Bell, Hardy, and GHZ states the correct topological spaces
turn out to be the 3 and 7-spheres. And unlike many of their cousins these spheres have some very desirable
properties \ocite{experiment}. The most significant among these is the fact that they remain closed under multiplication,
and hence their points necessarily preserve the locality condition of Bell. More precisely, both of these spheres
are topological groups\footnote{Actually, 7-sphere, satisfying the octonionic multiplication rules, is not
associative, and hence it is only a topological quasi-group. A
quasi-group satisfies all the axioms of a group except the axiom of associativity. Thus a group, by definition, is an
associative quasi-group.}${\,}$\ocite{Munkres}, which are topological spaces tailor-made to preserve the locality
condition of Bell. To see this more clearly,
recall that a topological group ${G}$, by definition, is a group that is also a topological space, such that
the group operation map
\begin{equation}
{G}\times\,{G}\,\longrightarrow\,{G}\,,\;\;\;\;\;\;\;{\rm sending}\;\;\;(x,\,y)\,\longmapsto\,x\,y\,,
\label{propyone}
\end{equation}
\vspace{-0.65cm}
${\!\!}$and the inversion map
\begin{equation}
\;{G}\,\longrightarrow\,{G}\,,\;\;\;\;\;\;\;\;{\rm sending}\;\;\;x\,\longmapsto\,x^{-1}\label{propytwo},
\end{equation}
are both continuous maps. Consequently, any point of the space ${G}$ can be factorized into two or more points of
the same space, and hence such a space is a tailor-made target space for hosting the locality condition of Bell. In
addition to these two maps, of course, there exists an identity element ${e\in G}$ (as in any group) such that
\begin{equation}
x\,e\,=\,e\,x\,=\,x\;\;\;\;{\rm and}\;\;\;\;x^{-1}\,x\,=\,x\,x^{-1}=\,e\,\;\;\;\forall
\;\,x,\,x^{-1}\in\,G\,.\label{iden-grop}
\end{equation}

With these observations, and drawing from the four examples we discussed in the previous sections, we now proceed to
show that a local-realistic completion of any arbitrary entangled state is always guaranteed within our framework.${\;}$To
this end, consider an arbitrary quantum state ${|\Psi\rangle\in{\cal H}}$, where ${\cal H}$ is a Hilbert
space of arbitrary dimensions, which may or may not be finite. It is important to note that we impose no restrictions
on either ${|\Psi\rangle}$ or ${\cal H}$, apart from their usual quantum mechanical meanings. In particular, the state
${|\Psi\rangle}$ can be as entangled as one may like, and the space ${\cal H}$ can be as large or
small as one may like. Next consider a self-adjoint operator
${{\cal\widehat O}({\bf a},\,{\bf b},\,{\bf c},\,{\bf d},\,\dots\,)}$ on this Hilbert${\;}$space, parameterized by
a number of local parameters, ${{\bf a},\,{\bf b},\,{\bf c},\,{\bf d},}$ etc., with their usual
contextual meaning \ocite{Contextual}, as in any Bell-type setup \ocite{Clauser-Shimony}. The quantum mechanical
expectation value of this observable in the state ${|\Psi\rangle}$ is then given by:
\begin{equation}
{\cal E}_{{\!}_{Q.M.}}({\bf a},\,{\bf b},\,{\bf c},\,{\bf d},\,\dots\,)\,
=\,\langle\Psi|\;{\cal\widehat O}({\bf a},\,{\bf b},\,{\bf c},\,{\bf d},\,\dots\,)\,|\Psi\rangle\,.\label{ourobse}
\end{equation}

Our goal now is to show that this expectation value can always be reproduced within our local-realistic framework.
To this end, our first task would be to determine the correct topological space of the corresponding EPR elements of
reality. This would be the space composed of all possible measurement results, and our task would be to determine the
correct topology of this space. Depending on how complicated the state ${|\Psi\rangle}$ is, in practice this could
be a formidable task. In the cases of Bell, Hardy, and GHZ states, however, we were able to determine this space
without difficulty. Thus there is no in-principle reason why it cannot always be determined in practice. In fact,
whether determinable in practice or not, all that is required for the locality condition of Bell to hold in general
is that this space is a topological space that satisfies the properties (\ref{propyone}) to (\ref{iden-grop}).
For the Bell, Hardy, and GHZ states the correct topological spaces do indeed turn out to be the spaces that satisfy
these properties, and, as we shall see, those are not exceptions.

To appreciate this fact, recall how we determined the topological spaces in our four examples
(cf. Eqs.${\,}$(\ref{ignore-1}), (\ref{ignore-2}), and (\ref{ignore-3})). Although we did not spell this out before,
in each case we began by noting that a Hilbert space in general is a topological vector space whose topology is
given by a norm. Then, by using the normalization condition on its elements we recognized---say, for the
two-level system---that the corresponding Hilbert space has the topology of a 3-sphere. Then, recalling the
EPR argument from Section II, we recognized that there is a one-to-one correspondence between the points of this
3-sphere and the corresponding EPR elements of reality for the system. The 3-sphere, however, is a topological
group. That is to say, it is a topological space that is also a group, satisfying the properties (\ref{propyone})
to (\ref{iden-grop}). And any Hilbert space---being a vector space---is also a topological group, with addition of
its vectors as the group operation. Moreover, at least in the case of the 3-sphere, the multiplicative group of
the EPR elements of reality is homomorphic to the additive group of the corresponding Hilbert space. That is to
say, if ${|\,\Psi_{\bf a}\,\rangle}$ and ${|\,\Psi_{\bf b}\,\rangle}$ are two vectors in ${\cal H}$ and
${{\mathscr A}_{\bf a}}$ and ${{\mathscr B}_{\bf b}}$ are two points of ${S^3}$, then there exist a homomorphism
${h:{\cal H}\rightarrow{S^3}}$ such that
\begin{equation}
h(\,|\,\Psi_{\bf a}\,\rangle+|\,\Psi_{\bf b}\,\rangle\,)\,=\,
h(\,|\,\Psi_{\bf a}\,\rangle\,)\;h(\,|\,\Psi_{\bf b}\,\rangle\,)\,=\,
{\mathscr A}_{\bf a}\,{\mathscr B}_{\bf b}\,\in\,{S^3}.\label{grouphomo}
\end{equation}
Thus---since any such group homomorphism is designed to preserve the group structure---we see that in this case the
group properties (\ref{propyone}) to (\ref{iden-grop}) of ${S^3}$ (apart from non-commutativity)
are inherited from the group properties of ${\cal H}$ itself.
%(viewed as an additive group). 
Similar observations hold also for the cases of the three and four-level systems,
except that the 7-sphere of these cases is not a group, but a topological quasi-group${{}^4}$. Nevertheless, even in
this case the basic properties (\ref{propyone}) to (\ref{iden-grop}) continue to hold, since they are
inherited from the additive group of the corresponding Hilbert spaces themselves.

Generalizing from these lessons, we now identify the topological spaces of the EPR elements of reality for arbitrary
quantum systems as follows. If ${|\,\Psi_{\bf a}\,\rangle}$ and ${|\,\Psi_{\bf b}\,\rangle}$ are two vectors in a Hilbert
space ${\cal H}$ representing a quantum system, and ${{\mathbb A}_{\bf a}}$ and ${{\mathbb B}_{\bf b}}$ are two points
of a topological space ${\Omega}$, then ${\Omega}$ is the topological space of the corresponding EPR elements of reality
if there exists a morphism ${m:{\cal H}\rightarrow{\Omega}}$ (in the concrete category of topological spaces) such that
\begin{equation}
m(\,|\,\Psi_{\bf a}\,\rangle+|\,\Psi_{\bf b}\,\rangle\,)\,=\,
m(\,|\,\Psi_{\bf a}\,\rangle\,)\;m(\,|\,\Psi_{\bf b}\,\rangle\,)\,=\,
{\mathbb A}_{\bf a}\,{\mathbb B}_{\bf b}\,\in\,{\Omega}\,.
\end{equation}
We need not go into the technical details of the category theory to understand the basic properties of this morphism
(but here is a convenient reference:${\;}$\ocite{Category}). All we need to know for our purposes is that,
(1) given a Hilbert space ${\cal H}$, one can always find a topological space ${\Omega}$ and a morphism
${m:{\cal H}\rightarrow{\Omega}}$ such that the above equation is satisfied;${\;}$and (2) the points of the so-defined
topological space---whether it is a group or not---would necessarily satisfy the condition (\ref{propyone}). That is
to say, since by definition any Hilbert space remains closed under addition of its vectors, the space ${\Omega}$
will necessarily be a topological space closed under multiplication of its points. In other words, the space of the EPR
elements of reality (or measurement results) identified by the above morphism will automatically satisfy the locality
condition of Bell. In particular, if---with the intention of reproducing (\ref{ourobse})---we consider
a collection of local maps
\begin{equation}
{\mathbb A}_{\bf a}(\lambda): {\rm I\!R}^3\!\times\Lambda\longrightarrow {\Omega_{\scriptscriptstyle {\cal H}}}, \;\;\;
{\mathbb B}_{\bf b}(\lambda): {\rm I\!R}^3\!\times\Lambda\longrightarrow {\Omega_{\scriptscriptstyle {\cal H}}}, \;\;\;
{\mathbb C}_{\bf c}(\lambda): {\rm I\!R}^3\!\times\Lambda\longrightarrow {\Omega_{\scriptscriptstyle {\cal H}}}, \;\;\;
{\mathbb D}_{\bf d}(\lambda): {\rm I\!R}^3\!\times\Lambda\longrightarrow {\Omega_{\scriptscriptstyle {\cal H}}},
\,\;\dots\,,\label{maps-s3ewm}
\end{equation}
then the joint beable
${({\mathbb A}_{\bf a}\,{\mathbb B}_{\bf b}\,{\mathbb C}_{\bf c}\,{\mathbb D}_{\bf d}\dots)(\lambda)}$
corresponding to the operator ${{\cal\widehat O}({\bf a},\,{\bf b},\,{\bf c},\,{\bf d},\dots)}$ can
always be factorized${\;}$as
\begin{equation}
({\mathbb A}_{\bf a}\,{\mathbb B}_{\bf b}\,{\mathbb C}_{\bf c}\,{\mathbb D}_{\bf d}\,\dots\,)(\lambda)\,=\,
{\mathbb A}_{\bf a}(\lambda)\,{\mathbb B}_{\bf b}(\lambda)\,{\mathbb C}_{\bf c}(\lambda)\,
{\mathbb D}_{\bf d}(\lambda)\,\dots\;\in\,\Omega_{\scriptscriptstyle {\cal H}}\,,\label{lity}
\end{equation}
\vspace{-0.75cm}
${\!\!}$such that
\begin{equation}
[\,{\mathbb A}_{\bf a}(\lambda)\,{\mathbb B}_{\bf b}(\lambda)\,{\mathbb C}_{\bf c}(\lambda)\,
{\mathbb D}_{\bf d}(\lambda)\,\dots\,]\,:\,\Omega_{\scriptscriptstyle {\cal H}}\times\,\Omega_{\scriptscriptstyle {\cal H}}
\times\,\Omega_{\scriptscriptstyle {\cal H}}\times\,\Omega_{\scriptscriptstyle {\cal H}}\,\dots\,
\longrightarrow\,\Omega_{\scriptscriptstyle {\cal H}}\,\ni\,({\mathbb A}_{\bf a}\,
{\mathbb B}_{\bf b}\,{\mathbb C}_{\bf c}\,{\mathbb D}_{\bf d}\,\dots\,)(\lambda)\,.
\end{equation}
It is now straightforward to calculate the local-realistic counterpart of the quantum mechanical prediction
(\ref{ourobse}) as
\begin{equation}
{\cal E}_{{\!}_{L.R.}}({\bf a},\,{\bf b},\,{\bf c},\,{\bf d},\,\dots\,)\,=\int_{\Lambda}
({\mathbb A}_{\bf a}\,{\mathbb B}_{\bf b}\,{\mathbb C}_{\bf c}\,{\mathbb D}_{\bf d}\,\dots\,)(\lambda)
\,\;d\rho(\lambda)\,=\int_{\Lambda}
{\mathbb A}_{\bf a}(\lambda)\,{\mathbb B}_{\bf b}(\lambda)\,{\mathbb C}_{\bf c}(\lambda)
\,{\mathbb D}_{\bf d}(\lambda)\,\dots\,\;d\rho(\lambda)\,.\label{prob-f}
\end{equation}

Suppose now the space ${\Omega_{\scriptscriptstyle {\cal H}}}$ is a ${k}$-dimensional space,
say ${\Omega^k}$. Then this space can be embedded into
the space${\;{\rm I\!R}^{k+1}}$, say by a map ${i:\Omega^k\hookrightarrow {{\rm I\!R}^{k+1}}}$, and normalized to unity,
so that the points of its set can be enumerated as binary numbers, ${\pm\,1}$, just as in the cases of the spheres ${S^k}$.
And again in analogy with the spheres (cf. Eqs.${\,}$(\ref{21}) and (\ref{q-mappq})), the integrand of the above equation
can be represented in terms of the parameters of the embedding space ${{\rm I\!R}^{k+1}}$ as
\begin{equation}
{\mathbb A}_{\bf a}(\lambda)\,{\mathbb B}_{\bf b}(\lambda)\,{\mathbb C}_{\bf c}(\lambda)
\,{\mathbb D}_{\bf d}(\lambda)\,\dots\,=\,
f({\bf a},\,{\bf b},\,{\bf c},\,{\bf d},\,\dots\,)\,+\,
{\mathbb P}_{\bf N}(\lambda)\,\;g\,({\bf a},\,{\bf b},\,{\bf c},\,{\bf d},\,\dots\,)\,.
\end{equation}
Here ${f({\bf a},\,{\bf b},\,{\bf c},\,{\bf d},\,\dots\,)}$ and ${g\,({\bf a},\,{\bf b},\,{\bf c},\,{\bf d},\,\dots\,)}$
are scalar functions in ${\rm I\!R}$, ${{\mathbb P}_{\bf N}(\lambda)=\pm\,1}$ is a binary point on the ``equator''
${\Omega^{k-1}}$
of the space ${\Omega^k}$, and ${{\bf N}({\bf a},\,{\bf b},\,{\bf c},\,{\bf d},\,\dots\,)}$ is a
${k}$-dimensional vector in ${{\rm I\!R}^{k}}$. Note that both sides of this equation simply represent a point of the
space ${\Omega^k}$, which is either ${+1}$ or ${-1}$, but in two different parameterizations. The left hand side is
parameterized in terms of the vectors ${{\bf a},\,{\bf b},\,{\bf c},\,{\bf d},}$ etc.${\;}$in ${{\rm I\!R}^{3}}$,
whereas the right hand side is parameterized in terms of the coordinates of the embedding space ${{\rm I\!R}^{k+1}}$.
That such a decomposition of a point of ${\Omega^k}$ can always be achieved is clear enough from the
elementary vector analysis, but it is also amply exemplified by the four examples we discussed in the previous
sections. What may be less obvious is the fact that one can always choose local maps${\;}$(\ref{maps-s3ewm})
so that the scalar functions ${f({\bf a},\,{\bf b},\,{\bf c},\,{\bf d},\,\dots\,)}$ and
${g\,({\bf a},\,{\bf b},\,{\bf c},\,{\bf d},\,\dots\,)}$ of the above decomposition are rendered independent of
the complete state ${\lambda}$, and
the 3-vector ${\bf n}$ in ${{\rm I\!R}^{3}}$ contained in ${\bf N\in{\rm I\!R}^{k}}$ is rendered exclusive to
the experimental directions ${{\bf a},\,{\bf b},\,{\bf c},\,{\bf d},}$ etc.${\;}$in ${{\rm I\!R}^{3}}$.
In other words, for any arbitrary quantum state ${|\Psi\rangle}$ and a self-adjoint operator
${{\cal\widehat O}({\bf a},\,{\bf b},\,{\bf c},\,{\bf d},\,\dots\,)}$, one can always choose the local maps
(\ref{maps-s3ewm}) representing the EPR elements of reality such that the expectation functional
(\ref{prob-f}) is simplified to the canonical form exemplified by the Eqs.${\,}$(\ref{q-prob-q}),
(\ref{firsgonee}), (\ref{expeconepoint}), and (\ref{notrealonepoint}):
\begin{equation}
{\cal E}_{{\!}_{L.R.}}({\bf a},\,{\bf b},\,{\bf c},\,{\bf d},\,\dots\,)\,=\,
f({\bf a},\,{\bf b},\,{\bf c},\,{\bf d},\,\dots\,)\int_{\Lambda}d\rho(\lambda)
\;\,+\;\,g\,({\bf a},\,{\bf b},\,{\bf c},\,{\bf d},\,\dots\,)\int_{\Lambda}\,{\mathbb P}_{\bf N}(\lambda)
\,\;d\rho(\lambda)\,. \label{generic-prob-q}
\end{equation}
Moreover, since ${{\mathbb P}_{\bf N}(\lambda)}$ is a binary point on ${\Omega^{k-1}}$ about the
direction ${\bf N}$ exclusive to the experimental directions, the second term on the right hand side of this
equation---just as in our explicit examples---will necessarily vanish for more than one reason. Consequently, for
any normalized distribution ${\rho(\lambda)}$, the above equation can always be reduced to
\begin{equation}
{\cal E}_{{\!}_{L.R.}}({\bf a},\,{\bf b},\,{\bf c},\,{\bf d},\,\dots\,)\,=\,
f({\bf a},\,{\bf b},\,{\bf c},\,{\bf d},\,\dots\,)\,.
\label{eric-prob-q}
\end{equation}
Thus, regardless of what the complete state ${\lambda}$ or its distribution ${\rho(\lambda)}$ is, and without having
to use any averaging procedure, we can always reproduce the quantum mechanical expectation value (\ref{ourobse}) for
any state ${|\Psi\rangle}$, provided we make judicious choices for the maps (\ref{maps-s3ewm})---based on the symmetries
and asymmetries of the system---to ensure that
\begin{equation}
f({\bf a},\,{\bf b},\,{\bf c},\,{\bf d},\,\dots\,)\,=\,
\langle\Psi|\;{\cal\widehat O}({\bf a},\,{\bf b},\,{\bf c},\,{\bf d},\,\dots\,)\,|\Psi\rangle\,.\label{ourobsenewtk}
\end{equation}
For example, for the four-particle GHZ state we were able to identify the correct points of a 7-sphere to ensure that
\begin{equation}
f({\bf n}_1,\,{\bf n}_2,\,{\bf n}_3,\,{\bf n}_4)\,=\,
\cos\theta_1\,\cos\theta_2\,\cos\theta_3\,\cos\theta_4\,-\,\sin\theta_1\,
\sin\theta_2\,\sin\theta_3\,\sin\theta_4\,\cos\,\left(\,\phi_1\,+\,\phi_2\,-\,\phi_3\,-\,\phi_4\,\right)\,.
\end{equation}

Of course, this example, as well as the examples of the three-particle GHZ state, the Hardy State, and the Bell state,
are all trivial examples compared to how nontrivial a general quantum state can be. Hence, to follow through the above
local-realistic framework in general would clearly be a formidable task. But it {\it can} be followed trough in
principle, for any quantum state ${|\Psi\rangle}$, and then the result would be a {\it bona fide} local-realistic
completion of that${\;}$quantum state, without requiring explicit knowledge of the corresponding complete state.
Thus, as suspected all along by Einstein, Podolsky, and Rosen \ocite{EPR}\ocite{Einstein-1948}, quantum mechanical
stochasticity and entanglement are both entirely dispensable. And although any future theory of physics may well
have to respect contextuality in general \ocite{Contextual}, it need not respect {\it remote} contextuality, for
we have just shown that the latter can be understood as a purely classical, topological effect.

\section{Concluding Remarks}

When a bizarre result occurs at the end of an experiment, one usually suspects that a systematic error has slipped-in
during the course of its execution. Occurrence of a systematic error in theoretical work is rather rare, but that is
precisely what has happened in the case of Bell's theorem. What is astonishing, however, is that, despite the fact that
the error in this case is in plain sight---in the very first equation of Bell's paper---it remains unrecognized, even
after being exposed explicitly \ocite{Christian}\ocite{Further}\ocite{experiment}. To be sure, the error is considerably
obscured by the probabilistic
narratives of Bell's theorem, but it becomes plainly obvious as soon as one leaves behind the
insularity of these narratives, goes back to the original EPR argument, and recognizes the correct topological structure
of the EPR elements of reality. It then becomes obvious that a belief in Bell's theorem is no better than the beliefs
of those Linelanders \ocite{Flatland} who were unaware of the second dimension (let alone the third). Once the
``Sphereland'' is glimpsed and the fictitiousness of the ``Lineland'' is recognized, however, it only takes a few
lessons in elementary topology to rectify Bell's error. Without justification,
Bell presumed incorrect topology for the EPR elements of reality, and thus forfeited his game from the start. The same is
true of all of the variants and spinoffs of Bell's theorem,
such as, for example, the GHZ or Hardy type theorems. With complete disregard for topology,
all such theorems represent EPR elements of reality as points of the real line, and consequently begin their
arguments by blissfully presuming local maps such as
\begin{equation}
{A}_{\bf a}(\lambda): {\rm I\!R}^3\!\times\Lambda\longrightarrow{\cal I}\subseteq{\rm I\!R}\,,\;\;\;{\rm with}\;\;\;
({A}_{\bf a}\,{B}_{\bf b})(\lambda):\,{\cal I}\times\,{\cal I}\,\longrightarrow\,{\cal I}\,.
\end{equation}
But these maps are a pure fiction. The EPR elements of reality have nothing whatsoever to do with the points of the real
line. They are not ``lined up'', as it were, forming a real line. More precisely, the topology of the EPR elements of
realty is anything but the order topology of the real line${{}^1}$. Hence what is usually called ``local realism'' in the
literature is a pure fiction. And what are usually called ``classical correlations'' have nothing whatsoever to do with
the classical reality. These are straw men, erected just to be knocked off. As is evident from the Hopf fibrations
of ${S^3}$ and ${S^7}$ \ocite{experiment}\ocite{7-Hopf}, classical reality has far deeper topological structure than
what Bell theorists seem to recognize. You may search the vast literature on Bell's theorem in vain, but you will not
find the slightest sensitivity to this deep structure. If, in an {\it ad hoc} manner, this structure is replaced with
the fictitious structure of the real line, then Bell's theorem follows; but why would any physicist be interested in
such a fictitious theorem? Put differently: just as Newtonians---with their Kantian
commitments to ``non-locality''---believed that spacetime is always flat and rigid regardless of what matter fields
are present, Bell theorists believe that EPR elements of reality
are always lined up as points of the real line, regardless of which system is being considered. But it is
abundantly clear from the argument of EPR that the elements of reality they argued for are ordered as points of a 2-sphere
(in the case of the singlet state), not lined up as the real line. More generally, in the cases considered by Hardy and
GHZ, the elements of reality are ordered as points of a 3 or 7-sphere. And since both of these spheres remain as closed
under multiplication as the real line, they respect the locality condition of Bell just as strictly as the real line.
Consequently, any physically meaningful Bell type theorem ought to begin by representing the EPR elements of reality
by local maps such as
\begin{equation}
{\mathscr A}_{\bf a}(\lambda): {\rm I\!R}^3\!\times\Lambda\longrightarrow S^3\,,\;\;\;{\rm with}\;\;\;
({\mathscr A}_{\bf a}\,{\mathscr B}_{\bf b})(\lambda):\,S^3\times\,S^3\,\longrightarrow\,S^3.
\end{equation}
And instead of comparing the quantum correlations with the correlations between the points of the real line, it
ought to be comparing them with the correlations between the points of a 3 or 7-sphere, as we have done in this paper
and elsewhere \ocite{Christian}\ocite{Further}\ocite{experiment}. When this is done correctly, no incompatibility
between the predictions of quantum mechanics and local realism arises. In fact, as we demonstrated above,
an exact, deterministic, local, and realistic model for the EPR correlations is easy to construct, with
natural extensions to the cases of rotationally non-invariant entangled states considered by GHZ and Hardy.
In particular, within our local-realistic framework we are able to reproduce
\begin{enumerate}
\item[(1)] the {\it exact} quantum mechanical expectation value for the singlet state:
${{\cal E}({\bf a},\,{\bf b})\,=\,-\,{\bf a}\cdot{\bf b}}$\,;
\item[(2)] the {\it exact} violations of Bell-CHSH inequalities:
${\,-\,2\sqrt{2}\,\;\leq\;{\cal E}({\bf a},\,{\bf b})+{\cal E}({\bf a},\,{\bf b'})+
{\cal E}({\bf a'},\,{\bf b})-{\cal E}({\bf a'},\,{\bf b'})\;\leq\;+\,2\sqrt{2}}$\;;
\item[(3)] all sixteen predictions of the Hardy state, such as
${\;\,\langle\Psi_{\bf z}\,|\,{\bf a'},\,+\rangle_1\,\otimes\,|{\bf b}\,,\,+\rangle_2\,=\,0}$\,,
\item[]${\,\;\;\;\;\;\;\;\;
\;\;\;\;\;\;\;\;\;\;\;\;\;\;\;\;\;\;\;\;\;\;\;\;\;\;\;\;\;\;\;\;\;\;\;\;
\;\;\;\;\;\;\;\;\;\;\;\;\;\;\;\;\;\;\;\;\;\;\;\;\;\;\;\;\;\;\;\;\;\;\;\;
\langle\Psi_{\bf z}\,|\,{\bf a}\,,\,+\rangle_1\,\otimes\,|{\bf b'},\,+\rangle_2\,=\,0}$\,,
\item[]${\,\;\;\;\;\;\;\;\;\;\;\;\;\;\;\;\;\;\;\;\;\;\;\;\;\;\;\;\;\;\;\;\;\;
\;\;\;\;\;\;\;\;\;\;\;\;\;\;\;\;\;\;\;\;\;\;\;\;\;\;\;\;\;\;\;\;\;\;\;\;
\;\;\;\;\;\;\;\;\;\;\;\langle\Psi_{\bf z}\,|\,{\bf a}\,,\,-\rangle_1\,\otimes\,|{\bf b}\,,\,-\rangle_2\;=\,0}$\,,
\item[]${\,\;\;\;\;\;\;\;\;\;\;\;\;\;\;\;\;\;\;\;\;\;\;\;\;\;\;\;\;\;\;\;\;\;
\;\;\;\;\;\;\;\;\;\;\;\;\;\;\;\;\;\;\;\;\;\;\;\;\;\;\;\;\;\;\;\;\;\;\;\;
\;\;\;{\rm but}\,\;\;\langle\Psi_{\bf z}\,|\,{\bf a'},\,+\rangle_1\,\otimes\,|{\bf b'},\,+\rangle_2\,=\,
\frac{\,\sin\theta\,\cos^2\theta}{\sqrt{1\,+\,\cos^2\theta\,}\,}\,\not=\,0}$\,;
\item[(4)] the {\it exact} quantum mechanical expectation value for the three-particle GHZ state:
\item[]
${\;\;\;\;\;\;{\cal E}({\bf n}_1,\,{\bf n}_2,\,{\bf n}_3)\,=\,
\cos\alpha\,\cos\theta_1\,\cos\theta_2\,\cos\theta_3\,+\,\sin\alpha\,\sin\theta_1\,
\sin\theta_2\,\sin\theta_3\,\cos\,\left(\,\phi_1\,+\,\phi_2\,+\,\phi_3\,+\,\delta\,\right)}$\,; \;and
\item[(5)] the {\it exact} quantum mechanical expectation value for the four-particle GHZ state:
\item[]
${\;\;\;\;\;\;{\cal E}({\bf n}_1,\,{\bf n}_2,\,{\bf n}_3,\,{\bf n}_4)\,=\,
\cos\theta_1\,\cos\theta_2\,\cos\theta_3\,\cos\theta_4\,-\,\sin\theta_1\,
\sin\theta_2\,\sin\theta_3\,\sin\theta_4\,\cos\,\left(\,\phi_1\,+\,\phi_2\,-\,\phi_3\,-\,\phi_4\,\right)}$.
\end{enumerate}
These results should not be surprising to anyone who is familiar with classical mechanics. In particular, they would
not have been surprising to Hamilton, Hopf, or Grassmann. For the first three of these results simply express
classical correlations among the points of a 3-sphere, whereas the last two express those among the points of a 7-sphere.
In fact, since we have been able to derive these results without specifying what the complete state ${\lambda}$ is or
the distribution ${\rho(\lambda)}$ is, and without employing any averaging procedure, they show that the correlations
in each of the above cases are purely topological effects. They are simply the local, realistic, and deterministic
correlations among certain points of the two topological spaces---namely, ${S^3}$ and ${S^7}$. In particular, they have
nothing to do with the monstrosities like ``non-locality'' or ``non-reality.'' What the quantum mechanical descriptions
of the Bell, GHZ, and Hardy states are providing us is nothing more than a useful shortcut to the local
correlations between the EPR elements of reality. Thus the conclusion of EPR stands as firmly today as it did in 1935.
In fact, there is nothing special about the above four entangled states. As we saw in the previous section, exact
local-realistic completion of any arbitrary entangled state can be easily achieved within our framework, at least in
principle, by employing local maps of the from
\begin{equation}
{\mathbb A}_{\bf a}(\lambda): {\rm I\!R}^3\!\times\Lambda\longrightarrow \Omega_{\scriptscriptstyle {\cal H}}\,
\;\;\;{\rm and}\;\;\;({\mathbb A}_{\bf a}\,{\mathbb B}_{\bf b})(\lambda):\,\Omega_{\scriptscriptstyle {\cal H}}\times\,
\Omega_{\scriptscriptstyle {\cal H}}\,\longrightarrow\,\Omega_{\scriptscriptstyle {\cal H}}\,,
\end{equation}
where ${\Omega_{\scriptscriptstyle {\cal H}}}$
is an arbitrary topological space composed of possible measurement results. Consequently, it is clear that
all any Bell type theorem can possibly prove is that correlations between the points of a sphere---or a more general
topological space---cannot be reproduced by the correlations between the points of the real line. But any eighteenth
century student would have known that long before the advent of quantum mechanics. Indeed, it is hard to find claims
in physics more vacuous than those of Bell-type theorems. In fact, they sound a lot like the impossibility
claims of those Linelanders \ocite{Flatland}, whose entire totality was the real line. But what may be
impossible for the Linelanders is a child's play for the Flatlanders. And what may be miraculous to the Flatlanders
is a mundane tedium to the Spacelanders. Hence the follies of a one-dimensional world can hardly be taken as proofs of
universal impossibility.

\acknowledgments

I am grateful to Abner Shimony, Lucien Hardy, David Hestenes, Michael Seevinck, Carsten Held,
Derek Abbott, and James John Bohannon for their comments on the arguments put forward
in Refs.${\,}$\ocite{Christian}, \ocite{Further}, and \ocite{experiment}.

\vspace{-0.16cm}

\renewcommand{\bibnumfmt}[1]{\textrm{[#1]}}

\end{document}